\documentclass[structabstract]{aa}  
\usepackage{graphicx}
\usepackage{txfonts}
\usepackage{natbib}
\bibpunct{(}{)}{;}{a}{}{,} 
\begin{document}
   \authorrunning{Jakub\'{\i}k et al.}  \titlerunning{Simulations of
   the formation of ice giants} \title{Considerations on the accretion
   of Uranus and Neptune by mutual collisions of planetary embryos in
   the vicinity of Jupiter and Saturn}


   \author{M. Jakub\'{\i}k \inst{1}
          \and
          A. Morbidelli \inst{2}
          \and
          L. Neslu\v{s}an \inst{1}
          \and
          R. Brasser \inst{2,3}
          }

   \institute{Astronomical Institute, Slovak Academy of Science,
             05960 Tatransk\'{a} Lomnica, Slovakia\\
             \email{mjakubik@ta3.sk; ne@ta3.sk}
             \and
             D\'{e}partement Cassiop\'{e}e, University of Nice - Sophia Antipolis, CNRS, Observatoire de la C\^{o}te d'Azur; Nice,
France
	     \and 
	Institute for Astronomy and Astrophysics, Academia Sinica, P.O. Box 
23-141, Taipei 106, Taiwan
\\
             \email{morby@oca.eu; brasser\underline{~}astro@yahoo.com}
             }

   \date{Received July 12, 2011; accepted February 9, 2012}

  \abstract
   {Modeling the formation of the ice giants Uranus and Neptune is a
long-lasting problem in planetary science. Due to gas-drag,
collisional damping, and resonant shepherding, the planetary embryos
repel the planetesimals away from their reach and thus they stop growing 
(Levison et al. 2010).
This problem persists independently of whether the accretion took
place at the current locations of the ice giants or closer to the Sun.}
   {Instead of trying to push the runaway/oligarchic growth of planetary
embryos up to 10$-$15 Earth masses, we envision the possibility that the
planetesimal disk could generate a system of planetary embryos of only
1$-$3 Earth masses. Then we investigate whether these embryos could have
collided with each other and grown enough to reach the masses of current
Uranus and Neptune.}
   {We perform several series of numerical simulations. The
dynamics of a considered set of embryos is influenced by the presence
of Jupiter and Saturn, assumed to be fully formed on non-migrating orbits 
in 2:3 resonance, and
gravitational interactions with the disk of gas.}
   {Our results point to two major problems. First, there is typically
a large difference in mass between the first and the second most massive
core formed and retained beyond Saturn. Second, in many simulations 
the final planetary system has more than two objects beyond Saturn. 
The growth of a major planet from a system of embryos requires strong damping of eccentricities and
inclinations from the disk of gas. But strong damping also favors embryos
and cores to find a stable resonant configuration, so that systems
with more than two surviving objects are found. In addition to these problems,
 in order to have substantial mutual accretion among embryos,
it is necessary to assume that the surface density of the gas was several times
higher than that of the minimum-mass solar nebula. However this contrasts with
the common idea that Uranus and Neptune formed in a gas-starving disk,
which is suggested by the relatively small amount of hydrogen and helium
contained in the atmospheres of these planets.}
   {Only  one of our simulations ``by chance'' successfully reproduced the structure of the
outer Solar System. However, our work has the merit to point out that models
of formation of Uranus and Neptune have non-trivial problems, which cannot be ignored 
and have to be addressed in future work.}


   \maketitle

\section{Introduction}

The accretion of Uranus and Neptune is a long-standing problem in
planetary science. \cite{safronov_1969} was the first to point out that the
accretion of these two planets from a planetesimal disk at their
current locations would have taken implausibly long timescales. This
problem was confirmed by \cite{2001Icar..153..224L} using modern numerical simulations.  
\citet{2004ApJ...614..497G,2004ARA&A..42..549G} claimed that the
in-situ formation of Uranus and Neptune could have been possible in a
planetesimal disk strongly dominated by collisional damping. This
claim, however, is not correct because, as showed by 
\cite{2007Icar..189..196L}, planetary cores in a disk with strong collisional
damping simply open gaps in the planetesimal distribution around their
own orbits and stop accreting.

There is now a consolidated view that the giant planets were closer to
each other in the past (probably all within $12\,$AU from the Sun) and
that they moved to their current orbits after their formation
\citep{1984Icar...58..109F,1993Natur.365..819M,1995AJ....110..420M,
1999AJ....117.3041H,1999Natur.402..635T,2005Natur.435..459T,2007AJ....134.1790M,
2010ApJ...716.1323B}. Thus, it is no longer necessary to
construct a model capable of explaining the formation of Uranus and
Neptune at their current, remote locations.

Forming Uranus and Neptune within 12$-$15$\,$AU from the Sun is in
principle easier than forming them at 20$-$30$\,$AU because the density
of solid material was probably higher and the dynamical timescale
(i.e. the orbital period) was shorter. However, forming 10$-$15 Earth
mass (M$_{\oplus}$) cores from a planetesimal disk turns out to be
difficult at {\it any} location. In fact, \cite{2010AJ....139.1297L} 
showed that, when planetary embryos achieve a mass of 1$-$3$\,$M$_{\oplus}$, they
tend to scatter the remaining planetesimals away rather than accreting
them. With the help of gas-drag, collisional damping and resonant
shepherding, the embryos repel the planetesimals away from their
reach and thus they stop growing.

In this paper we explore another possible venue for the formation of
Uranus and Neptune. Instead of trying to push the runaway/oligarchic
growth of planetary embryos up to 10$-$15$\,$M$_{\oplus}$, we envision
the possibility that the planetesimal disk could generate a system of
planetary embryos of only 1$-$3$\,$M$_{\oplus}$; then we investigate with
numerical simulations whether these embryos could have collided with
each other because they converged at specific orbital radii where
their radial migration in the gas-disk was stopped by the presence of
Jupiter and Saturn. This growth mode by giant collisions would explain 
in the natural way the large obliquities of Uranus and Neptune.

More specifically, our scenario is based on the consideration that
Jupiter and Saturn presumably got caught in their mutual 2:3 mean
motion resonance (MMR), which prevented them from migrating further
towards the Sun. Instead, after being trapped in the 2:3 MMR, Jupiter
and Saturn either migrated outwards or stayed roughly at steady
locations \citep{2001MNRAS.320L..55M,2007Icar..191..158M,
2007AJ....134.1790M,2008A&A...483..633P}.
To date, this is the only explanation we have for why our giant planets 
did not migrate permanently into the inner Solar System.

The presence of Jupiter and Saturn on orbits not migrating towards the
Sun would have acted as an obstacle against the inward Type-I migration
\citep{1980ApJ...241..425G} of the planetary embryos from the outer 
Solar System. More precisely, any planetary embryo migrating towards 
the Sun would have been, sooner or later, trapped and halted 
in a mean motion resonance with Saturn.
Then, the accumulation of embryos in these resonances could in principle
have boosted their mutual accretion. This paper aims at investigating
this possibility with numerical simulations. 

We are aware of the new result according to which the real migration
of planetary embryos is very different from the classical Type-I
migration envisioned in ideal, isothermal disks 
\citep{2006A&A...459L..17P,2008ApJ...678..483B,2008A&A...485..877P,
2008A&A...487L...9K,2011MNRAS.410..293P}.
In particular, in disks with realistic cooling times, migration is
expected to be outward in the inner part of the disk and inward in its
outer part \citep{2009A&A...497..869L}. This generates a region in between the
inner and the outer parts of the disk where Type-I migration is
basically inhibited. Planetary embryos are expected to be in/close to
this no-migration zone, which seems to invalidate our assumption that
embryos migrated towards the giant planets until they got captured in
resonances.

However, \cite{walsh_2011}, from constraints provided by 
the terrestrial planet system and the asteroid belt, argued strongly
that Jupiter and Saturn migrated outwards over a range of several
AUs. What is important for our purposes is the relative motion of
Jupiter/Saturn and the embryos. It does not really matter whether
Jupiter/Saturn are on fixed orbits and the embryos tend to migrate
towards the Sun, or the embryos do not migrate while Jupiter/Saturn
move outwards. In fact, in both cases the embryos approach the giant
planets until they are captured in MMRs, which may act like a privileged
site for embryo clustering and mutual accretion. Thus, in our
simulations, for simplicity we assume that Jupiter/Saturn are on
non-migrating resonant orbits while the embryos are affected by inward
migration, with different migration speeds from one simulation to
another. This migration speed can be interpreted as an actual inward
migration speed of the embryos (most likely reduced relative to the
classical Type-I migration speed in iso-thermal disks), or the outward
migration speed of Jupiter/Saturn or a combination of the two.

Two caveats related to our work need to be stated up-front. First,
our study assumes that Jupiter and Saturn are fully formed, while the
accretion of these planets is by itself an unsolved problem that we
do not address here. This may sound strange. However, there is a
consensus that Uranus and Neptune formed after Jupiter and Saturn,
because they did not accrete nearly as much of gas. This leads to two
considerations: (a) Jupiter and Saturn existed already when Uranus and
Neptune formed, so that the former should/may have influenced the
accretion process of the latter and (b) whatever mechanism allowed the
formation of Jupiter (obviously not the presence of a pre-existing
giant planet!), it did not work for Uranus and Neptune, otherwise they 
would have formed nearly at the same time as Jupiter. 

The second caveat is that, because our model is based on migration of
planets and embryos in a gas-disk, the study should be performed with
hydro-dynamical simulations. These simulations, though, are too slow
to treat the evolution of tens of bodies for a few millions of years,
as we need to perform in this study. The paper by \cite{2008A&A...478..929M} 
is, to date, the only attempt to study 
the accretion of the cores of giant planets using hydro-dynamical 
simulations, and it shows all the limitations of this technique. 
Thus, for this study we use N-body simulations, 
with artificial forces exerted onto the embryos to
mimic the migration and tidal damping forces exerted from the disk.
This approach also has its own limitations as it does not account for
indirect mutual perturbations that the embryos may exert onto each other
through the modifications that they induce in the density distribution
of the gas-disk.

The structure of this paper is as follows. In Sect.~2, we explain our
simulation methods. In Sect.~3 we illustrate some basic ingredients of
the dynamics of embryos and giant planets. In particular, we discuss
the concept of resonance trapping with Saturn, resonant trapping in
mutual embryo-embryo resonances, resonance loading, onset of a global
dynamical instability and possible mutual accretion. Just for
illustrative purposes, we do this by introducing one embryo at the
time at a large distance from Saturn, even though this is NOT how we
think the real evolution proceeded. Then, in Sect.~4 we move to more
``realistic'' simulations, where multiple $3\,$M$_{\oplus}$ embryos
are introduced at the beginning of the simulation over the
10$-$35$\,$AU range. We test the dependence of the results on the
total amount of gas in the disk, the inward migration rate and the
total number of embryos. Having realized that several embryos are lost
by having close encounters with Jupiter and Saturn which either eject
them onto distant orbits or inject them into the inner Solar System,
in Sect.~5 we show how the coorbital corotation torque exerted at the
edge of Saturn's gap can act like a planet trap \citep{2006ApJ...642..478M,
2008A&A...483..633P} and prevent large mass losses. In Sect.~6 we discuss
how the results depend on the initial mass of the embryos. In Sect.~7
we address the role of turbulence in the disk and Sect.~8 collects the
conclusions and considerations that we derive from this study.

We anticipate that our study is not ``successful'', in the sense that
our simulations do not typically lead to the formation of only two
planets with masses close to those of Uranus and Neptune. However it
shows interesting dynamical mechanisms and intriguing consequences that
will need to be addressed in detail in future studies, most likely using
hydro-dynamical simulations.

Note on nomenclature: hereafter we call embryos the objects of 
1-$3\,$M$_{\oplus}$ that our simulations start with and core any 
object that is formed during the simulation by merging at least two embryos.

\section{Simulation Methods}

For our simulations, we use the integration software {\it Symba}
developed in \cite{1998AJ....116.2067D}, that we modified 
in order to take into account the planet-gas gravitational interactions. 

\begin{figure}
\centering
   \includegraphics[width=6cm, angle=-90]{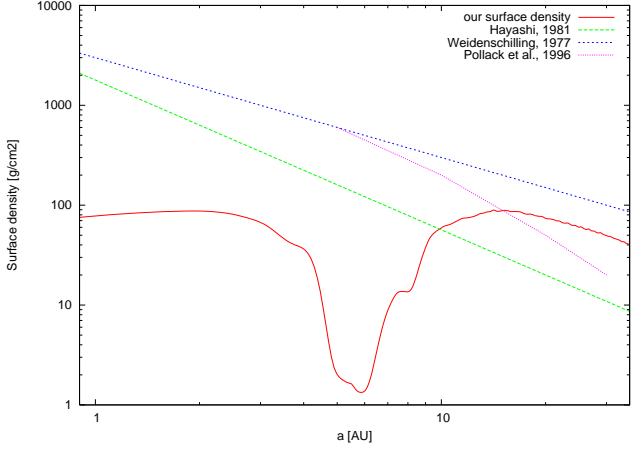}
\caption{Gas surface densities according to different works. The red
  curve shows the result of a hydro-dynamical simulation by 
\cite{2007Icar..191..158M}, with Jupiter and Saturn in the mutual 2:3
  resonance. This is the profile that we assume in this paper, 
  possibly scaled by a factor $f_d$. The blue line is
  from \cite{1977Ap&SS..51..153W} and is proportional to $1/r$. 
The green line is from \cite{1981PThPS..70...35H}
and is proportional to $1/r^{3/2}$. The magenta curve reports 
the amount of gas required in \cite{1996Icar..124...62P}
  model of giant planet accretion.}
\label{density}
\end{figure}

The gas density profile that we consider is taken from the
hydro-dynamical simulations of \cite{2007Icar..191..158M}, which
accounted for Jupiter and Saturn in their mutual 2:3 resonance; it is
shown in Fig.~\ref{density} (red curve). Notice the gap opened around
the position of Jupiter at $5.2\,$AU and the ``plateau'' at the right
hand-side of the gap, which is due to the presence of Saturn at
$\sim$$7\,$AU. The figure also compares this density profile to those of
three classical {\it Minimal Mass Solar Nebul\ae} (MMSN). In the
10$-$35$\,$AU range, the surface densities are comparable, within an order
of magnitude. In particular, our surface density falls in between the
estimates from \cite{1977Ap&SS..51..153W} and \cite{1981PThPS..70...35H}. 
Instead, inside of the orbit of Jupiter, the radial profile of our surface
density is flat and significantly lower than those expected
from the authors above. This is because the presence of Jupiter opens
a partial cavity inside its orbit, by limiting the flow of gas from
the outer part of the disk \citep[see][]{2007A&A...461.1173C}.
In \cite{2007Icar..191..158M} the considered disk was narrow, with 
an outer boundary at $35\,$AU. Here, we extend its surface density 
profile beyond $35\,$AU assuming a $r^{-3/2}$ radial decay.

The concept of MMSN was historically introduced assuming that giant
planet formation was 100\% efficient. In reality, there is growing
evidence that much more mass is needed to grow the giant planets, even
in the most optimistic scenarios \citep{2003Icar..161..431T}.
Therefore, in our simulations we multiply the assumed initial surface density
profile by a factor $f_d$, which will be specified from simulation to
simulation. All simulations are run for $5\,$My and for simplicity we assume that 
the gas density is constant over this simulated timescale. The role of the $5\,$My 
parameter will be discussed in Sect. \ref{formUN}. 
In our code, the surface density profile of the gas is used to compute
the migration and damping forces acting onto the embryos, namely the
so-called ``type-I torques''. The analytic formulae that we use are
those reported in \cite{2008A&A...482..677C}; they depend on the
local surface density of the disk and on the embryos' eccentricities
and inclinations. Because inward migration can be significantly
slower in realistic disks with radiative transfer than in ideal,
isothermal disks, we give ourselves the possibility of multiplying the
forces acting on the embryo's semi-major axes by a factor $1/f_{I}$,
which will be specified below for each simulation.

However, in Sect.~\ref{sectTrap}, we use a more sophisticated formula
for the radial migration torque $\Gamma$, which accounts for the
radial gradient of the surface density \citep{2010MNRAS.401.1950P,
2010ApJ...715L..68L}:

\begin{equation}
\Gamma=\Gamma_0 (-0.85 -\alpha -0.9\beta)\ ,
\label{torque}
\end{equation}
where $\Gamma_0=(M/h)^2\Sigma r^4\Omega^2$, $M$ is the mass of the
embryo relative to the Sun, $h$ is the scale-height of the disk 
normalized by the semimajor axis,
$\Sigma$ stands for the local surface density, and $\Omega$ is the
orbital frequency. In (\ref{torque}) 
\begin{equation}
\alpha=-{{d\log\Sigma}\over{d\log r}}\ ,\quad
\beta=-{{d\log T}\over{d\log r}}\ ,
\end{equation}
where $T$ is the local temperature of gas.

The radial migration torque in \cite{2008A&A...482..677C}, at zeroth 
order in eccentricity and inclination, corresponds to (\ref{torque}) for
$\beta=1$ and $\alpha=0$, i.e. it is valid for a flat disk with constant
scale height. The inclusion of the $\alpha$-dependence in (\ref{torque})
stops inward migration where $\alpha=-1.75$, that is where there is a
steep positive radial gradient of the surface density. With the surface
density profile shown in Fig.~\ref{density}, this happens at
$\sim$$10\,$AU. This location acts like a planet trap \citep{2006ApJ...642..478M}: 
an embryo migrating inwards from the outer disk, 
if not trapped in a mean motion resonance with Saturn, is ultimately trapped
at this location. 

In the simulations that mimic turbulent disks (see Sect.~\ref{turb}),
we simply apply stochastic torques to the embryos following the
recipe extensively described in \cite{2007Icar..188..522O}; see Sect.~2.2 of that
paper). The only difference is that the total number of Fourier modes
in the torque spectrum is not $m=50$, as in Ogihara et al. (2007), but
is $m=50/\log 4 \times log(r_{\rm out}/r_{\rm in})$ which, given
$r_{\rm out}=24\,$AU and $r_{\rm in}=8\,$AU in our case, makes
$m=40$. This functional form for $m(r_{\rm out},r_{\rm in})$ is
necessary in order to make the results independent of the simulated
size of the disk. In fact, if one used a fixed number of Fourier
modes, the effect of turbulence would be stronger in a narrow disk
than in an extended disk. With our recipe for the number of modes, the stochastic
migration of planetesimals observed in the full MRI simulations of
\cite{2010MNRAS.409..639N} is reproduced with a ``turbulent strength''
parameter $\gamma$ (see eq. 6 in \cite{2007Icar..188..522O}) equal 
to $3\times 10^{-3}$.

A final technical note concerns the treatment that we reserve to
Jupiter and Saturn. The migration of these two planets is a two-planet
Type-II-like process \citep{2001MNRAS.320L..55M,2007Icar..191..158M}, 
and therefore cannot be described with the Type-I torques reported above. 
As stated in the Introduction, we assume that
the disk parameters are such that Jupiter and Saturn do not migrate
(see \cite{2007Icar..191..158M}, for the identification of the
required conditions). Consequently, one could think that we should
apply no fictitious forces to these planets. However, if we did so,
the system would become unstable. In fact, as soon as an embryo is
trapped in a resonance with Saturn, it would push Saturn inwards and
in turn also Jupiter (because Jupiter and Saturn are locked in
resonance). This would also increase the orbital eccentricities of the
two major planets. This behavior is obviously artificial, because we
do not consider the forces exerted by the gas onto Jupiter and Saturn,
which would stabilize these planets against the perturbations from the
much smaller embryo. To circumvent this problem, we apply damping
forces to Jupiter and Saturn so that $(de/dt)/e=(di/dt)/i=10^{-5}$/y, together
with a torque that tends to restore the initial semi-major axes of
their orbits in $10^5$ years. Tests show that, with this recipe,
Jupiter and Saturn are stable under the effect of embryos piling up in
resonances and pushing inwards. Moreover the eccentricities of Jupiter
and Saturn attain finite but non-zero limit values. However strong scattering 
events can remove Jupiter and Saturn from their 2:3 resonance 
(see Sect.~\ref{formUN}).

\section{Basic dynamical mechanisms in mutual migration of embryos and
  giant planets}

To illustrate the interplay between migration, resonance trapping and
mutual scattering, in this section we do simple experiments, where we
introduce in the system one embryo at the time. The time-span between
the introduction of two successive embryos is not fixed: we let the
system relax to a stable configuration before introducing a new embryo
in the simulation. Each embryo has initially $3\,$M$_{\oplus}$.

\begin{figure}
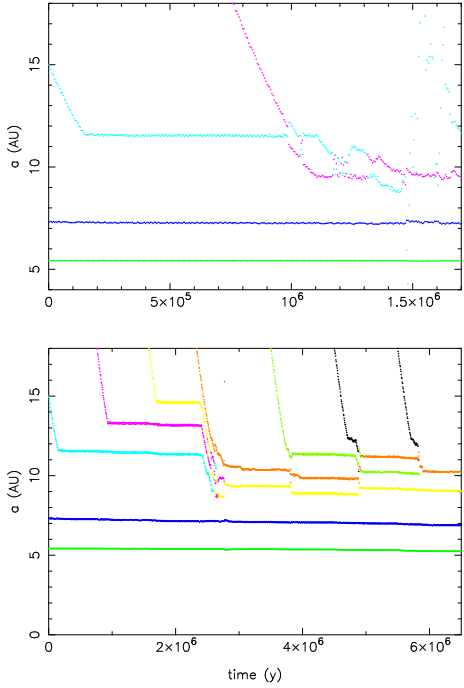

\centering{
   \includegraphics[width=6cm]{2emb.ps}}\\
\vspace*{0.5cm}
\centering{
   \includegraphics[width=6cm]{D2.ps}}
\caption{Simulations of the dynamical evolution of embryos in the disk
  outside of Saturn's orbit. Top: a case with parameters
  $f_d=f_I=1$. Bottom, a case with $f_d=f_I=2$. The green and blue
  curves at $\sim$$5.4$ and $\sim$$7.3\,$AU show the semi-major axis
  evolution of Jupiter and Saturn, respectively. The other curves of
  various colors illustrate the evolution of the semi-major axes of
  the embryos. See text for a description of the dynamical evolution
  and the accretion events.}
\label{2emb}
\end{figure}

We start by assuming the nominal values of the parameters ($f_d=1$,
$f_I=1$). The top panel of Fig.~\ref{2emb} shows the evolution. The
first embryo is introduced at the beginning of the simulation at
$15\,$AU. It migrates inwards (cyan dotted curve) until it is trapped in
resonance with Saturn (precisely the 1:2 MMR) at $t \sim 150,000\,$y.
Thus, it stops migrating. Resonance trapping converts the force acting
onto the semi-major axis into an eccentricity excitation. Thus, the
eccentricity of the embryo grows to about 0.07 but then it stops (not
shown in the figure). This happens because a balance is reached between the
eccentricity excitation from the resonance and the direct damping from
the disk. Thus, the three-planet system (Jupiter, Saturn and the embryo)
reaches a stable, invariant configuration at about $t=200,000\,$y.

At $t=750,000\,$y we introduce a second embryo in the system, initially
at $18\,$AU. This embryo also migrates inward (magenta dotted line). 
Because of the relatively large eccentricity of the cyan embryo, its outer 
MMRs are not stable enough to capture the incoming magenta embryo.
Thus the latter comes down to $\sim$$12\,$AU and
starts to have close encounters with the cyan embryo. The system becomes
unstable. The eccentricities and inclinations of the embryos become very
large, up to 0.6$-$0.7 and 10 degrees respectively. The cyan embryo has
even close encounters with Saturn and Jupiter. It is clear that this
phase of violent scattering is not very favorable for embryo-embryo
accretion.

We then present another simulation, where we increase the gas surface
density by a factor of two ($f_d=2$), which has the effect of
increasing both the eccentricity/inclination damping and the radial 
migration by the same factor. However, we also assume $f_I=2$, so
that the migration rate of the semi-major axis is in fact the same as
in the previous simulation. The result of this new simulation is
illustrated in the bottom panel of Fig.~\ref{2emb}. As one can see,
the magenta embryo is now trapped in resonance with the cyan embryo
(precisely, the 4:5 MMR). So, it also stops migrating and a stable
four-planet configuration is achieved. Given that the only difference
with respect to the previous simulation is the eccentricity damping,
this illustrates the crucial role of this parameter on the dynamics.

A third embryo (yellow) is introduced at $t=1.6\,$My. It also migrates
until it stops, trapped in the 6:7 MMR with the cyan embryo. The fully
resonant system is, again, stable. 

A fourth embryo is then released at $t=2.3\,$My (orange). Now there are
too many embryos to form a stable, resonant system. So, when the orange
embryo comes in, resonance locking is broken. All embryos move inward
and start to have encounters with each other. Because of the stronger
damping from the disk, the eccentricities and inclinations do not
become as large as in the previous experiment. Thus, the conditions are
more favorable for mutual accretion. In fact, at $t=2.64\,$My the
yellow and cyan embryos accrete each other. Arbitrarily, we
assume that it is the yellow embryo that survives, with twice its
original mass and the cyan embryo disappears. The system of embryos,
however, is still too excited to be stable. It stabilizes only after
the ejection of the magenta embryo at $t=2.9\,$My. The system is now
made of two embryos, the yellow and orange ones, in their mutual 6:7
MMR. The yellow embryo is in the 2:3 MMR with Saturn. 

We proceed the experiment by introducing a new embryo (green) which,
after a short phase of instability, ends in the 4:5 MMR with the orange
embryo and forces the orange and yellow embryos to go to smaller
heliocentric distances: the yellow embryo ends in the 5:7 MMR with
Saturn. Given the high-order resonances involved, the system is
close to an instability. 

Thus, when the next embryo (black) is introduced and moves inwards,
the system becomes unstable. The new crisis is solved with the yellow
embryo accreting the black one.  At the end of the instability phase,
the yellow embryo is back into the 2:3 MMR with Saturn and the three
surviving embryos are in resonance with each other. 

The final embryo is introduced at $5.5\,$My (black again). 
The new embryo generates a new phase of instability during which it
first collides with the orange embryo, then the orange embryo accretes
also the green one. Therefore, this simulation ends with two embryos,
each with a mass of $9\,$M$_{\oplus}$, in a stable resonant
configuration: the yellow embryo is in the 2:3 MMR with Saturn
and in the 6:5 MMR with the orange one. 

These experiments, as well as other similar ones that we do not
present for brevity (with $f_d=f_I=3, 5$ and $f_d=1$, $f_I=3$), show
well the importance of mutual resonances in the evolution of the
embryos. For individual masses equal to $3\,$M$_{\oplus}$, embryos are
easily trapped in mutual resonances if $f_I\ge 2$. If the embryos are
smaller, $f_I$ needs to be larger, because the mutual resonant torques
are weaker. A system of several embryos, all in resonance with each
other, can be stable; if this is the case, mutual accretion is not
possible. However, when the embryos are too numerous, resonances
cannot continue to hold the embryos on orbits well separated from each
other. The system eventually has to become unstable; collisions or
ejections then occur. Finally, when the number of embryos is reduced,
a new, stable resonant configuration is achieved again. This sequence
of events (resonance loading, global dynamical instability, accretion
or ejection) can repeat cyclically, as long as mass is added to the
system.  Eccentricity and inclination damping from the disk also plays
a pivotal role in the evolution of the system. If damping is weak,
eccentricities and inclinations can become large and scattering
dominates over mutual accretion; eventually embryos encounter Jupiter
or Saturn and are ejected from the system. If damping is large, the
most likely end-state of a dynamical instability is the mutual
accretion of embryos, which leads to the growth of massive planets.

Of course, the experiments presented in this section are not
realistic, as embryos are introduced one by one. They are just intended
to illustrate the basic dynamical mechanisms at play. 
In the next section, we present more realistic simulations, with all
embryos simultaneously present from  $t=0$.

\section{A first attempt to form Uranus and Neptune}
\label{formUN}

We present a series of simulations with 14 embryos originally distributed
in semi-major axis from 10 to $35\,$AU. The initial orbits have low
eccentricities ($e\sim10^{-2}$) and inclinations ($i\sim10^{-2}\,$rad)
relative to the common invariable plane of the system. The initial mass
of each embryo is assumed to be $3\,$M$_{\oplus}$ and the mutual orbital
separation among embryos is 5 mutual Hill radii. We run simulations with
$f_{d} = 1$, 3, and 6, which progressively increase the effects of
eccentricity/inclination damping. As well, we divide the speed of Type-I
migration, by the factor $f_{I} = 1$, 3, or 6. For each of these 9 
combinations of $f_{d}$ and $f_{I}$, we perform 10 simulations 
with embryos having different initial, randomly generated, sets of
position and velocity vectors (but which satisfy the above-mentioned
distribution properties). Every simulation is run for $5\,$My, 
a typical lifetime for the disk, during which we keep constant the density
of the gas (i.e. the values of $f_{d}$ and $f_{I}$).

\begin{figure}
 \centering
    \includegraphics[width=8cm]{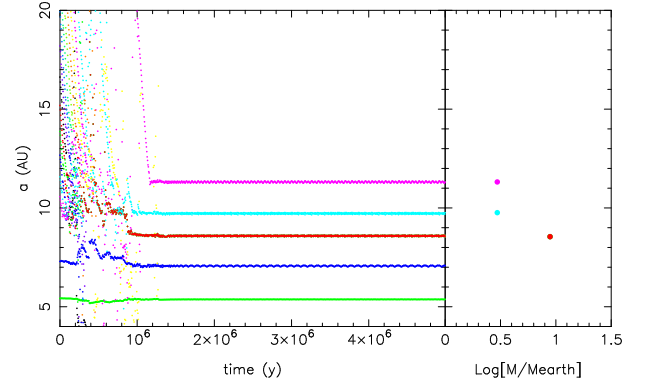}
\caption{
The evolution of the semi major axes of the embryos in simulation 
No. 7 with $f_{d} = f_{I} =3$. Each color represents a different
embryo. The system stabilizes in $\sim 1.2\,$My. 
The panel on the right shows the final masses of the embryos/cores, 
as a function of their semi major axes. 
}
\label{evol}
\end{figure}

Fig.~\ref{evol} shows evolution of the system for a simulation 
$f_{d} = f_{I} =3$. The embryos migrate towards Jupiter and Saturn 
and they have frequent encounters with each other,
which makes the evolution very chaotic. Several of them have also close 
encounters with Jupiter and Saturn and consequently are scattered onto 
distant elliptic orbits or into the inner Solar System. Through  
these scattering events and mutual collisions, the number of embryos beyond 
Saturn eventually decreases to 3 in $1.5\,$My and the surviving 
bodies find a stable resonant configuration. The most massive object has grown 
to $9\,$M$_{\oplus}$ whereas the two others have preserved their initial mass 
($3\,$M$_{\oplus}$). The three embryos that have been scattered into the inner 
Solar System eventually accrete with each other forming a 9 Earth-mass object.
Because the system remains ``frozen'', once a resonant stability is achieved, 
the lifetime of the gas-disk would not influence significantly the results. In the
case illustrated in this figure any lifetime longer than $1.5\,$My would produce 
the same final system. Instead, if the lifetime had been shorter than $1.5\,$My 
the system would not have reached the final state, leaving many embryos on eccentric 
and chaotic orbits.

\begin{figure*}[!ht]
\vspace{-0.5cm}
\centerline{\includegraphics[width=6cm, angle=-90]{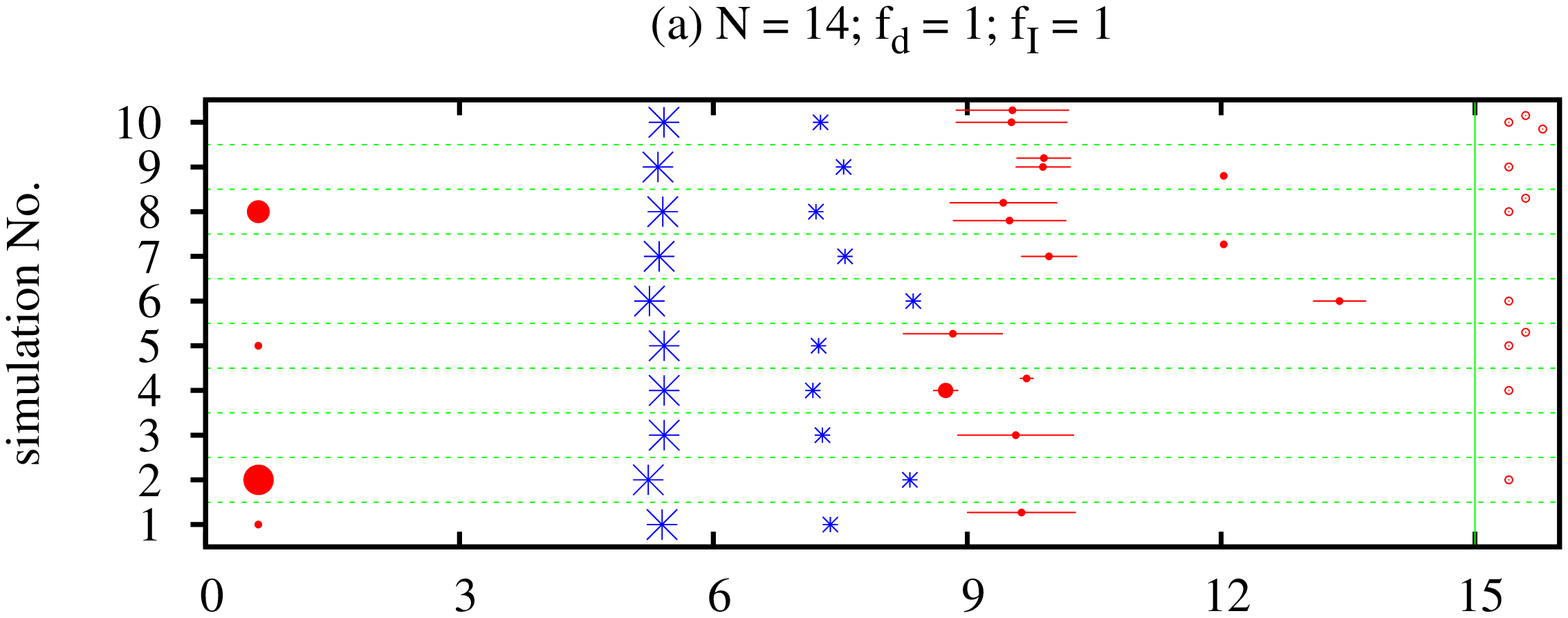}
    \includegraphics[width=6cm, angle=-90]{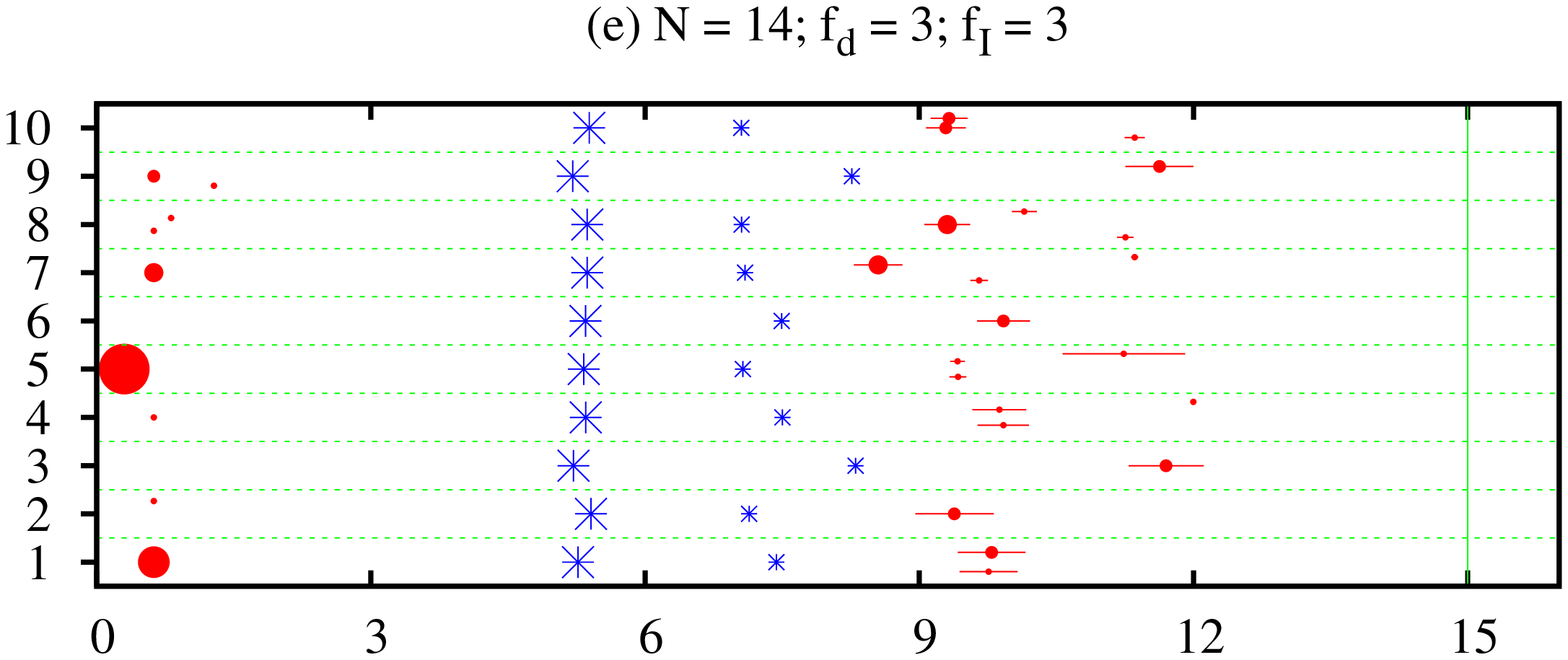}}
\vspace{-2.7cm}
\centerline{\includegraphics[width=6cm, angle=-90]{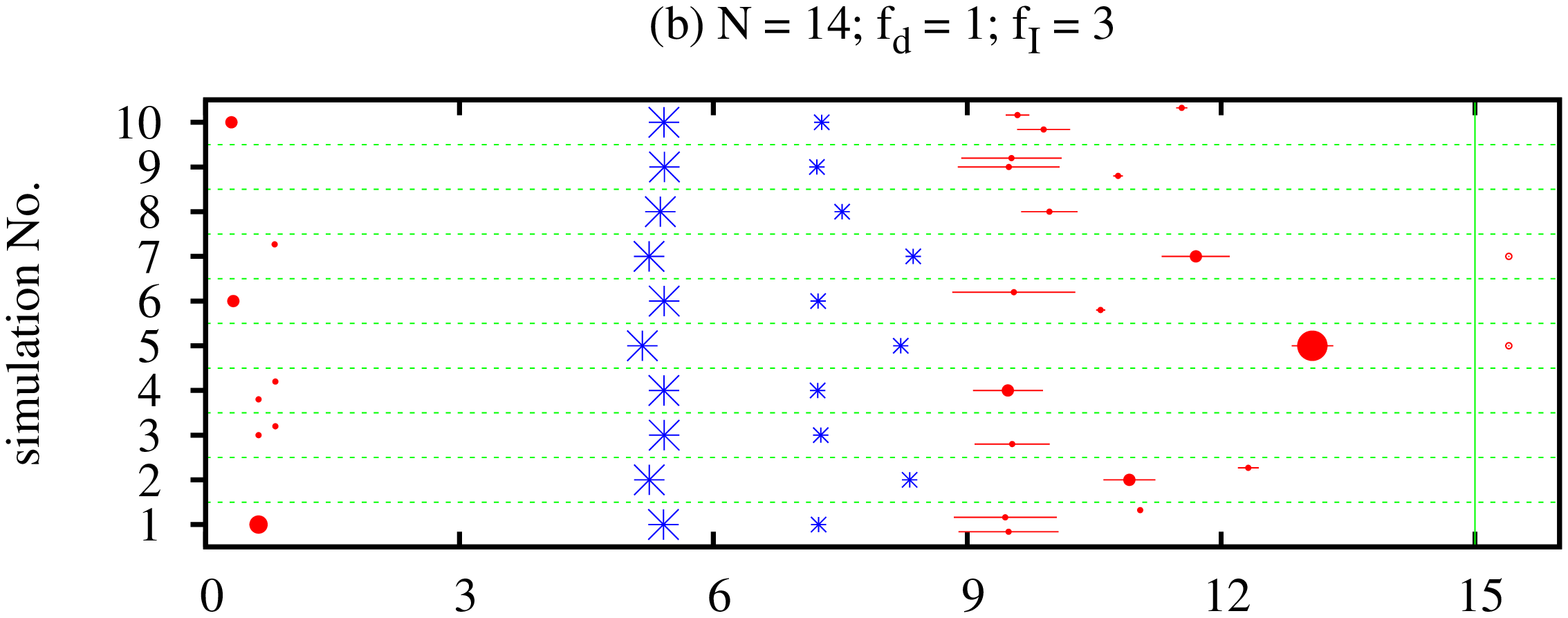}
    \includegraphics[width=6cm, angle=-90]{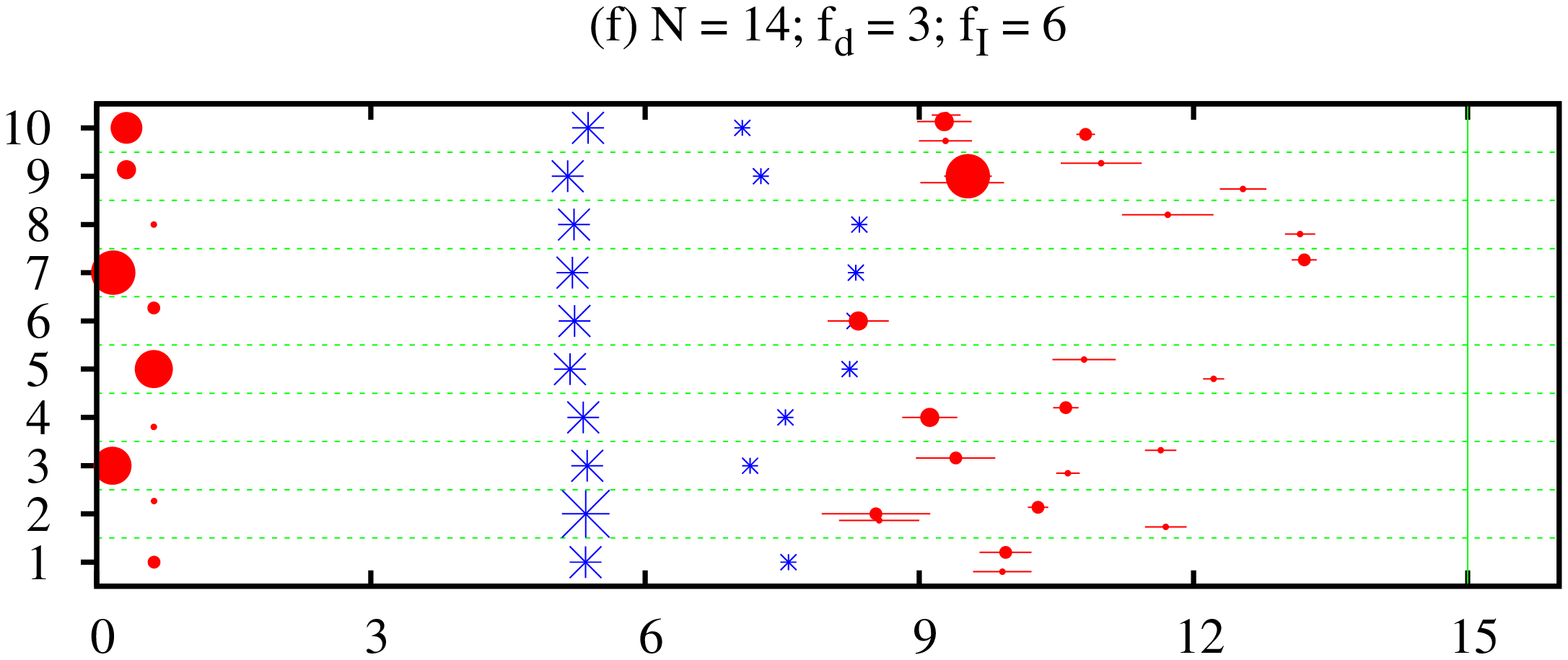}}
\vspace{-2.7cm}
\centerline{\includegraphics[width=6cm, angle=-90]{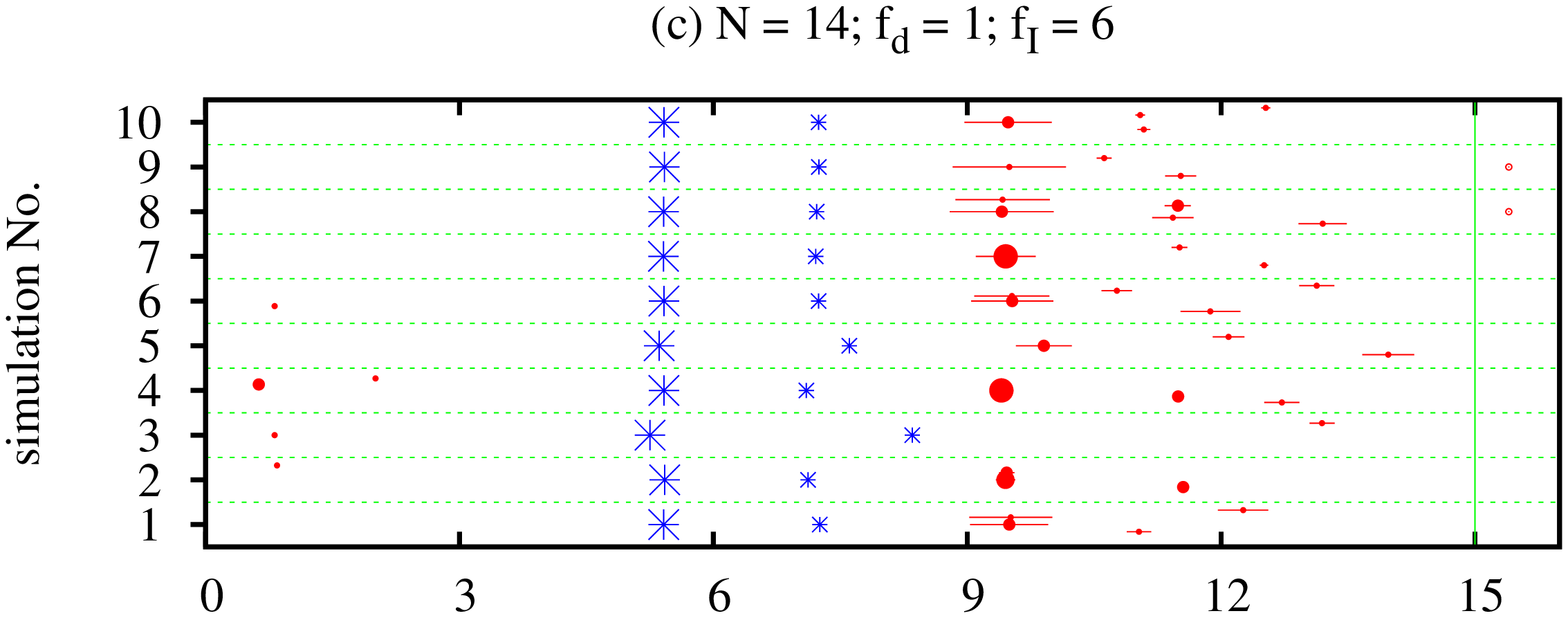}
    \includegraphics[width=6cm, angle=-90]{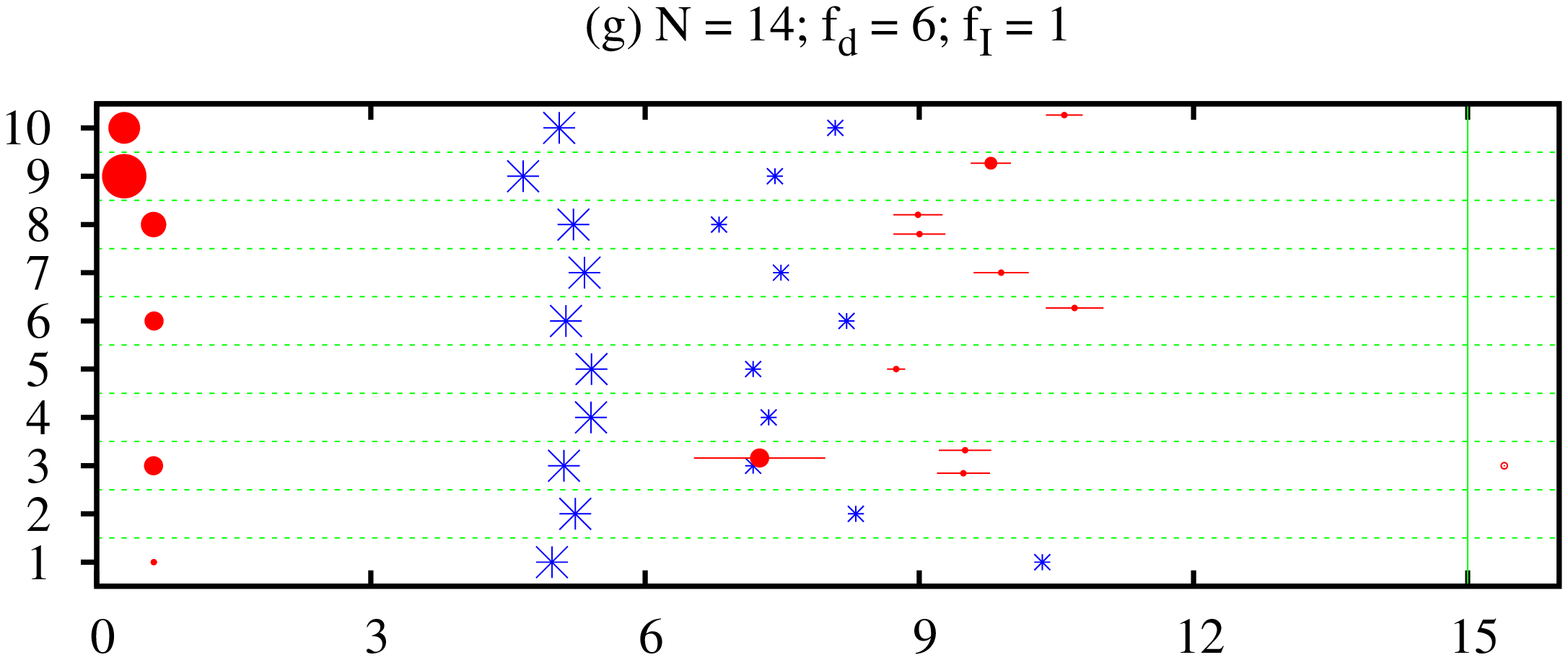}}
\vspace{-2.7cm}
\centerline{\includegraphics[width=6cm, angle=-90]{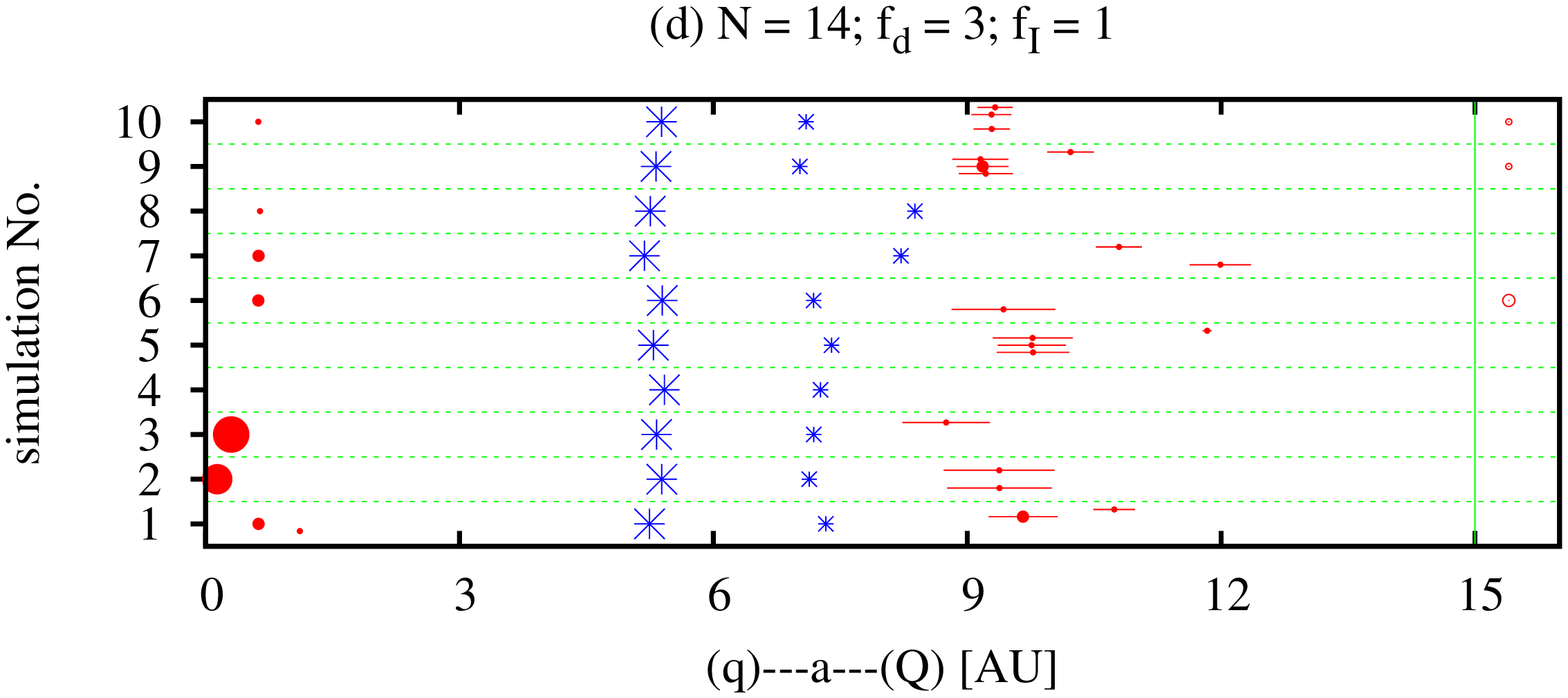}
    \includegraphics[width=6cm, angle=-90]{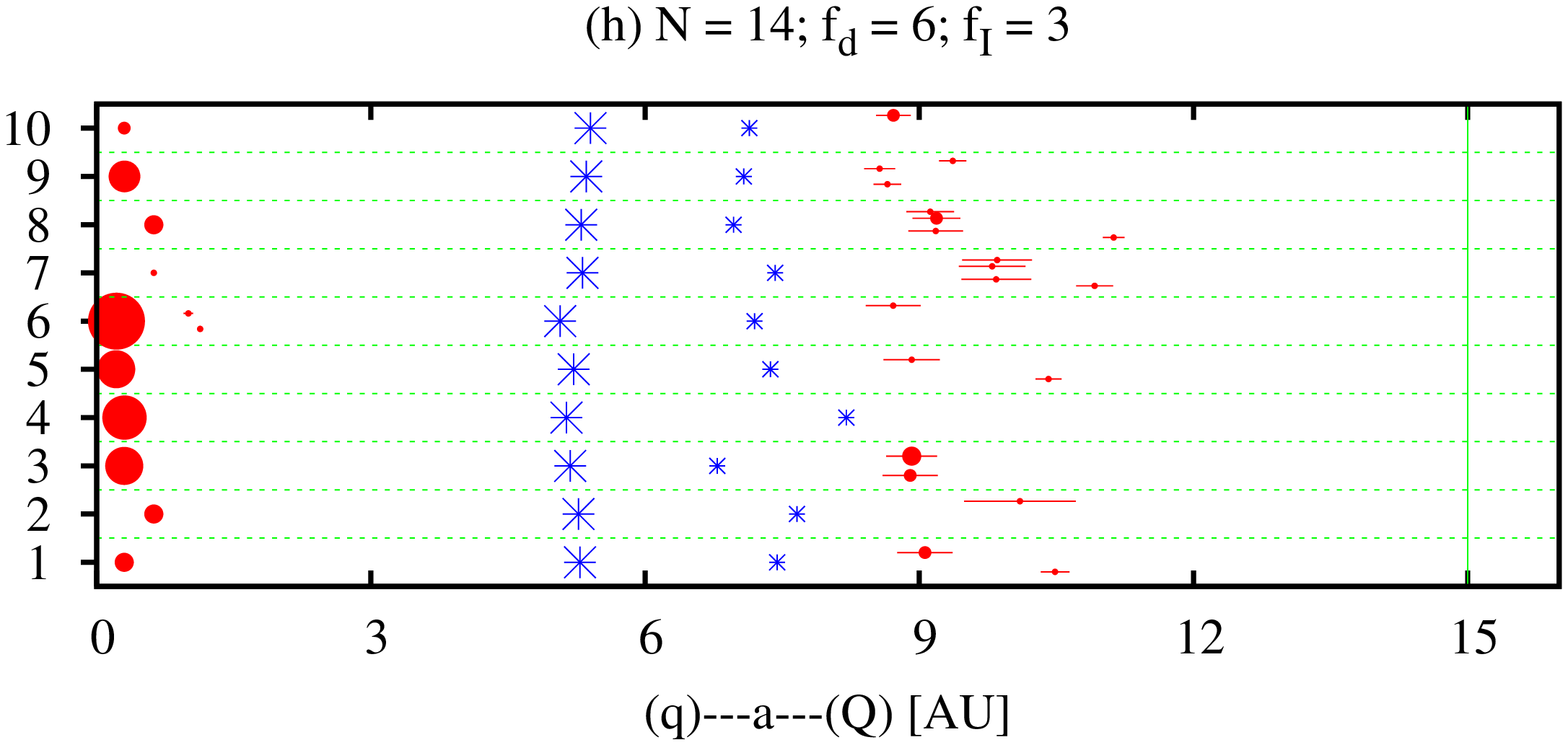}}
\vspace{-2.7cm}
\centerline{\hspace{-1cm}\includegraphics[width=6cm, angle=-90]{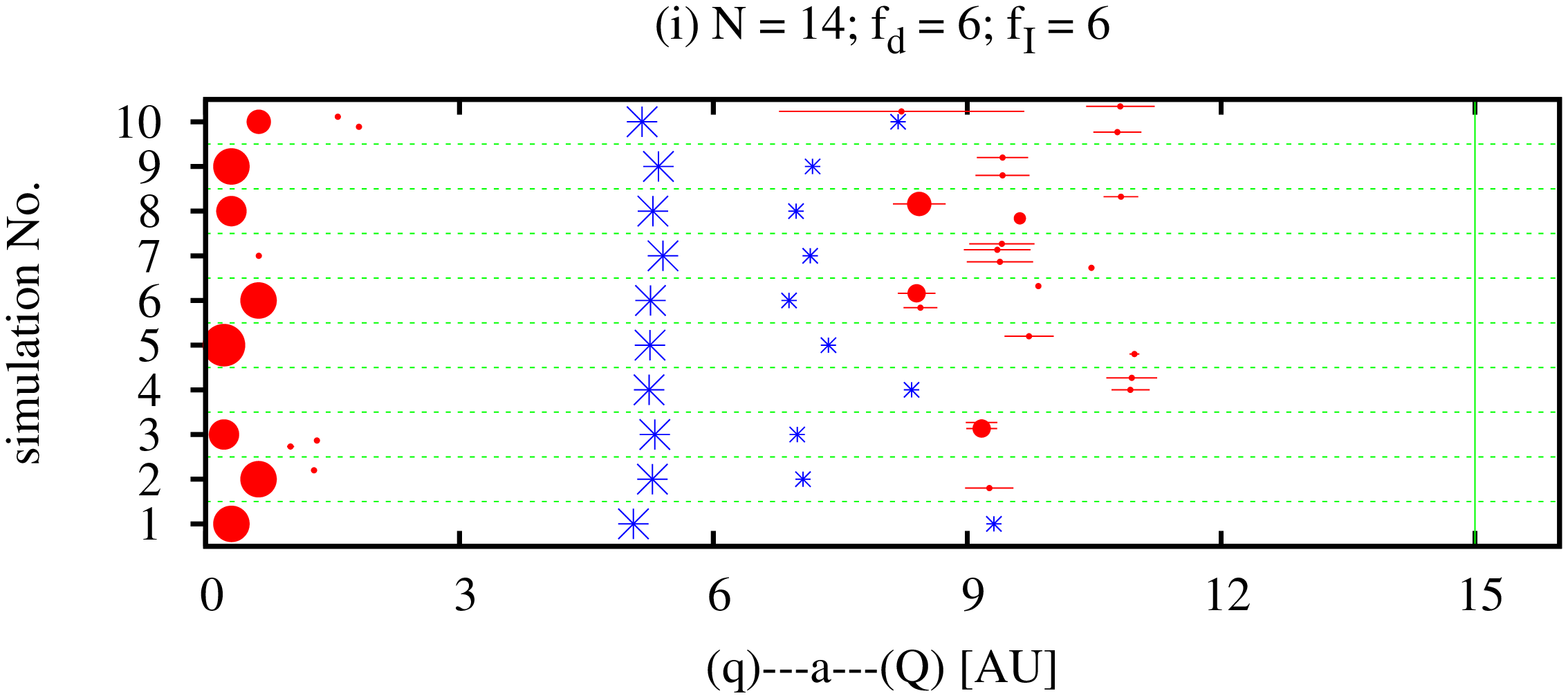}}
\vspace{-3.7cm}
\centerline{\includegraphics[width=6.5cm,angle=-90]{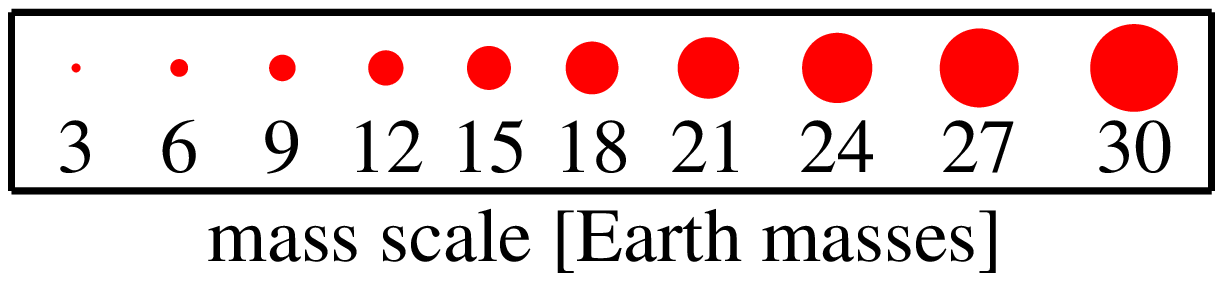}}
\vspace{-2cm}
\caption{The end-states (in $5\,$Myr) of the simulations starting
with 14 embryos, each with $3\,$M$_{\oplus}$. The simulations are
separated by horizontal dashed green lines. The surviving embryos/cores
are shown with filled red dot, whose size is proportional to the objects
mass. The scale is shown on the bottom panel. The red horizontal bar
shows the perihelion-aphelion excursion of these objects on their
eccentric orbits. The objects beyond $15\,$AU are plotted for simplicity
at $15.5\,$AU, beyond the vertical solid green line.
Jupiter and Saturn are shown as blue asterisks. The label on top of each
panel reports the number $N$ of embryos and the values of $f_I$ and $f_d$ adopted
in the simulations.}
\label{n14n28}
\end{figure*}

Fig. \ref{n14n28} shows the end-states of the simulations for all combinations of
$f_I = 1$, 3, 6  and $f_d = 1$, 3, 6.
 The number of bodies surviving beyond Saturn changes from 1 to 3 for one 
simulation to another. Similarly, the mass of the body in the inner 
Solar System changes from 3 to $27\,$M$_{\oplus}$. 
Notice that Saturn is not always in the same position relative to Jupiter, 
despite of the forces that we applied to keep these planets in the mutual 
2:3 resonance. In fact, in some cases Saturn moves out of the 2:3 
resonance while scattering an embryo inwards; then the forces drive it 
back towards Jupiter but Saturn is caught in a resonance other 
than 2:3 (mostly 1:2 or 1:3). This evolution depends on the magnitude 
of the migration forces that we impose. We have checked that, if we 
increase the strength of the migration forces by an order of magnitude, 
Saturn goes back to 2:3 resonance in all of the  simulations. Notice that 
the hydrodynamical simulations \citep{2008A&A...483..633P} show that in the MMSN 
disk Saturn always reaches 2:3 resonance with Jupiter.

\begin{figure*}[!ht]
\centerline{\includegraphics[width=6cm, angle=-90]{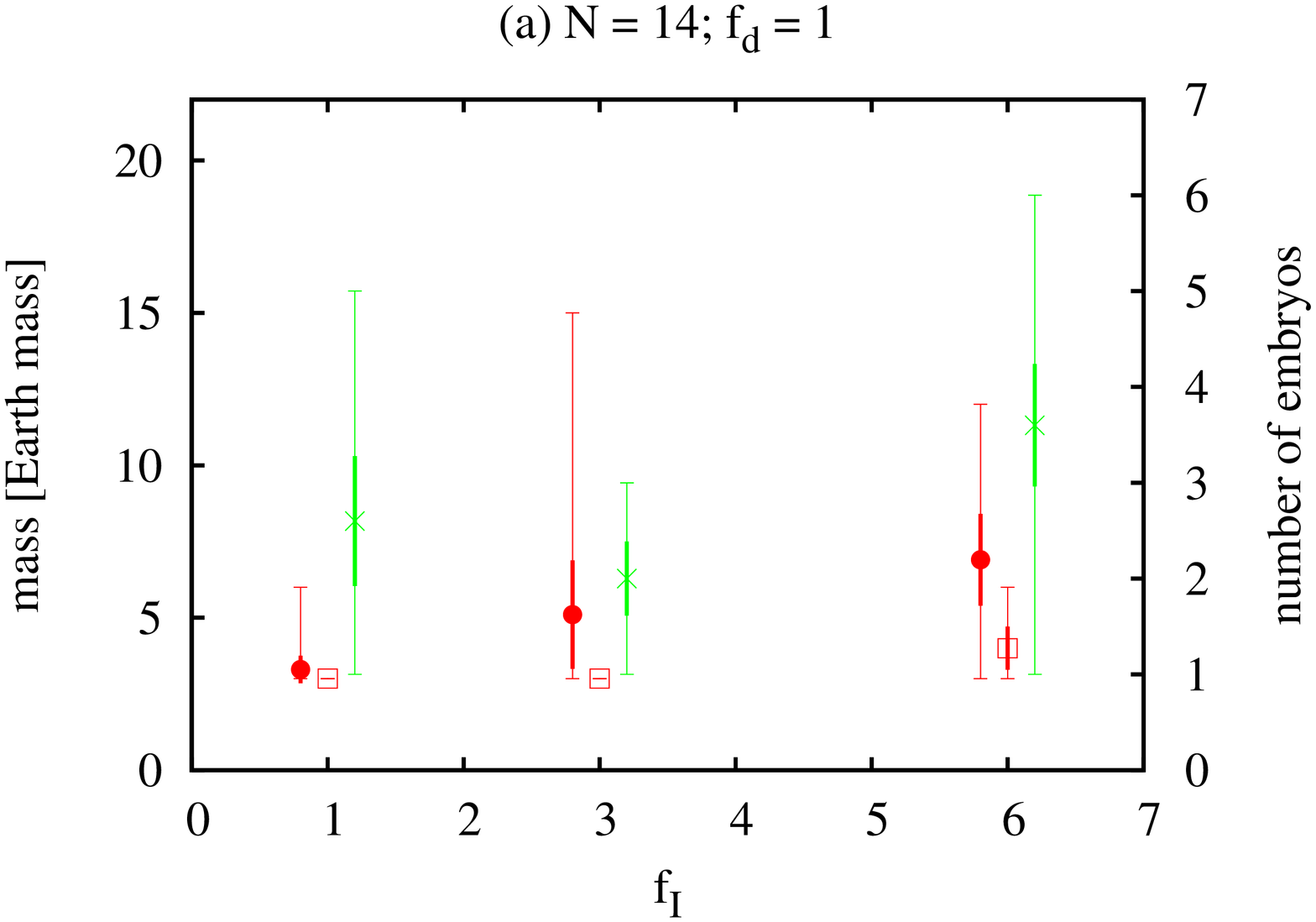}
    \includegraphics[width=6cm, angle=-90]{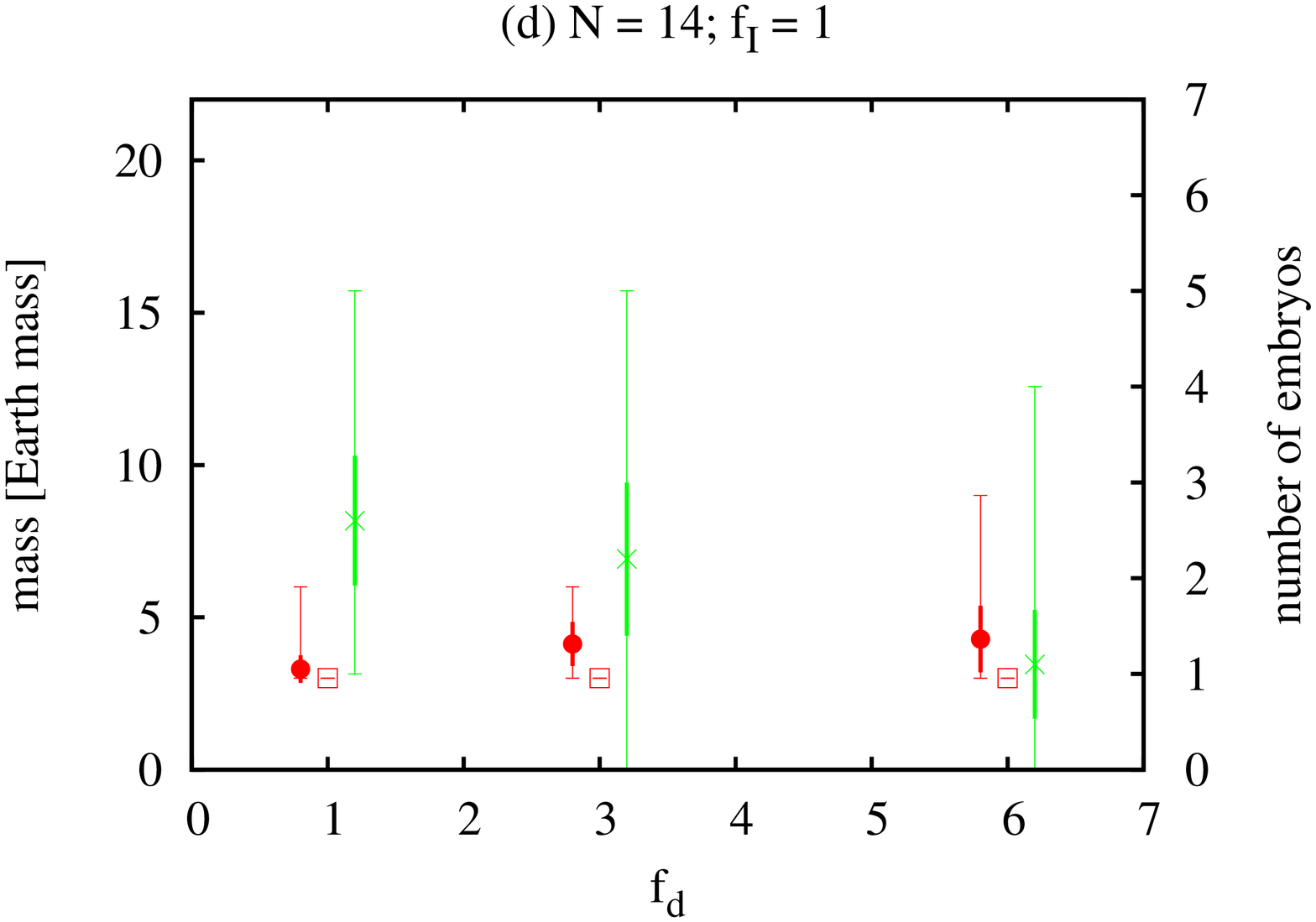}}
\centerline{\includegraphics[width=6cm, angle=-90]{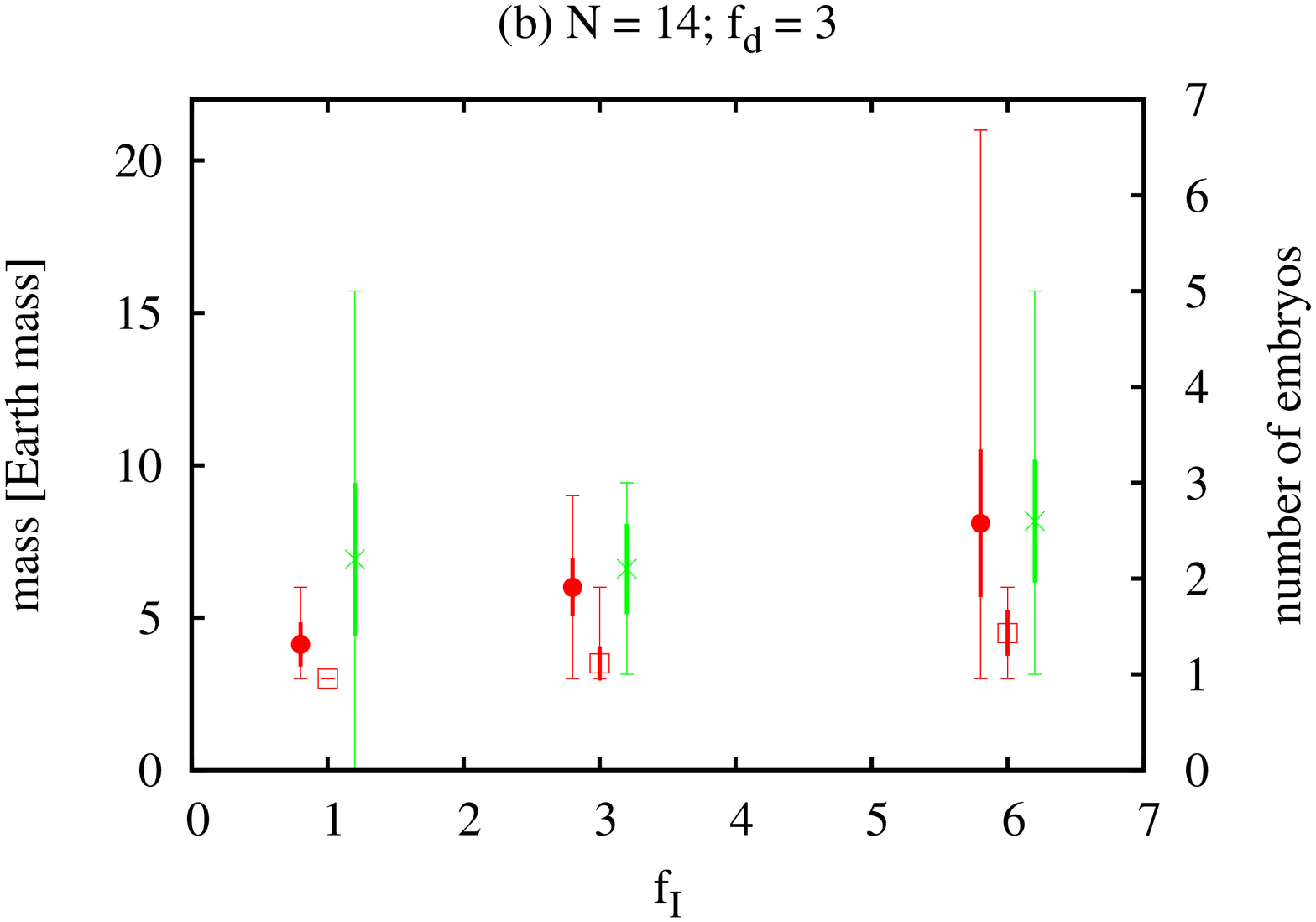}
    \includegraphics[width=6cm, angle=-90]{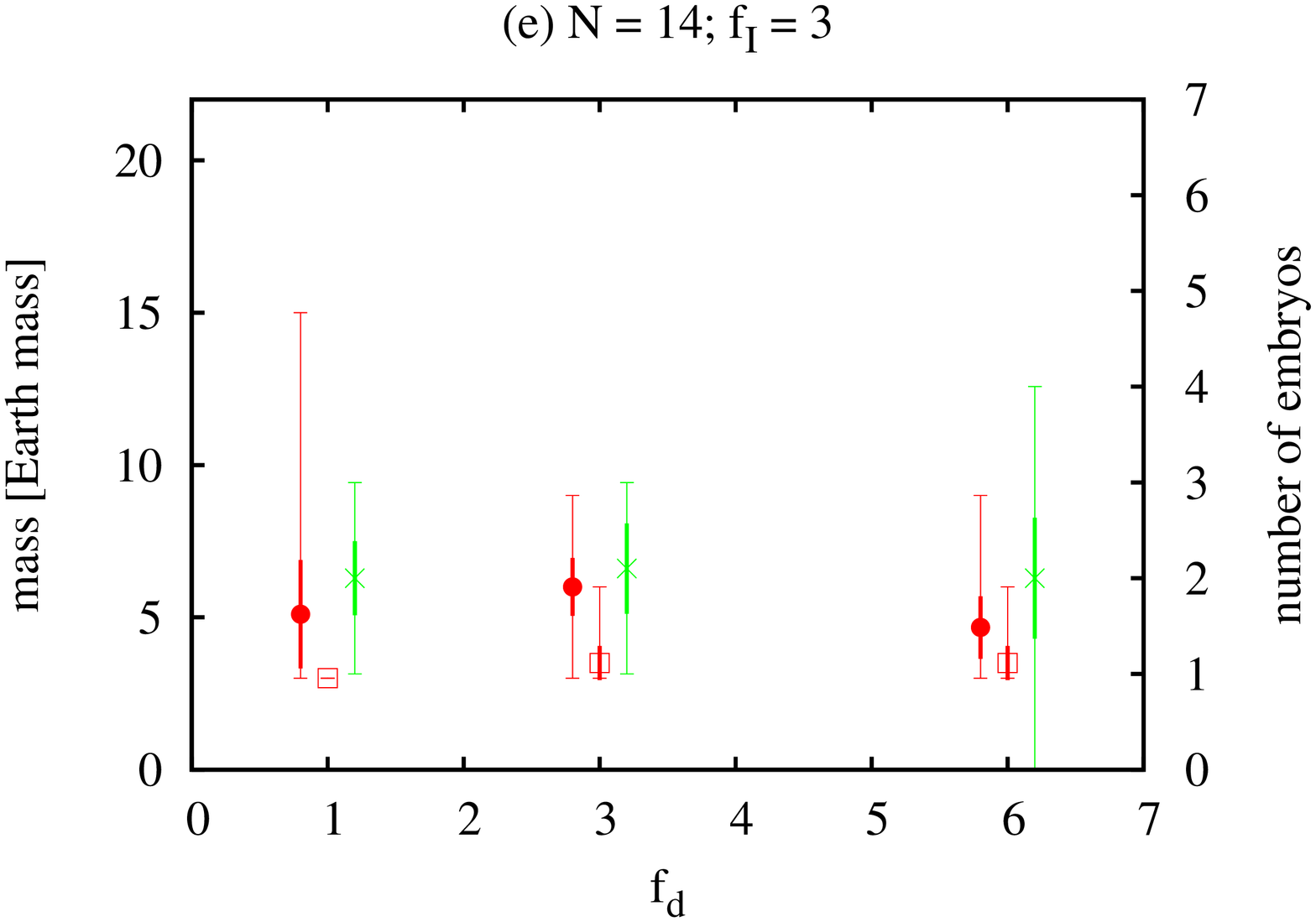}}
\centerline{\includegraphics[width=6cm, angle=-90]{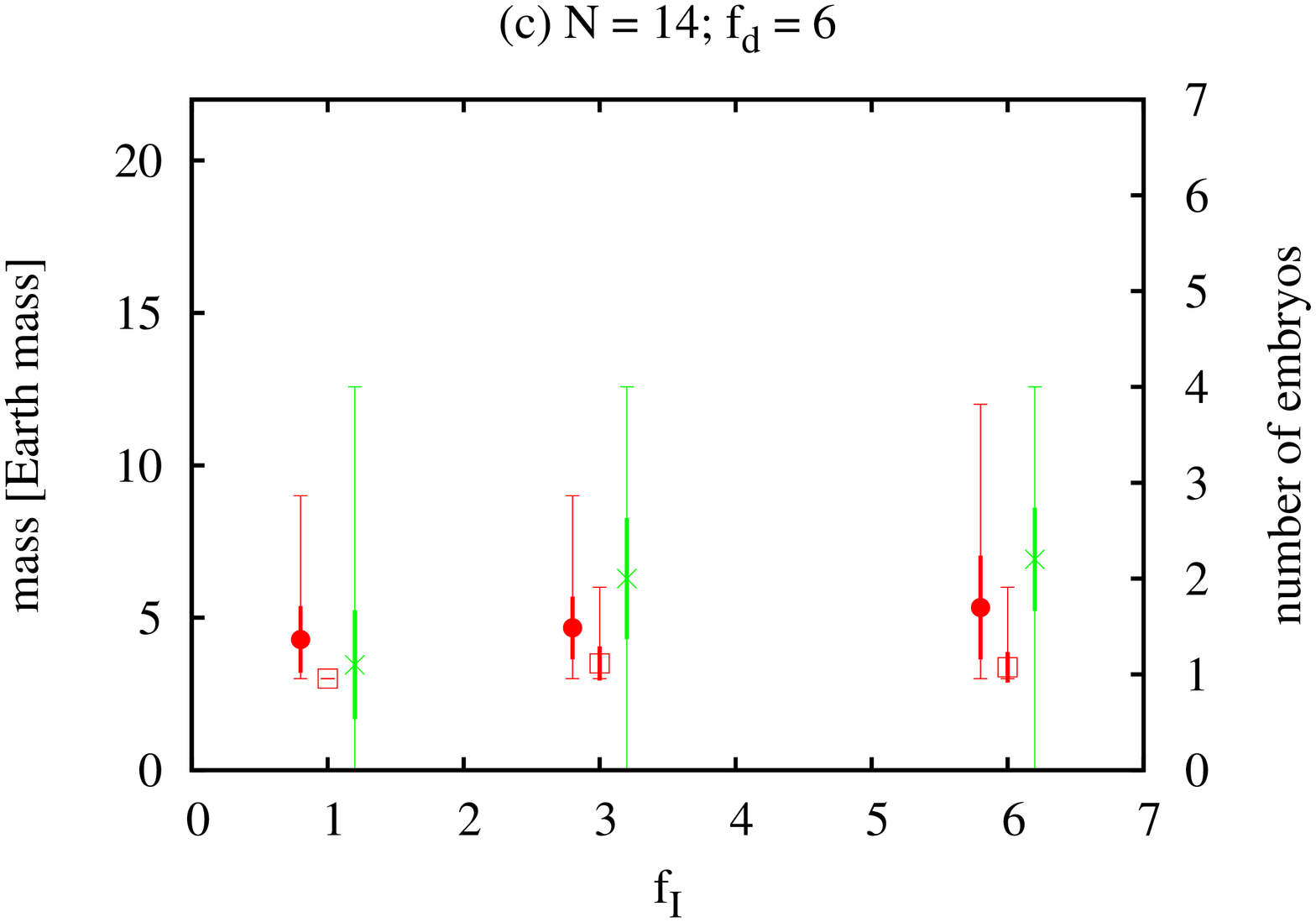}
    \includegraphics[width=6cm, angle=-90]{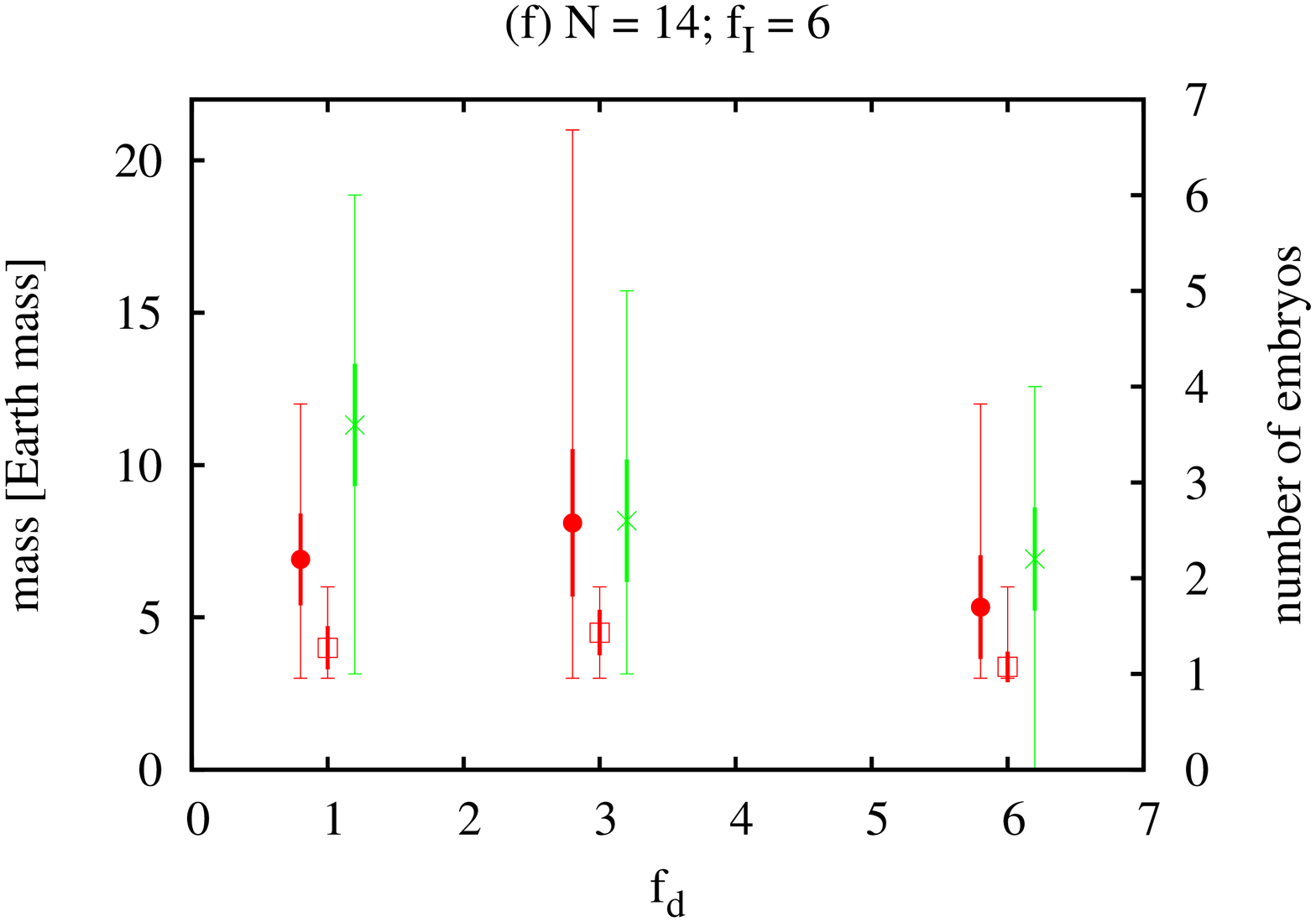}}
\caption{Statistical analysis of the results of the simulations starting
from 14 embryos. The $x$-axis of left plots reports the value of
$f_{I}$ at fixed $f_{d}$, given at the title of each plot. Similarly,
the $x$-axis of right plots reports the value of $f_{d}$ at fixed $f_{I}$.
The dot, slightly displaced to the left, is the mean mass of the largest core
 surviving beyond Saturn, computed over the corresponding 10 simulations.
The thick vertical bar is the rms deviation of this quantity.
The thin bar shows the excursion of the same quantity from
minimum to maximum. The square, in the middle, and related bars are
the same, but for the second largest core. The mass scale is reported
on the left of the diagram. The cross, slightly displaced to the right,
reports the mean number of embryos/cores surviving beyond Saturn,
to be read against the scale on the right hand side. Again, the thick bar
is for the rms deviation and the thin bar for the minimum-maximum quantity.
}
\label{n14n28stat}
\end{figure*}

We now discuss the statistical dependence of the results on the values of 
 $f_{d}$ and $f_{I}$. Fig.~\ref{n14n28stat} shows the number of bodies surviving
in the system as well as the masses of the first and second largest bodies beyond Saturn, 
as a function of these parameters. 

It is clear from the figure, that there are no statistically significant trends of 
the considered quantities with respect to the investigated disk parameters.
There is however a significant trend that is apparent from the panels in the
 Fig.~\ref{n14n28}: the mass of the body formed in the inner 
Solar System increases on average with $f_{d}$ and $f_{I}$. If these parameters are 
small several embryos are ejected from the system or can be found on distant orbit; 
instead if $f_{d}$ and $f_{I}$ are larger, no bodies are ejected and most of the objects 
scattered by Jupiter and Saturn end up in the inner Solar System.
 Notice that the injection of embryos inside the orbit of Jupiter (which happens in most
simulations) is probably inconsistent with the
current structure of the Solar System, but might explain the structure
of some extra-solar planetary systems, particularly those with a hot
Neptune and a more distant giant planet like HD 215497. Similarly, the
ejection of embryos onto distant, long-period orbits, might find one day
an analog in the extra-solar planet catalogs.

Overall our results concerning the formation of the cores of 
Uranus and Neptune are disappointing. The best cases are simulation 
No. 4 with $(f_{d},~f_{I}) = (1,~6)$ and No. 8 with $(f_{d},~f_{I}) = (6,~6)$, 
both producing beyond Saturn two cores of 12 and $6\,$M$_{\oplus}$, 
together with one leftover embryo. Thus, in general, although the mass of the first core 
is close to the mass of Uranus, the mass of the second biggest core is never large 
enough to correspond to Neptune.

One could think that our inability to produce two objects as massive as
Uranus and Neptune is due to an insufficient total mass in embryos in
the initial disk beyond Saturn. To investigate if this is true, we performed
another series of simulations with more embryos and, therefore, a
higher initial total mass. Specifically, we consider 28 embryos
distributed in the same range of heliocentric distances, from 10 to
$35\,$AU. The initial mass of each embryo is assumed to be
$3\,$M$_{\oplus}$, again. With this set-up, the orbital separation
between embryos is on average smaller than 5 mutual Hill radii. So, the embryos
are on orbits which are closer to each other than predicted by the
theory of runaway/oligarchic growth \citep{1998Icar..131..171K}. 
We nevertheless assume such initial configurations in order to explore 
how the results change by doubling the total initial mass. 

In the simulations with 28 embryos, the masses of the cores surviving beyond Saturn do not 
improve significantly, relative to the 14-embryos cases.
(The final masses and orbital characteristics of the
surviving objects in the 28-embryos simulations can be seen in Fig.~A1.) 
 The main effect of the doubled initial mass is 
in the mass of the cores formed in the inner Solar System, the most massive of which 
reaches $57\,$M$_{\oplus}$ (simulation No.~8 for $(f_{d},~f_{I}) = (6,~3)$). 

Statistically, no significant trend can be seen. (The statistical
dependence for all combinations of $f_{d}$ and $f_{I}$ is shown in Fig.~A2.) 
 There are no significant differences from the results 
obtained with half the initial mass. The mass in excess in the new runs has ended up 
in the large cores in the inner Solar System, but it does not contribute to form 
bigger cores beyond Saturn. The most massive core surviving beyond Saturn has 
$24\,$M$_{\oplus}$ (simulation No.~7 for $(f_{d},~f_{I}) = (3,~6)$) but there is no 
second core. The best system is that achieved in simulation 
No.~1 for $(f_{d},~f_{I}) = (1,~6)$, with cores of 15 and $9\,$M$_{\oplus}$ beyond 
Saturn (together with three leftover embryos). These best case results are better 
than those achieved starting from 14 embryos but still they do not successfully 
reproduce the real case of Uranus and Neptune.

\section{A ``planet trap'' at the edge of Saturn's gap}
\label{sectTrap}

In the previous section we have seen that a large fraction of the
initial embryo population is lost due to migration into the
Jupiter-Saturn region.  Most of the embryos that come too close to the
giant planets are eliminated because of collisions with Jupiter and
Saturn, ejection onto hyperbolic orbits or injection into the inner
Solar System.

This result may be due to the fact that the migration torque that we
implemented (from \cite{2008A&A...482..677C} see Sect.~2), does not take into account 
the so-called coorbital corotation torque \citep{2001ApJ...558..453M}.
This torque would stop inward-migrating embryos at a ``planet
trap'' just inward of the outer edge of the gap opened in the disk of
gas by Jupiter and Saturn \citep{2006ApJ...642..478M}. The planet trap would
prevent the embryos from coming too close to the giant planets. To test
what effects this would have on planetary accretion and evolution,
we have done simulations implementing a new formula from 
\cite{2010MNRAS.401.1950P}, which accounts for the co-orbital corotation 
torque, as explained in Sect.~2.

\begin{figure}
 \centering
    \includegraphics[width=8cm]{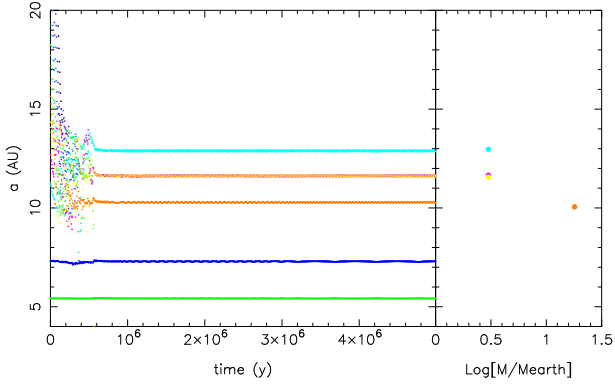}
\caption{
The same as Fig. \ref{evol} but for No.~3 with $f_{d} = 3$, $f_I=1$ and 
accounting for the planet trap.
}
\label{evol_trap}
\end{figure}

To reduce parameter-space we fixed $f_I=1$. This is legitimate, because in principle, 
there is no need to reduce Type-I migration speed as it automatically decreases
to zero approaching $\sim$$10\,$AU. We did three series of 10 simulations
for $f_{d} = 1$, 3, and 6. All started with a system of 10 embryos of $3\,$M$_{\oplus}$, 
originally distributed from 11 to $34\,$AU, with a mutual orbital separation of 7 
mutual Hill radii. (The end-states are illustrated in Fig.~A3.) 
 Fig.~\ref{evol_trap} gives an example of evolution taken from simulation 
No.~3 with $f_{d} = 3$. It is instructive to compare this figure with Fig.~\ref{evol}
above.
In this case we see much fewer embryos coming into the 
Jupiter-Saturn region and suffering scattering events. This is precisely the effect of the
planet trap. As a consequence, 90 percent of the initial embryos mass is retained 
in the system beyond Saturn, whereas only 38 percent of the mass was retained in 
the case of Fig.~\ref{evol}. Notice also that the accretion process lasts for a shorter time 
($0.6\,$My against $1.2\,$My, respectively), because the system of embryos/cores remains 
less dynamically excited.

Nevertheless, we see from Fig.~\ref{evol_trap} that only one core grows massive ($18\,$M$_{\oplus}$) 
and the rest of the mass ($9\,$M$_{\oplus}$ in total) is splitting in three leftover 
embryos, that are on stable orbits in resonance with the massive core and with each other.
Notice, the presence of two embryos in co-orbital configuration, at $\sim$12 AU. This result is typical as can be
checked in Fig.~A3. 
This is very different from the case of the Solar
System, with two major planets of comparable masses beyond Saturn (Uranus and Neptune) and 
no leftover embryos.

\begin{figure}
 \centering
    \includegraphics[width=6cm,angle=-90]{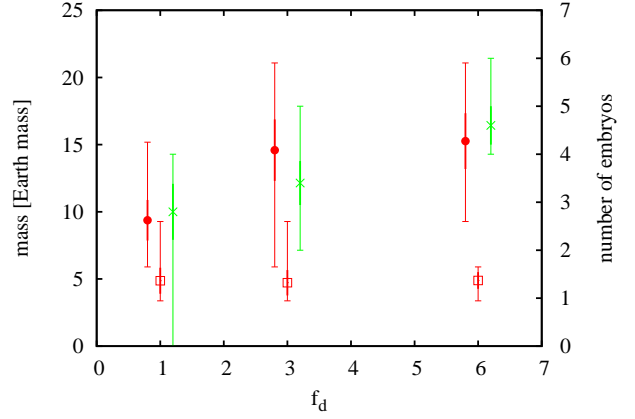}
\caption{
The same as Fig.~\ref{n14n28stat} but for the simulations with the planet trap.
}
\label{trap_stat}
\end{figure}

Fig.~\ref{trap_stat} shows the statistical analysis of our results as a function of the parameter 
$f_{d}$. We see a trend, although only marginally significant: the mass of the largest core and 
the number of bodies surviving beyond Saturn increases with $f_{d}$. However, 
the mass of the second largest object is remarkably independent of $f_{d}$ and it is small 
in all cases. We interpret this result as follows: A larger value of $f_d$ increases 
the eccentricity damping on the embryos. With smaller eccentricities, 
the embryos have a larger gravitational focusing factor \citep{1992Icar..100..440G}
 and therefore they can accrete each other more easily. 
However, because embryos tend to cluster at the location of the planet 
trap, basically only the innermost  embryo, i.e. the one located the 
nearest to the trap, gets the benefit of the situation.

Overall, these experiments show that the planet trap is very effective
in reducing the loss of mass during the embryo's evolution. Still, the
final results are not in good agreement with the structure of the
Solar System. In fact, only one major core grows and the rest of the mass 
remains stranded in too many embryos in stable resonant configuration.
Consequently, despite initially we have a total mass comparable to 
the combined mass of Uranus and Neptune, at the end only one massive planet 
is formed. 

\begin{figure}
 \centering
    \includegraphics[width=8cm]{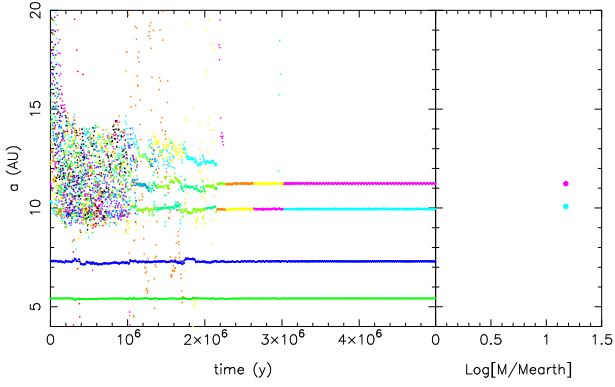}
\caption{
The same as Fig.~\ref{evol} but for the only simulation 
(No.~2) that successfully produces Uranus and Neptune.
}
\label{perfect}
\end{figure}

To test how these results would change with the total mass initially
available in embryos, we ran a series of 10 additional simulations
starting with 15 embryos of $3\,$M$_{\oplus}$. The orbits of the embryos
were initially separated by 5 mutual Hill radii and were distributed
from 10 to $35\,$AU. For simplicity, we fixed $f_d=3$. No statistically
significant trend with the initial total mass in the system appears (the end-states
are reported in Fig.~A3) 
, so that the problems discussed
above (too small mass  of the second core and too many resonant leftovers) remain.
This implies that  the additional mass is typically lost. Nevertheless, we have got 
one simulation whose result is about perfect relative to the Solar 
System structure, with only two cores surviving beyond Saturn, each with 
a mass of $15\,$M$_{\oplus}$ (see Fig.~\ref{perfect}).

\section{The initial mass of the embryos}

All the simulations presented up to this point in the paper started
from a system of embryos of individual mass equal to $3\,$M$_{\oplus}$.
In this section we study the influence of the individual mass of the
embryos on the final results. Following the structure of the paper, we
first present a series of simulations which do not account for the
presence of a planet trap at the edge of Jupiter/Saturn's gap, then we
focus on the impact of the planet trap.  

\subsection{Without a planet trap}
\label{wout_trap}

In this new series of simulations, we reduced the initial mass of
embryos from 3 to $1.5\,$M$_{\oplus}$. However, we increased the
number of embryos from 14 to 28, thus preserving the total
mass of the system used in the first part of Sect.~\ref{formUN}. 
The embryos are initially located in the
10$-$35-AU interval with mutual orbital separations of 3.8 mutual Hill radii.

As in Sect.~\ref{formUN}, we performed 10 simulations for each combination
 of $f_I = 1$, 3, 6  and $f_d = 1$, 3, 6. 
We find much less growth than in the equivalent simulations 
of Sect.~\ref{formUN}, starting with $3\,$M$_{\oplus}$. This is due to the migration 
and damping forces are linear in the mass of the embryos and therefore are weaker 
in this case. Thus, under their mutual encounters the embryos become more dynamically 
excited and their collision probability decreases dramatically. Most of the simulations 
with $f_d=1$ and $f_I = 3$, 6 did not reach a stable final configuration at $5\,$My and 
when many embryos are still on highly eccentric and mutually crossing orbits. 
We find the most significant growth in the runs with $f_d=3$ and $f_I=6$, but only one 
major core is produced, with a mean mass of $8\,$M$_{\oplus}$. The runs with 
$f_d=6$ and $f_I=6$ transfer most of the mass in the inner Solar System.
(The end-states of these simulations are summarized in the Fig.~A4.) 

The statistical analysis of our results (Fig.~A5) 
 reveals no correlation of the results with $f_d$ for any value of $f_I$. However, 
we find some correlation between the mass of the largest core and the value of $f_I$ for 
each value of $f_d$. Remember that in absence of a planet trap, embryos are driven by 
migration to encounter the giant planets, unless they are trapped in 
resonance with Saturn. The larger is $f_I$, the slower is migration and 
therefore the easier is for embryos to be stopped in resonance and 
(possibly) accrete with each other. This may explain the observed 
correlation.

\subsection{With a planet trap}

We did a series of 10 simulations, starting from 30 embryos, each of
$1\,$M$_{\oplus}$, thus preserving the total
mass of the system used in the first part of Sect. \ref{sectTrap}.
 They are initially distributed from 9 to $35\,$AU, 
with a mutual orbital separation of 4.5 mutual Hill radii. We fixed 
$f_{I} = 1$ and $f_d = 3$. The end-states are reported in Fig.~A3. 

The results are significantly different from those of the runs starting
from more massive individual embryos, reported in Sect.~\ref{sectTrap}. 
On average, the mass of the largest core beyond 
 Saturn drops by a factor of 2 and that of the second largest core drops 
by a factor of 1.5. These drops are statistically significant. The mean 
number of surviving cores/embryos is 4, with simulations producing up to 7 
objects. Therefore the results are very far from the structure of 
the current Solar System. More mass is lost in these simulations 
than in the corresponding simulations starting
from more massive embryos. As said above, this is probably because the
forces damping the eccentricities and inclinations of the embryos are
weaker. 

\section{The effect of turbulence in the disk}
\label{turb}

The problems that we experienced in the previous sections due to
embryos and cores being protected by
resonances from mutual collisions suggests that some level of
turbulence in the disk might promote accretion.
In fact, turbulence provides stochastic torques onto the bodies,
inducing a random walk of their semi-major axes \citep{2010MNRAS.409..639N}.
If these torques are strong enough, the bodies may be dislodged
from mutual resonances, which in turn would allow them to have mutual
close encounters and collisions. 

The problem is that it is not known a priori how strong turbulence should be in
the region of the disk where giant planets form. The strength of MRI
turbulence depends on the level of ionization of the disk 
\citep{2000ApJ...530..464F}, and it is possible that in the massive regions 
of the disk where giant planets form the ionization of the gas is very low 
because the radiation from the star(s) cannot penetrate down to the midplane 
\citep{1996ApJ...457..355G}. Several formation
models, therefore, argue that the planets form in a ``dead zone''
where, in absence of ionization, there is no turbulence driven by the
magneto-rotational instability. This justifies the assumption that we
made up to this point in the paper that the disk is laminar. 

For the sake of completeness, though, we test in this section how the
results change if strong or weak turbulence is assumed in the
disk. The experiments are conducted in the framework of the planet
trap case, with 10 embryos of $3\,$M$_\oplus$ (see Sect.~\ref{sectTrap}). 
As in previous section \ref{wout_trap} we assume $f_{I} = 1$ and $f_d = 3$.
We performed three sets of 10 simulations with three different values of 
the turbulent strength $\gamma$. We used $\gamma = 3\times 10^{-4}$, 
$1\times 10^{-3}$ and $3\times 10^{-3}$. The latter correspond to strong turbulence 
without dead-zone \citep{2010ApJ...709..759B}. The end-states are, again, reported 
in Fig.~A3. 

\begin{figure}
 \centering{
\centerline{\includegraphics[width=4.5cm, angle=-90]{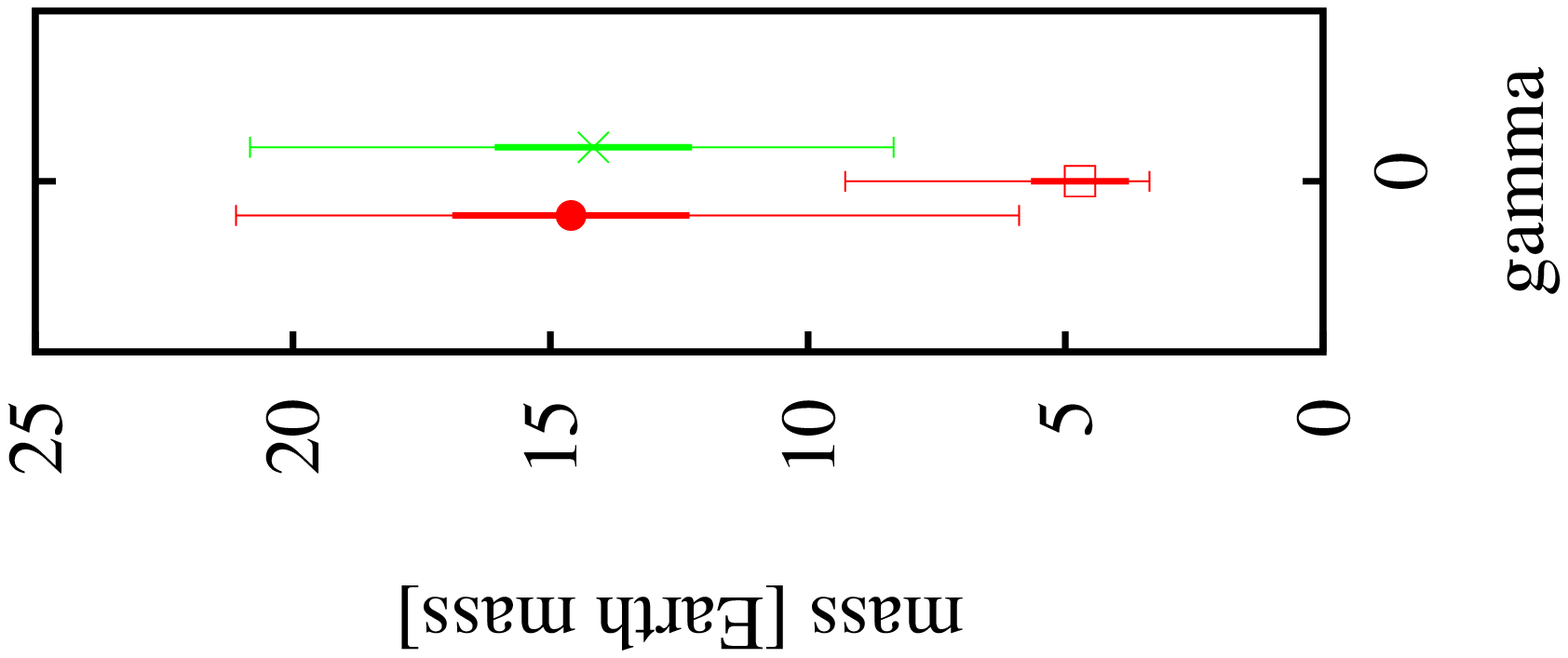}
\hspace{-11 mm} \includegraphics[width=4.5cm, angle=-90]{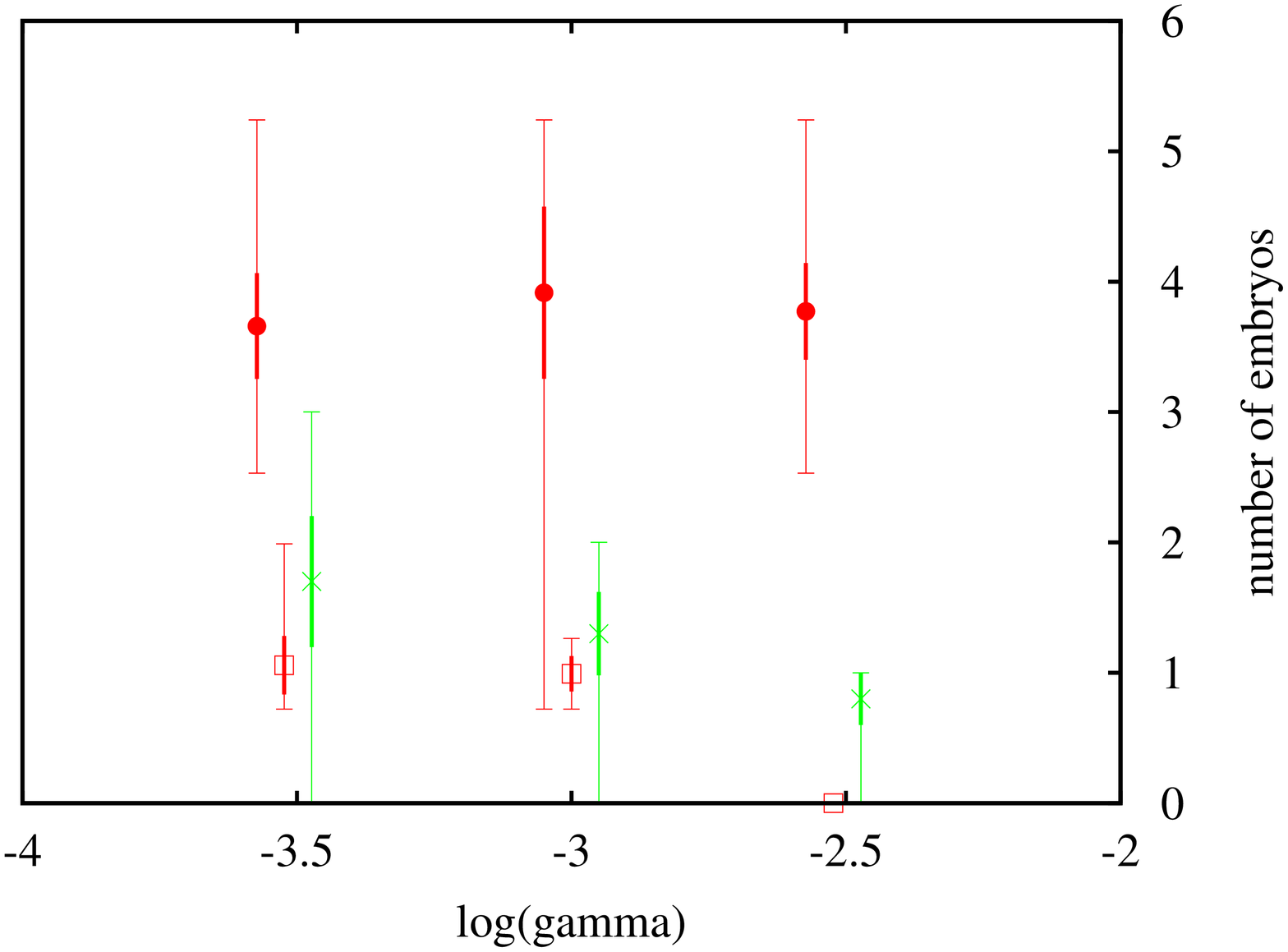}}
}
\caption{
The same as Fig.~\ref{n14n28stat} but for the series of simulations with
different turbulence strength $\gamma$. The series for $\gamma = 0$ (no-turbulence) 
is that presented in Sect.~\ref{sectTrap}.
}
\label{turb_stat}
\end{figure}

\begin{figure}
 \centering
    \includegraphics[width=8cm]{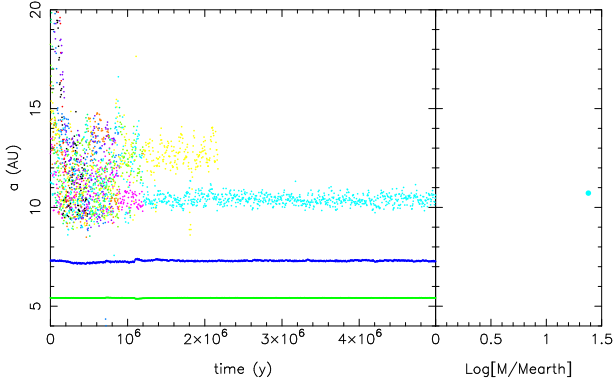}
\caption{
The same as Fig.~\ref{evol} but for a simulation with strong turbulence (No.~6 with
$\gamma = 3\times 10^{-3}$).
}
\label{turb_evol}
\end{figure}

Fig.~\ref{turb_stat} reports a statistical analysis of our results. There is a clear 
correlation, with the number of objects surviving beyond Saturn decreasing with increasing 
strength of the turbulence, as expected. The runs with $\gamma = 3\times 10^{-3}$ 
typically leave only one object beyond Saturn. The mass of the second largest core 
also decreases with increasing turbulence. 
Instead we do not see any statistical correlation between the mass of the largest core and 
the turbulence strength. 
This is a surprise, because we expected that, with fewer 
embryos stranded in resonant isolation, the largest core would have 
grown more massive. In reality, though, it seems that more mass is lost 
as the turbulence increases so that the largest core does not benefit 
from the situation. The reason for this is clearly that resonant configurations 
are not stable if turbulence is too strong. 
Fig.~\ref{turb_evol}  gives an illustration of this, showing the evolution 
of the system in simulation No.~6 with $\gamma = 3\times 10^{-3}$. 
We can see a stochastic behavior of the semi-major axis of the cores, that 
eventually forces the last two cores to collide with each other. The surviving core in this simulation
has the mass of $24\,$M$_\oplus$, which is the largest among all objects produced in our turbulence 
simulations. Typically, more objects interact with Jupiter and Saturn and are eventually lost.

We conclude from these tests that turbulence may be helpful to
break the isolation of planets and to avoid the formation of a too
crowded planetary system beyond Jupiter and Saturn. However, the
formation of two massive cores, with no left-over embryos seems
to be still a distant goal, as typically only one core is produced.

\section{Conclusions and Discussions}

The formation of Uranus and Neptune by runaway/oligarchic growth from
a disk of planetesimals is very difficult. The accretion rate is not
large enough; moreover the accretion stalls when the major bodies
achieve a few Earth masses \citep{safronov_1969,2001Icar..153..224L,
2007Icar..189..196L}.

In this paper, we have explored a mechanism for the formation of
Uranus and Neptune that is different from those explored before.  We
assume that runaway/oligarchic growth from a planetesimal disk
generated a system of embryos of 1$-$3 Earth masses, but not larger, in
agreement with the results of \cite{2010AJ....139.1297L}.
Then, we follow the dynamical evolution of such a system of embryos, accounting for
their mutual interactions, their interaction with a disk of gas and
the presence of fully formed Jupiter and Saturn on resonant,
non-migrating orbits. Because of computation speed, we do not perform
the simulations with an hydro-dynamical code. Instead, we use an
N-body code, and we apply fictitious forces to the embryos that mimic
the orbital damping and migration torques that the disk exerts on
the embryos. We use two prescriptions for the migration torque: one
that ignores the co-orbital corotation torque and one that takes it
into account. With the second prescription a ``planet trap'' 
\citep{2006ApJ...642..478M} appears at about $10\,$AU, just inward 
of the outer edge of the gap opened by Jupiter and Saturn in the disk 
of gas.

Our results show that the idea works in principle. In many runs,
particularly those accounting for a disk a few times more massive than
the MMSN and those accounting for the presence of the planet trap, 
there is significant mass growth.
In these cases, the major core beyond Saturn exceeds 10 Earth
masses. These results highlight the importance of the planet trap and
of the damping effect that the gas-disk has on the orbital
eccentricities and inclinations of planetary embryos.

Our simulations typically do not successfully reproduce the structure
of the outer Solar System. In the wide range of the end-states due to the 
stochasticity  of the accretion process, only one simulation is successful, 
with two cores of $15\,$M$_\oplus$ produced and surviving beyond Saturn and no 
rogue embryos. In general, our results point to at least two major
problems. The first is that there is typically a large difference in
mass between the first and the second most massive cores, particularly
in the simulations showing the most spectacular mass growth. This
contrasts with Uranus and Neptune having comparable masses.  The
second problem appears mostly if we account for the planet trap, so that 
more material is kept beyond the orbit of Saturn. The problem is that the final 
planetary system typically has many more than two embryos/cores. Several original
embryos, or partially grown cores, survive at the end on stable,
resonant orbits. In several cases bodies are found in coorbital
resonance with each other. This is in striking contrast with the outer
Solar System, where there are no intermediate-mass planets accompanying
Uranus and Neptune. It might be possible that these additional planets
have been removed during a late dynamical instability of the planetary
system, but the likelihood of this process remains to be proven.

Accounting for some turbulence in the disk alleviates this
last problem. However, when turbulence is strong enough to prevent 
the survival in resonance of many bodies, the typical result is the 
production of only one core. 

In addition to the problems mentioned above, there is another
intriguing aspect suggested by the results of our model. We have seen
that, in order to have substantial accretion among the embryos, it is
necessary that the parameter $f_d$ is large, namely that the surface
density of the gas is several time higher than that of the
MMSN. However this contrasts with the common idea that Uranus
and Neptune formed in a gas-starving disk, which is suggested by the
small amount of hydrogen and helium contained in the atmospheres of
these planets compared to those of Jupiter and Saturn. How to solve
this conundrum is not clear to us.

In summary, our work does not bring solutions to the problem of the
origin of Uranus and Neptune. However, it has the merit to point out
non-trivial problems that cannot be ignored and have to be addressed
in future work.

\begin{acknowledgements}
We thank the reviewer Eiichiro Kokubo for extremely useful comments, which 
have greatly improved this paper.
This work was supported by France's EGIDE and the Slovak 
Research and Development Agency, grant ref. No. SK-FR-0004-09. 
RB thanks Germany's Helmholz Alliance for funding through 
their 'Planetary Evolution and Life' programme. 
\end{acknowledgements}

\newpage
\renewcommand\figurename{Fig. A}

 \addtocounter{figure}{-10}
\begin{figure*}
\vspace{0.8cm}
\centerline{{\Large{\bf Appendix A: additional figures}}}
 \vspace{-1.2cm}
\centerline{\includegraphics[width=6.5cm, angle=-90]{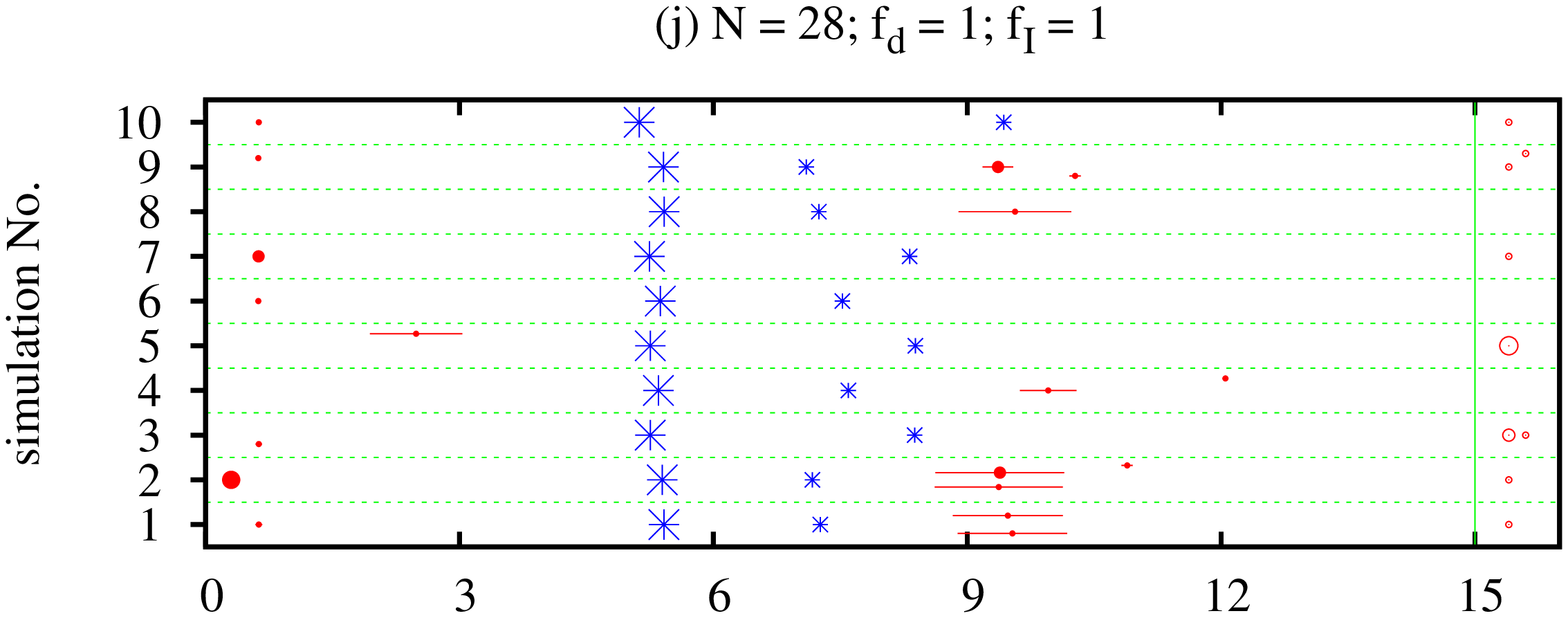}
    \includegraphics[width=6.5cm, angle=-90]{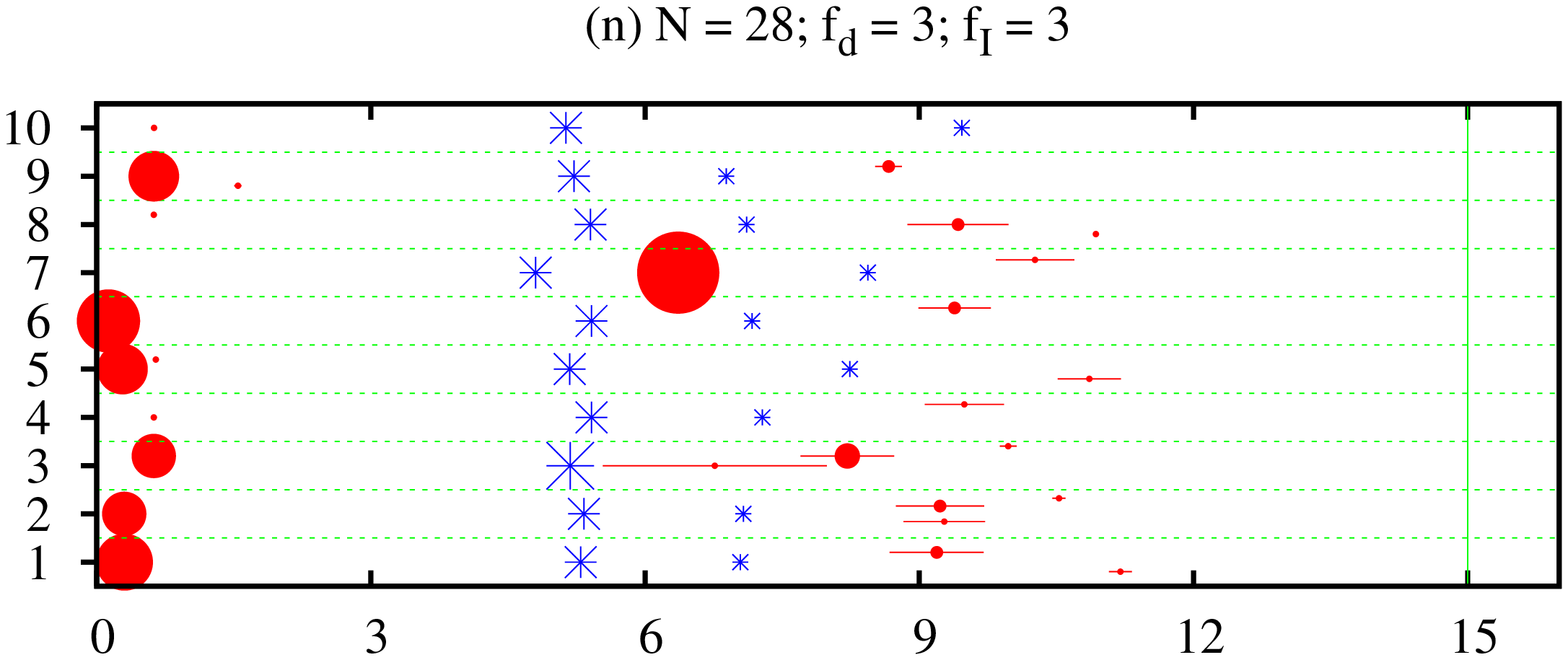}}
\vspace{-3cm}
\centerline{\includegraphics[width=6.5cm, angle=-90]{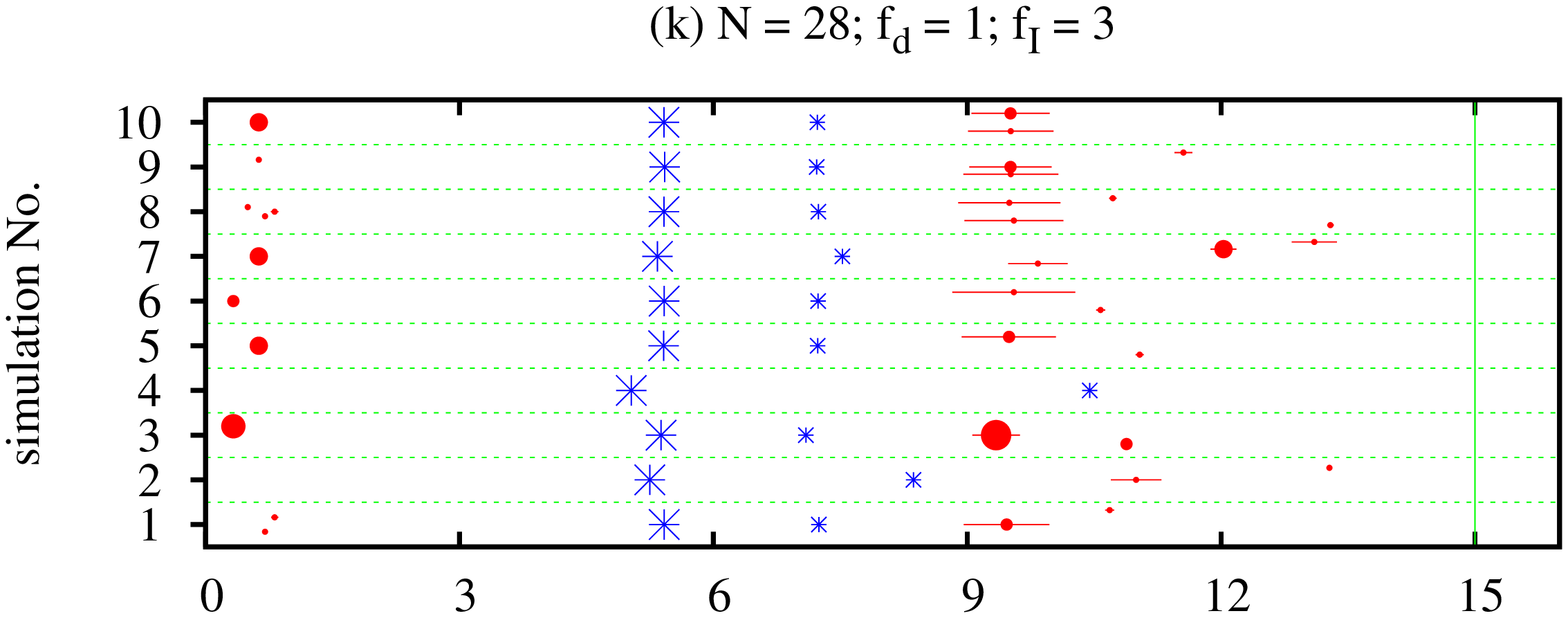}
    \includegraphics[width=6.5cm, angle=-90]{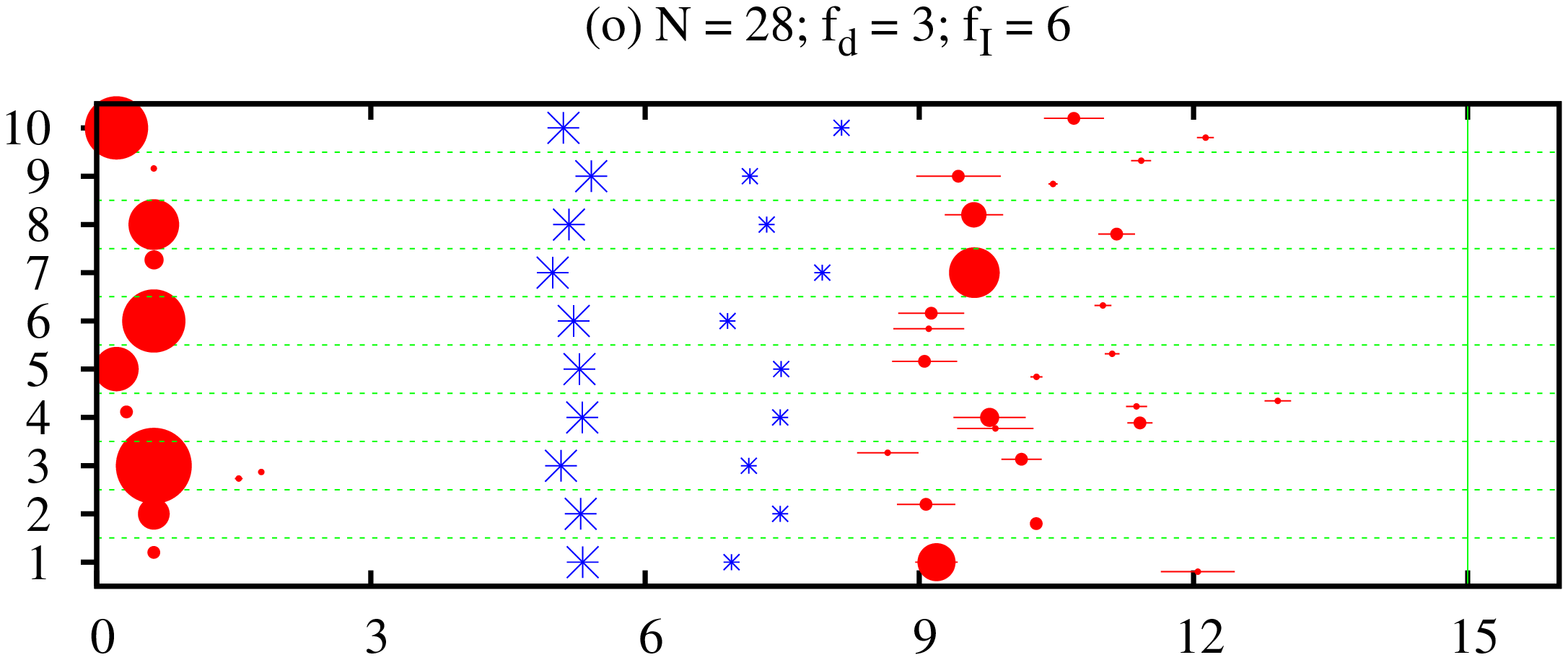}}
\vspace{-3cm}
\centerline{\includegraphics[width=6.5cm, angle=-90]{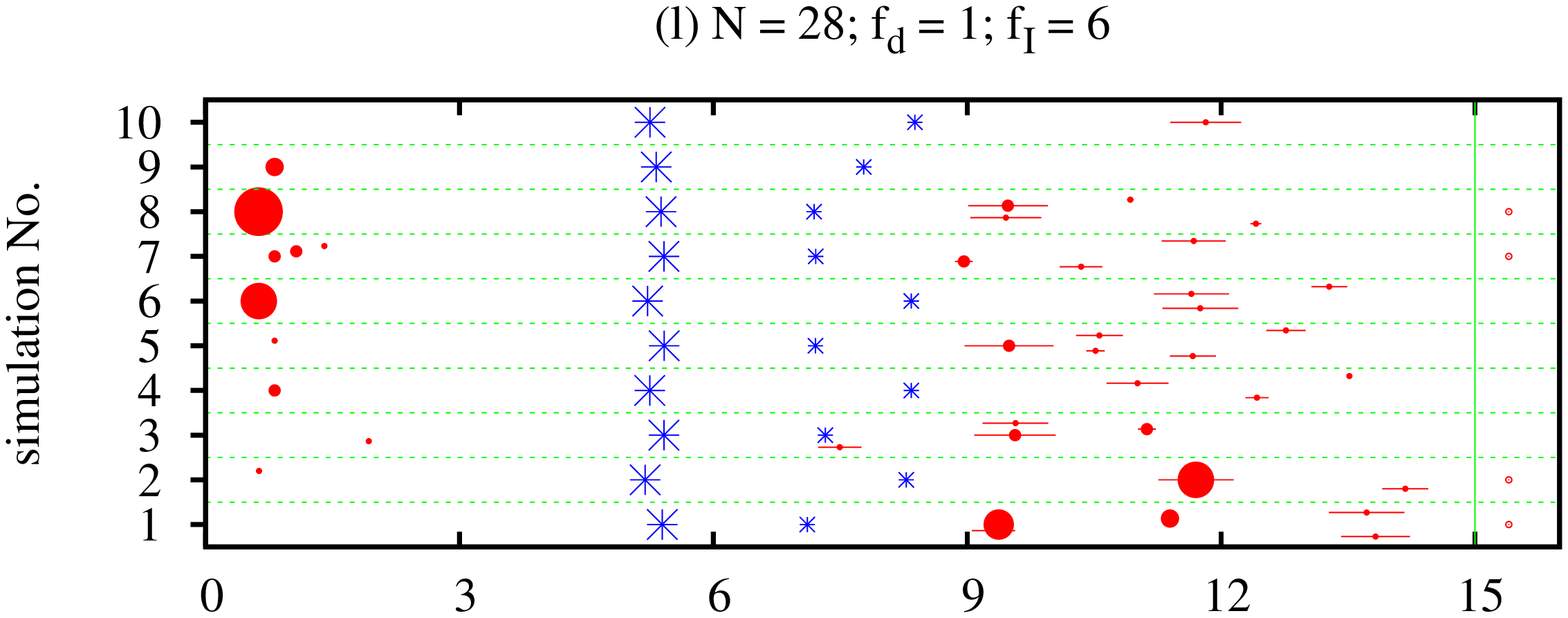}
    \includegraphics[width=6.5cm, angle=-90]{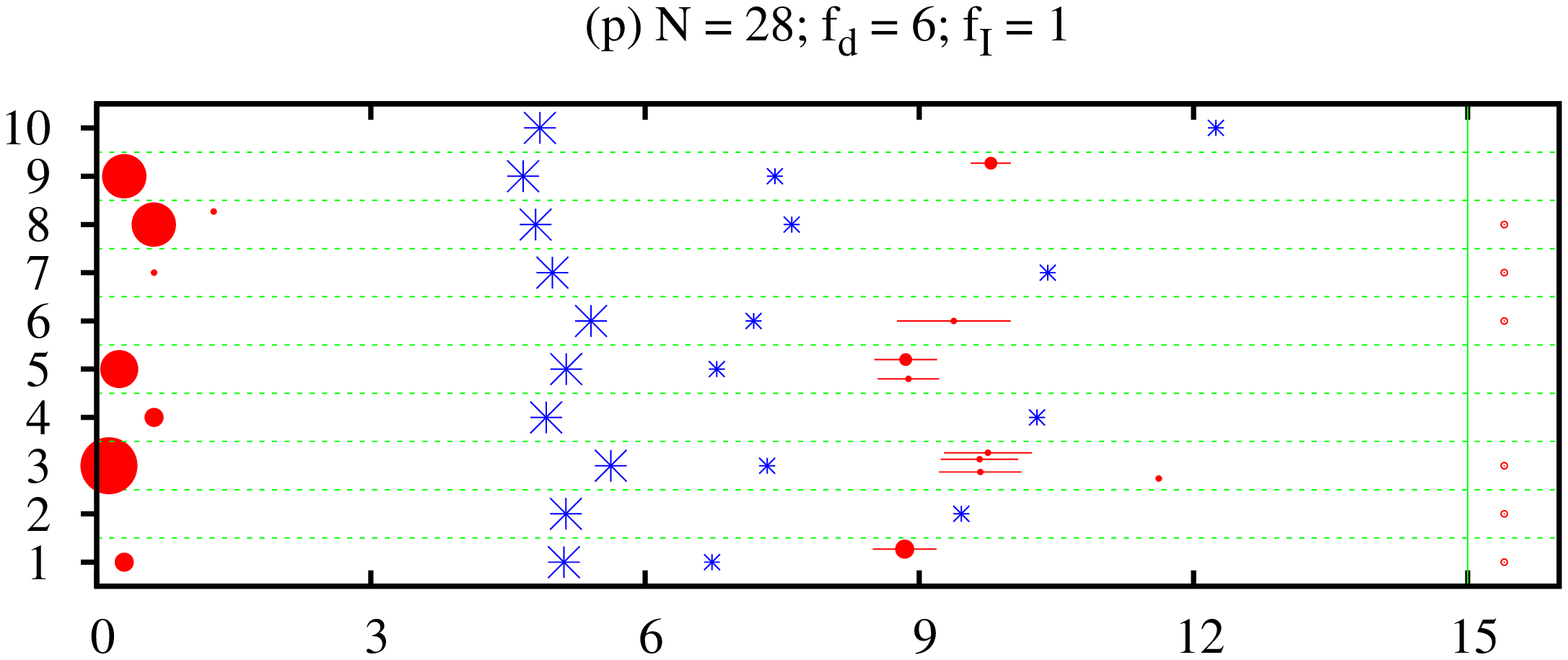}}
\vspace{-3cm}
\centerline{\includegraphics[width=6.5cm, angle=-90]{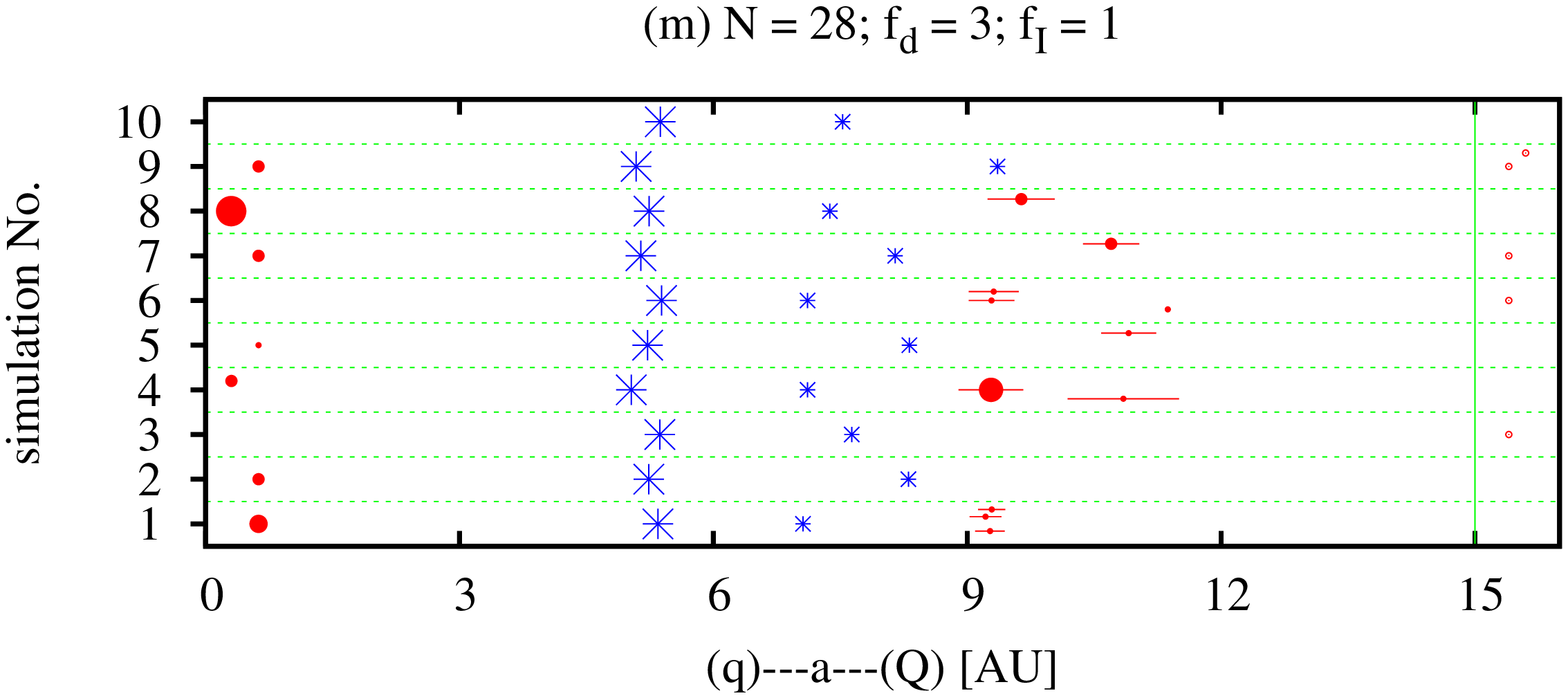}
    \includegraphics[width=6.5cm, angle=-90]{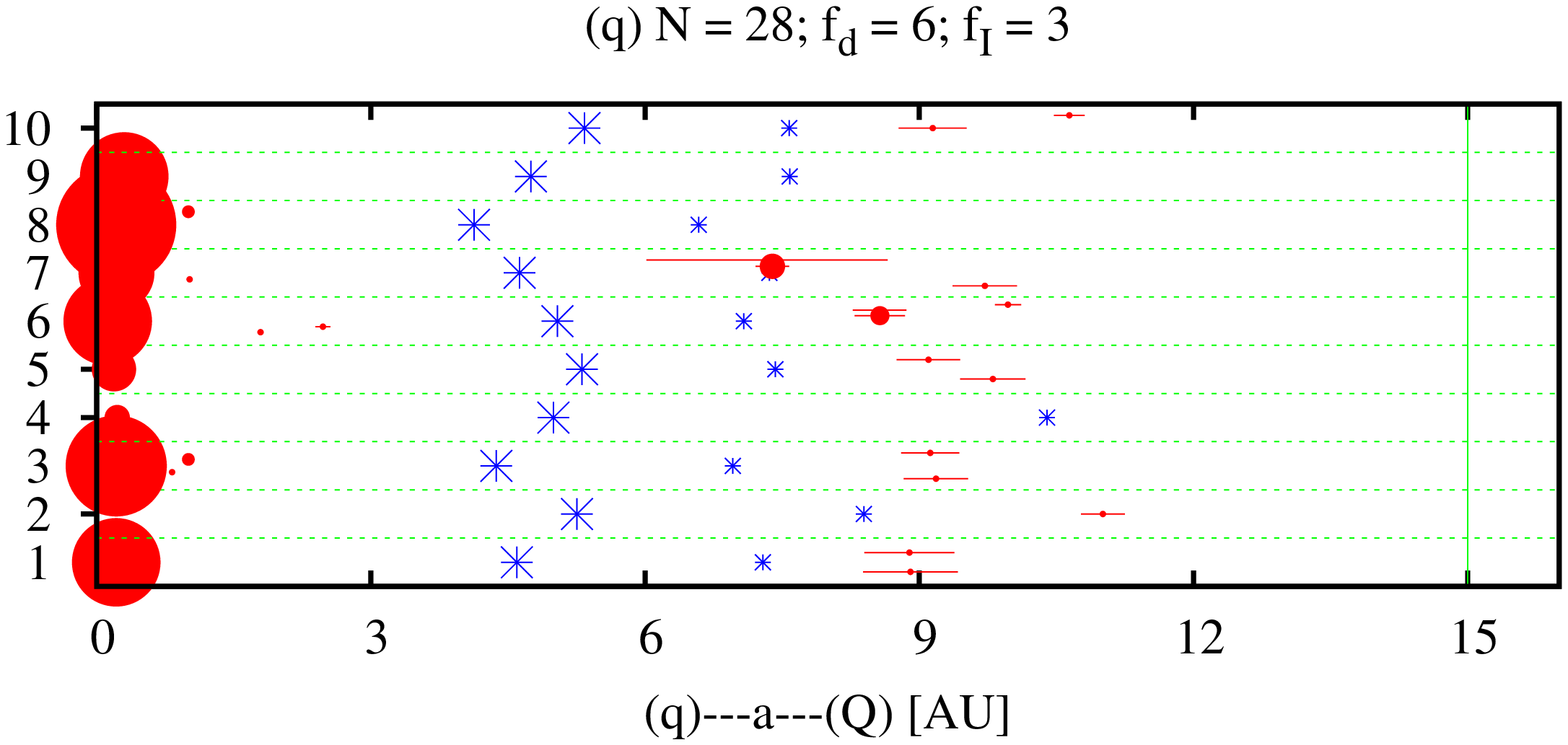}}
\vspace{-3cm}
\centerline{\includegraphics[width=6.5cm, angle=-90]{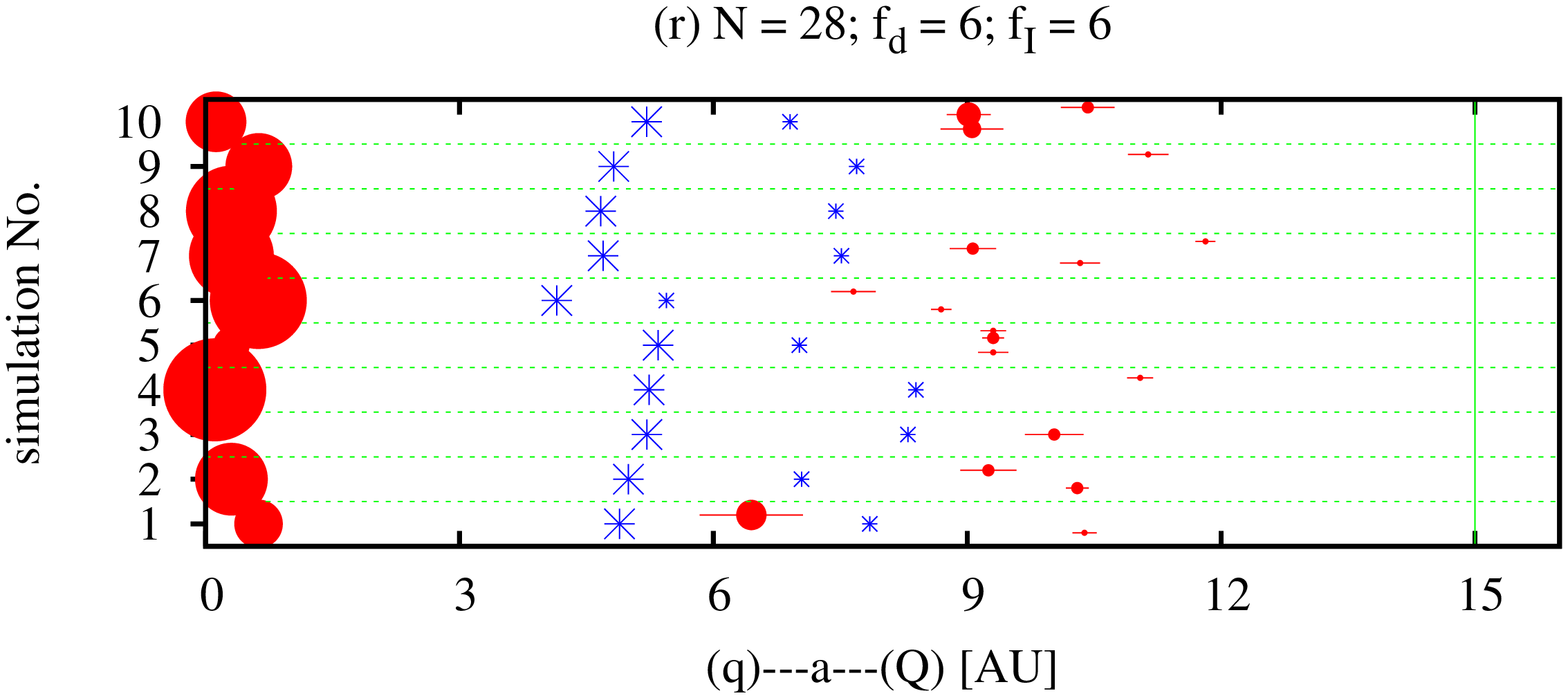}}
\vspace{-3.4cm}
\centerline{\includegraphics[width=7.0cm,angle=-90]{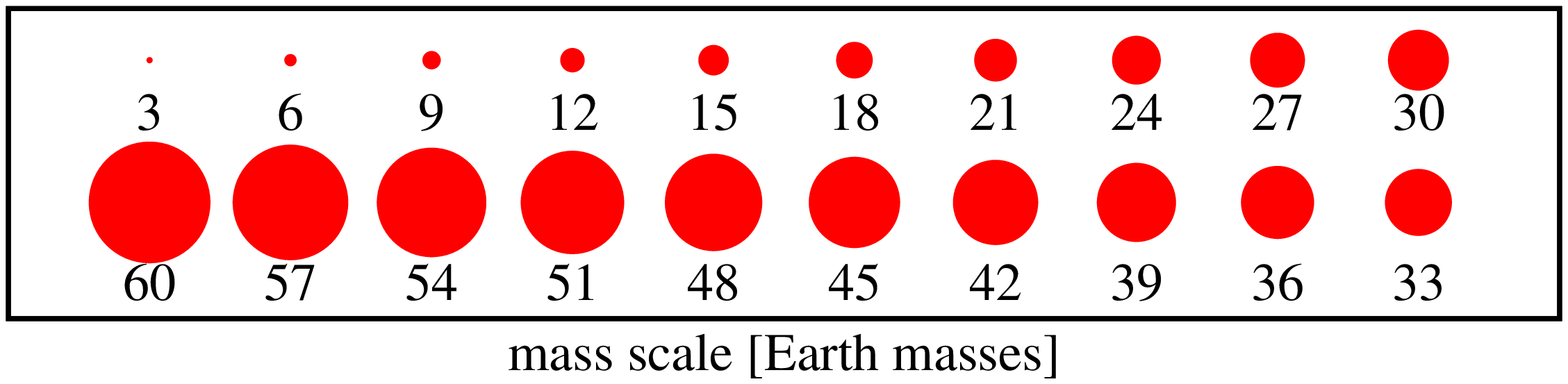}}
\vspace{-2cm}
\caption{The end-states (in $5\,$Myr) of the simulations starting
with 28 embryos, each with $3\,$M$_{\oplus}$. The simulations are
separated by horizontal dashed green lines. The surviving embryos/cores
are shown with filled red dot, whose size is proportional to the objects
mass. The scale is shown on the bottom panel. The red horizontal bar
shows the perihelion-aphelion excursion of these objects on their
eccentric orbits. The objects beyond $15\,$AU are plotted for simplicity
at $15.5\,$AU, beyond the vertical solid green line.
Jupiter and Saturn are shown as blue asterisks. The label on top of each
panel reports the number $N$ of embryos and the values of $f_I$ and $f_d$ adopted
in the simulations.
(For the discussion - see Sect.~4).
}
\label{smFIG1}
\end{figure*}

\newpage

\begin{figure*}
\centerline{\includegraphics[width=6.5cm, angle=-90]{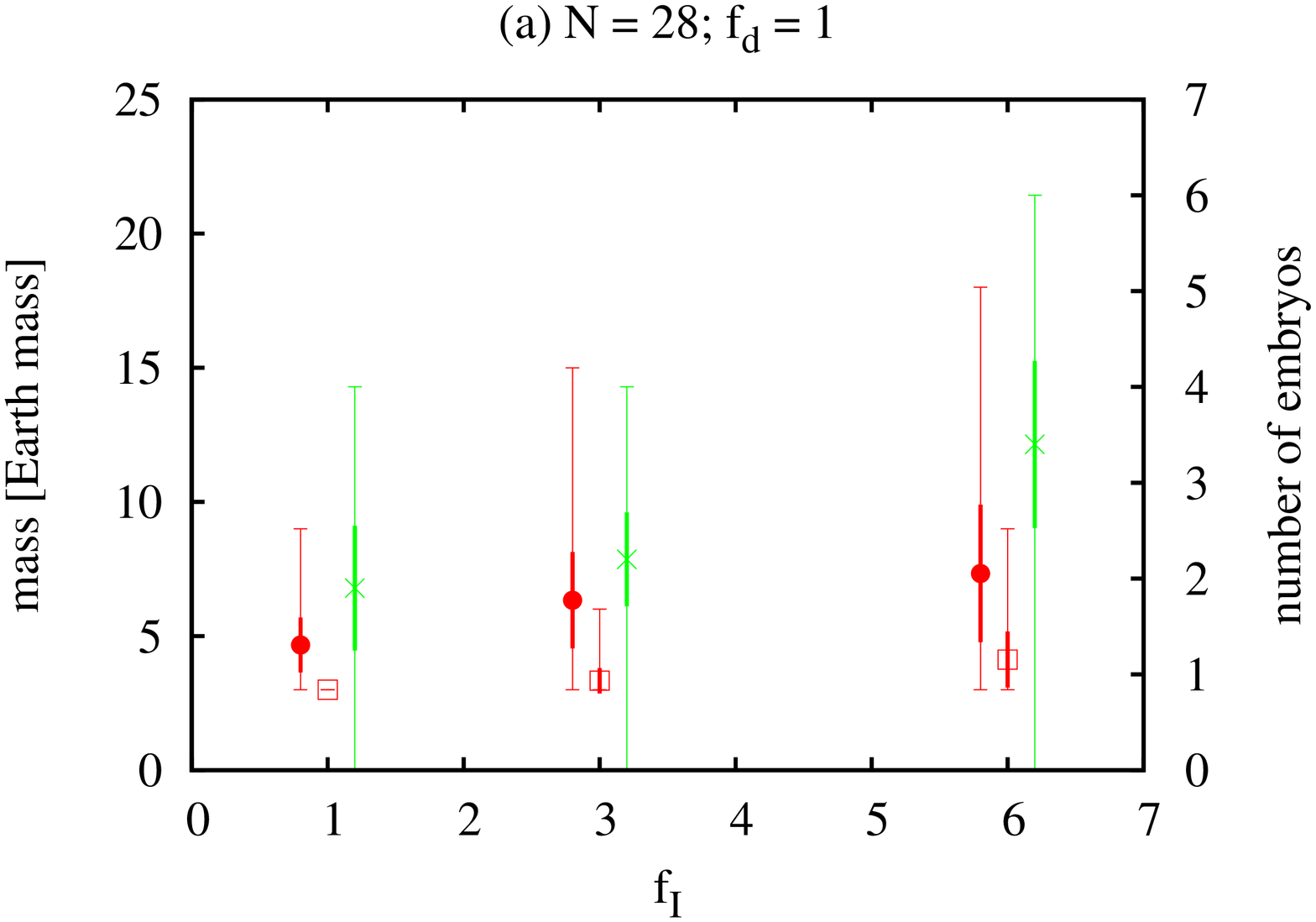}
    \includegraphics[width=6.5cm, angle=-90]{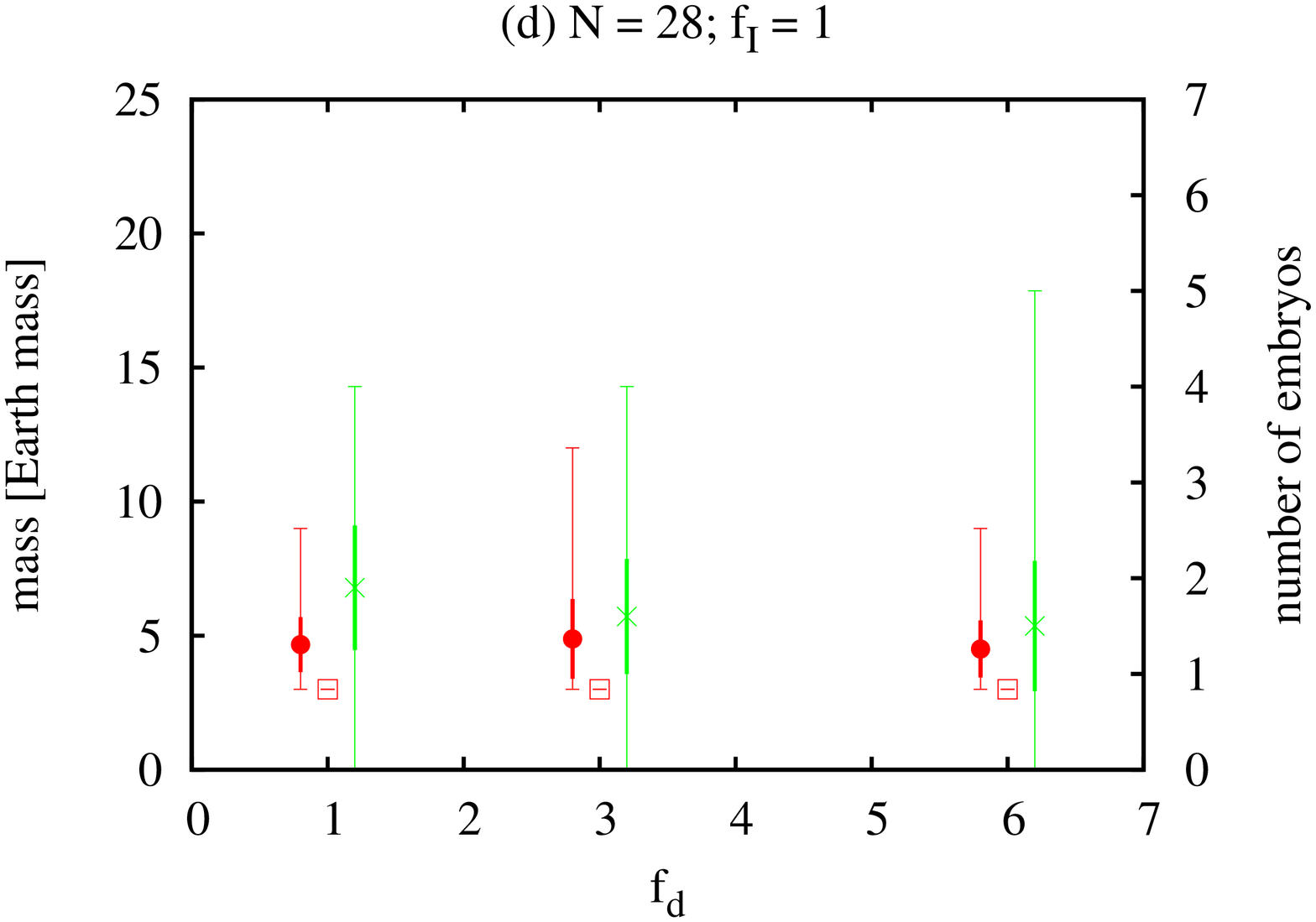}}
\centerline{\includegraphics[width=6.5cm, angle=-90]{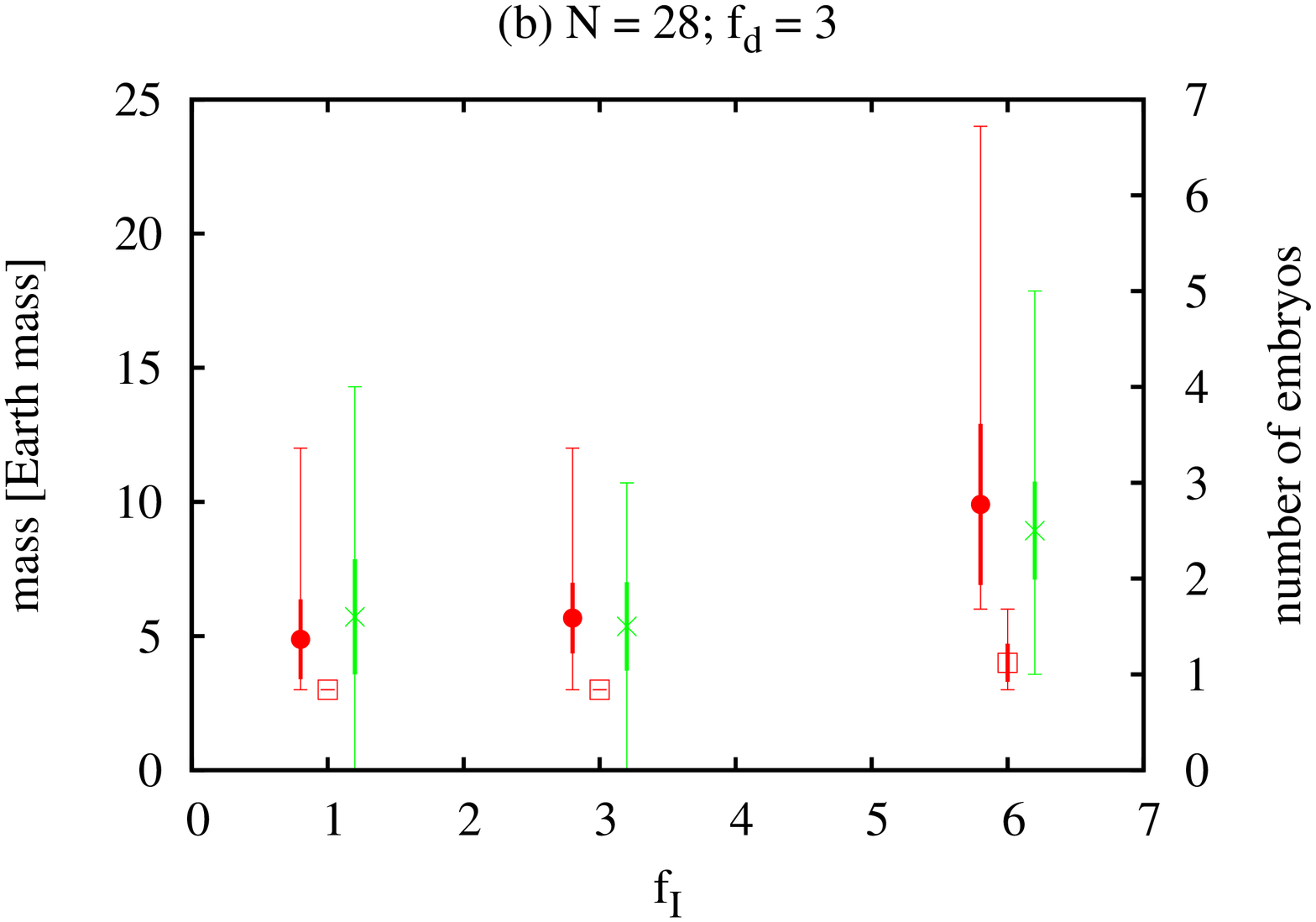}
    \includegraphics[width=6.5cm, angle=-90]{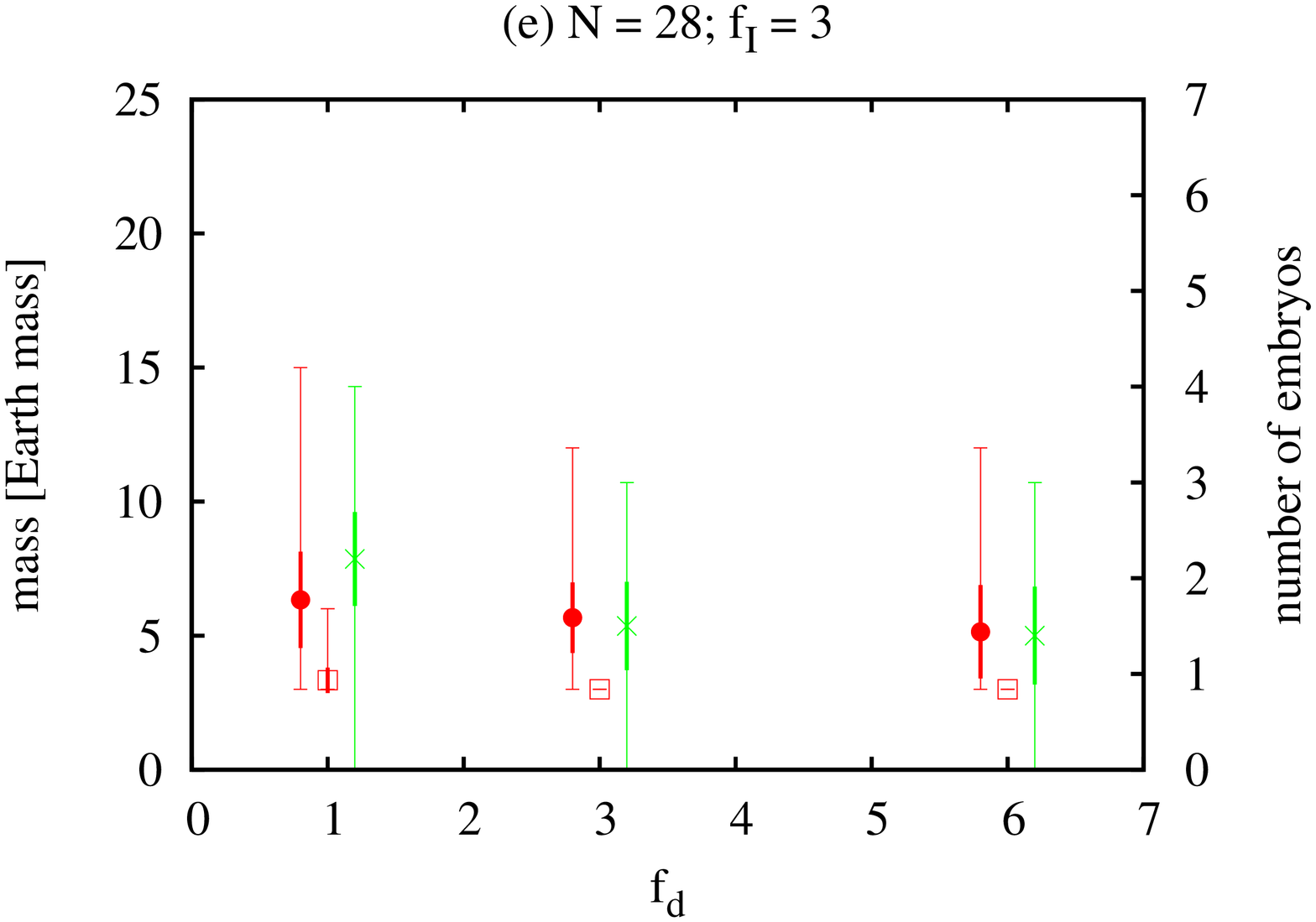}}
\centerline{\includegraphics[width=6.5cm, angle=-90]{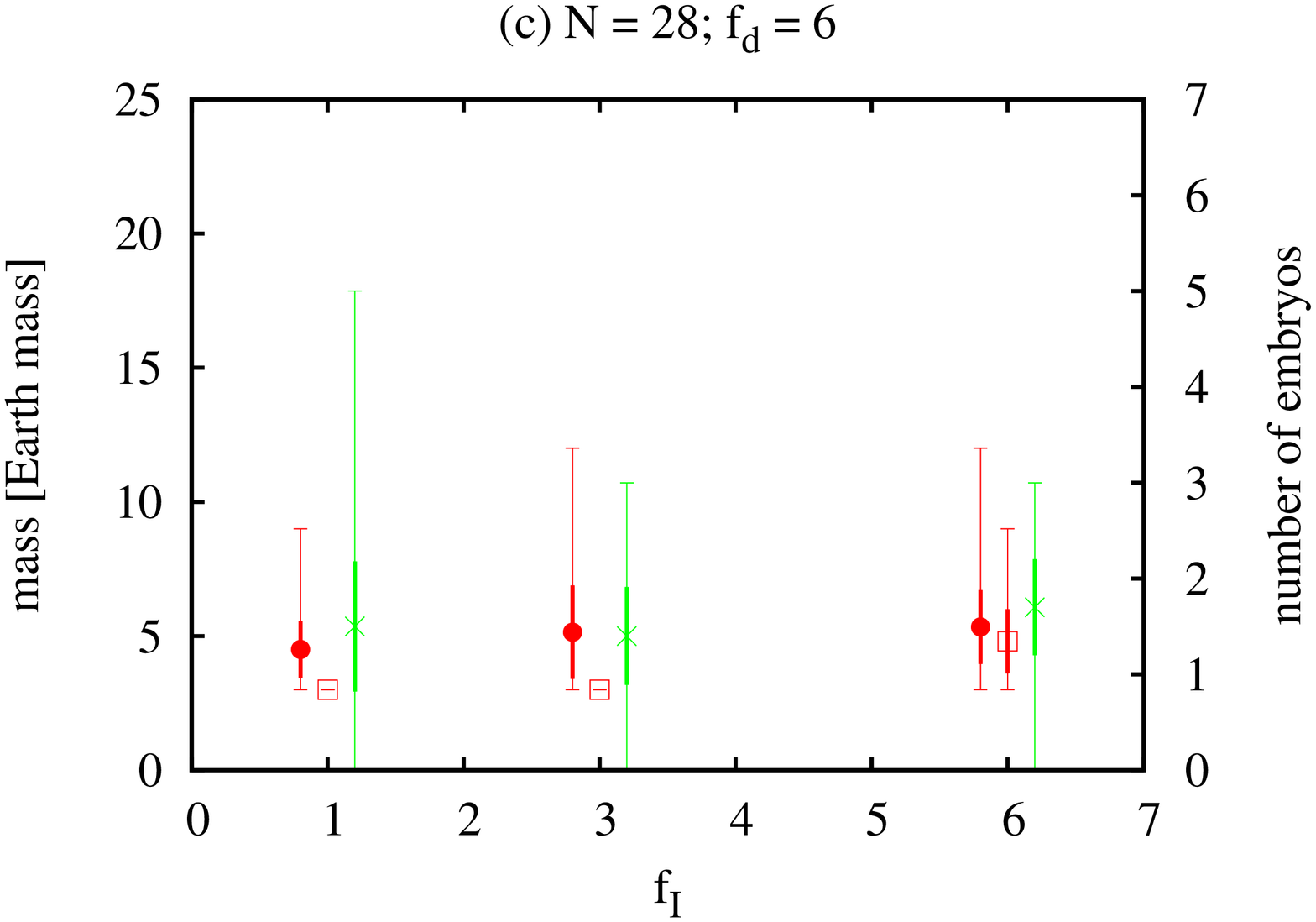}
    \includegraphics[width=6.5cm, angle=-90]{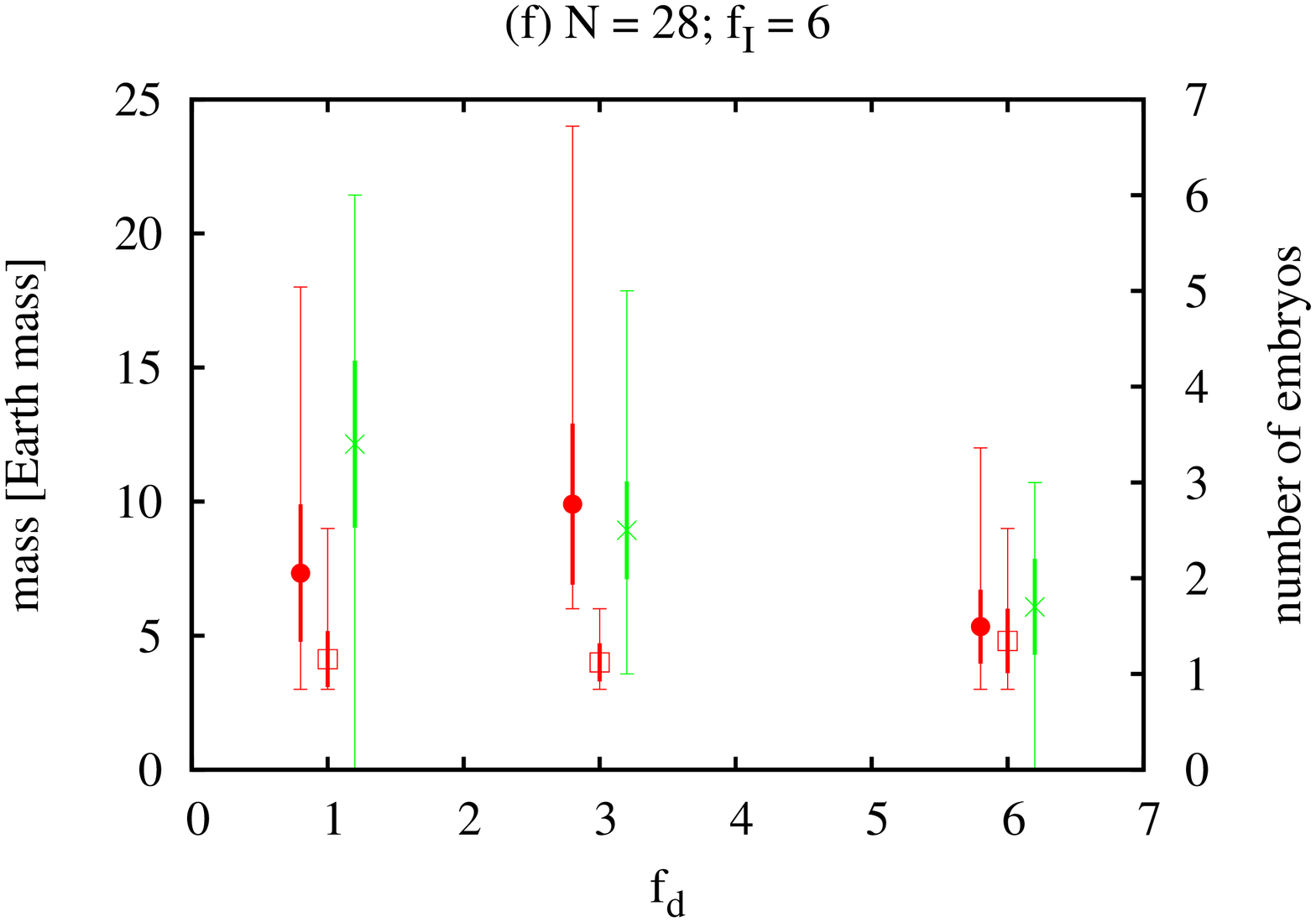}}
 \caption{
Statistical analysis of the results of the simulations starting
from 28 embryos, each with $3\,$M$_{\oplus}$. 
The $x$-axis of left plots reports the value of
$f_{I}$ at fixed $f_{d}$, given at the title of each plot. Similarly,
the $x$-axis of right plots reports the value of $f_{d}$ at fixed $f_{I}$.
The dot, slightly displaced to the left, is the mean mass of the largest core
 surviving beyond Saturn, computed over the corresponding 10 simulations.
The thick vertical bar is the rms deviation of this quantity.
The thin bar shows the excursion of the same quantity from
minimum to maximum. The square, in the middle, and related bars are
the same, but for the second largest core. The mass scale is reported
on the left of the diagram. The cross, slightly displaced to the right,
reports the mean number of embryos/cores surviving beyond Saturn,
to be read against the scale on the right hand side. Again, the thick bar
is for the rms deviation and the thin bar for the minimum-maximum quantity.
(For the discussion - see Sect.~4).}
\label{smFIG2}
\end{figure*}

\newpage

\begin{figure*}
\vspace{-0.5cm}
\centerline{\includegraphics[width=6.5cm, angle=-90]{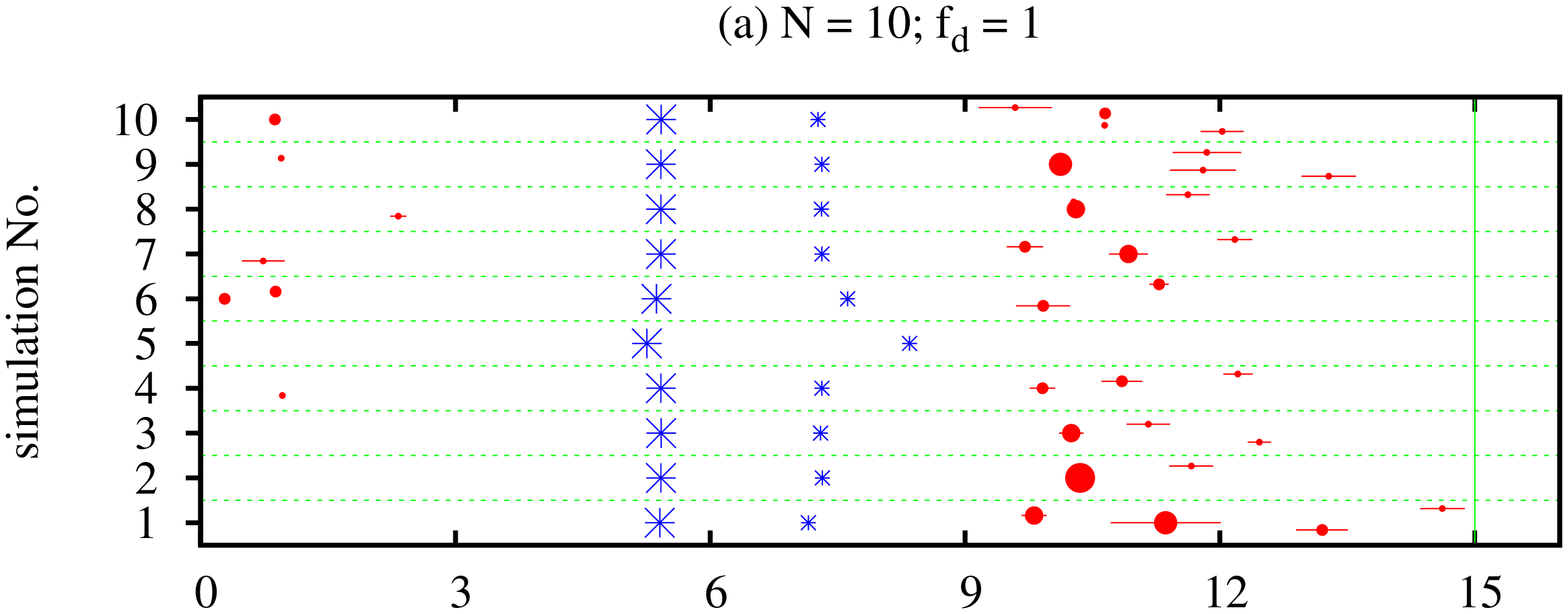}
    \includegraphics[width=6.5cm, angle=-90]{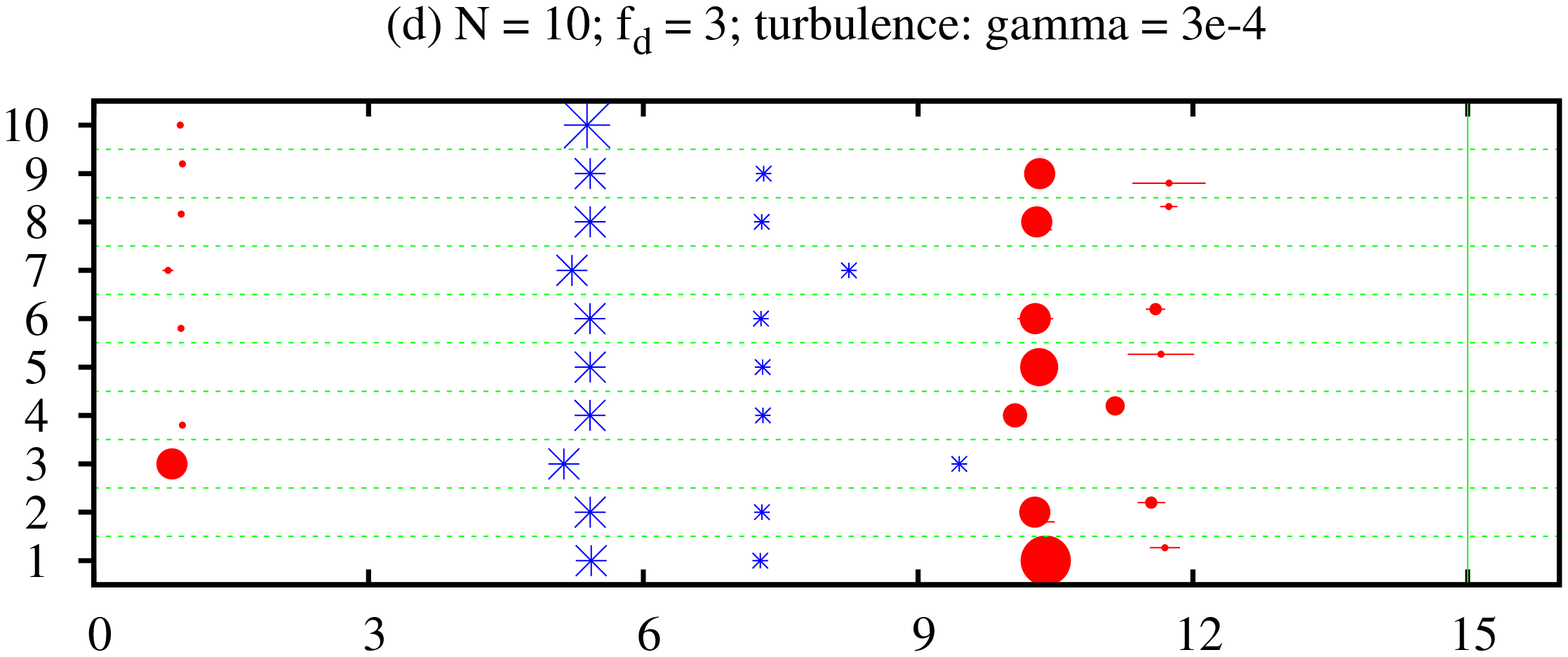}}
\vspace{-2.7cm}
\centerline{\includegraphics[width=6.5cm, angle=-90]{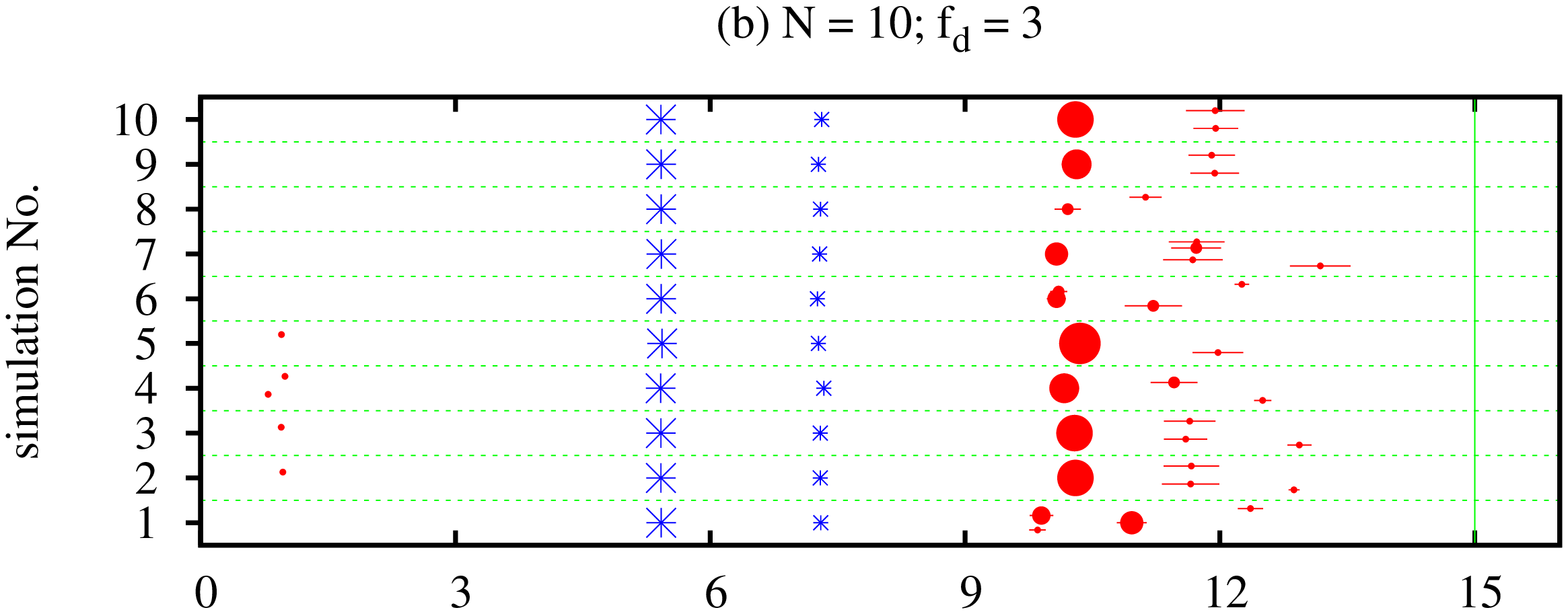}
    \includegraphics[width=6.5cm, angle=-90]{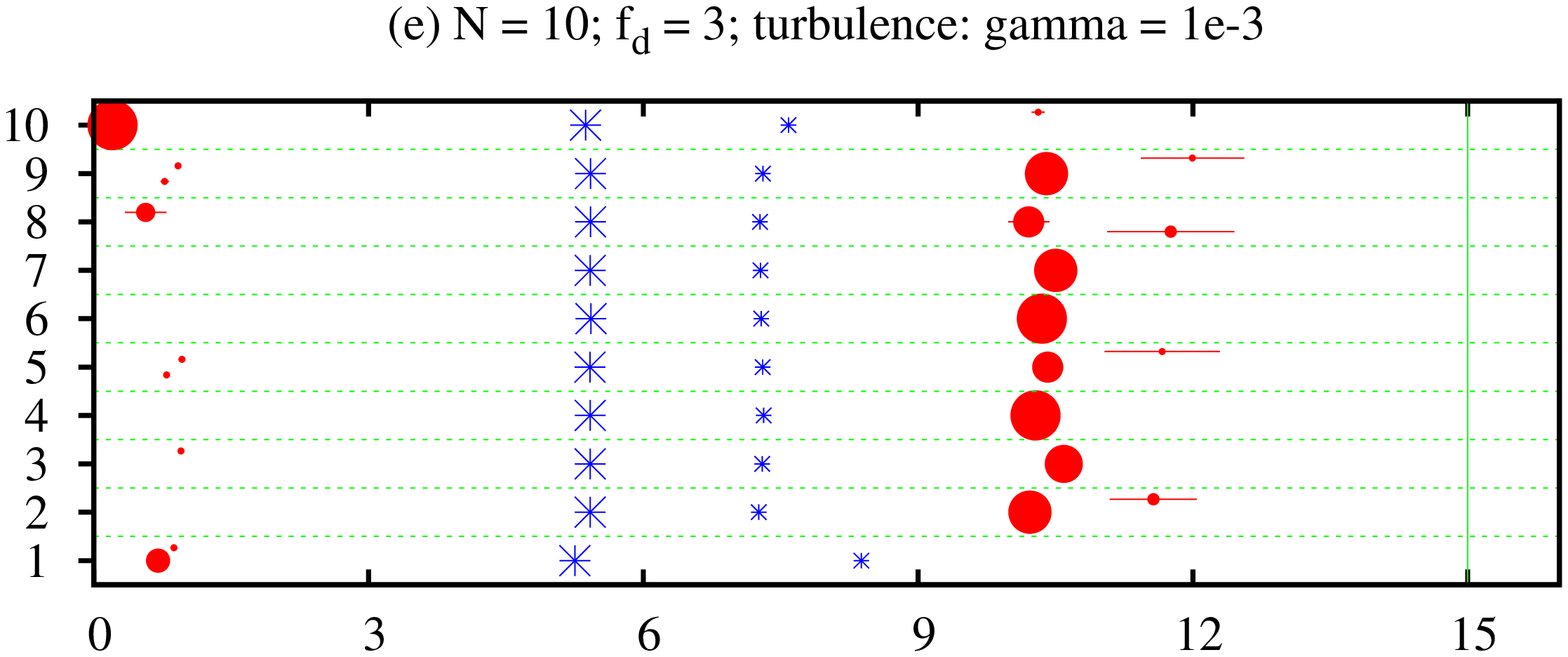}}
\vspace{-2.7cm}
\centerline{\includegraphics[width=6.5cm, angle=-90]{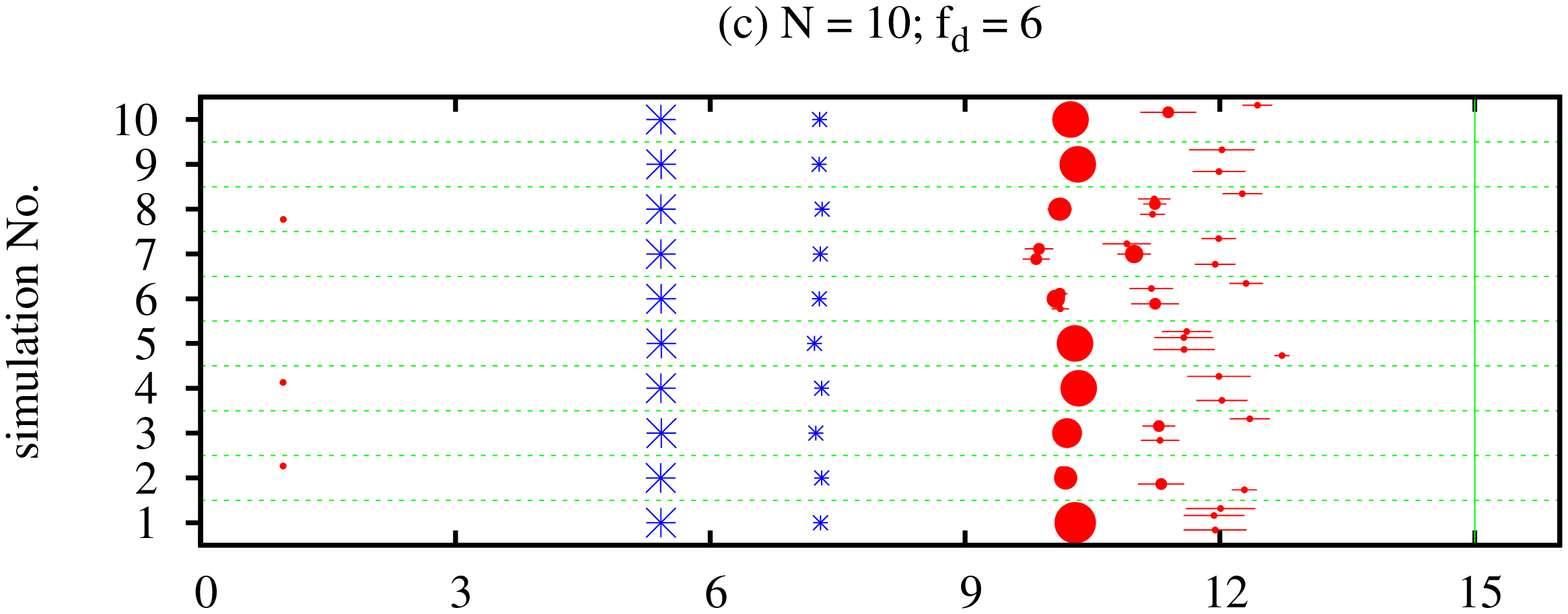}
    \includegraphics[width=6.5cm, angle=-90]{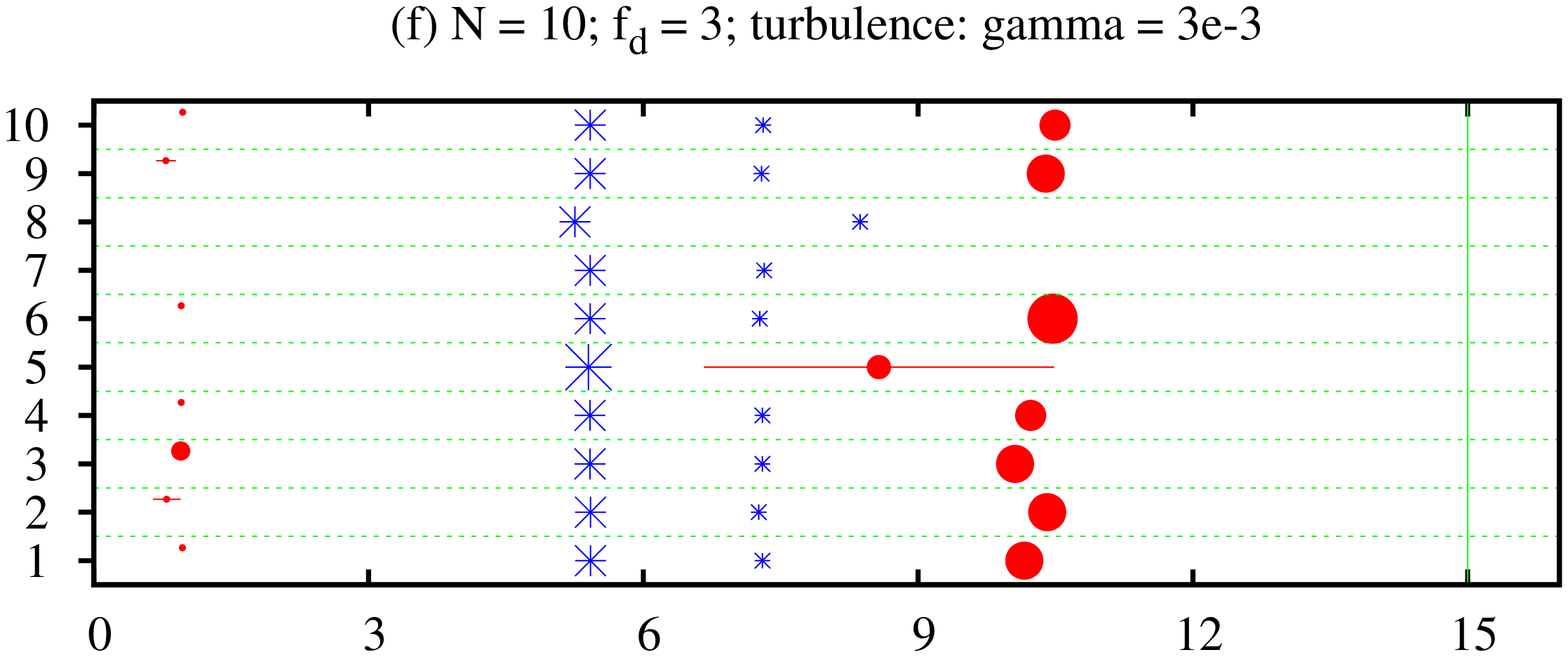}}
\vspace{-1.5cm}
\centerline{\includegraphics[width=6.5cm, angle=-90]{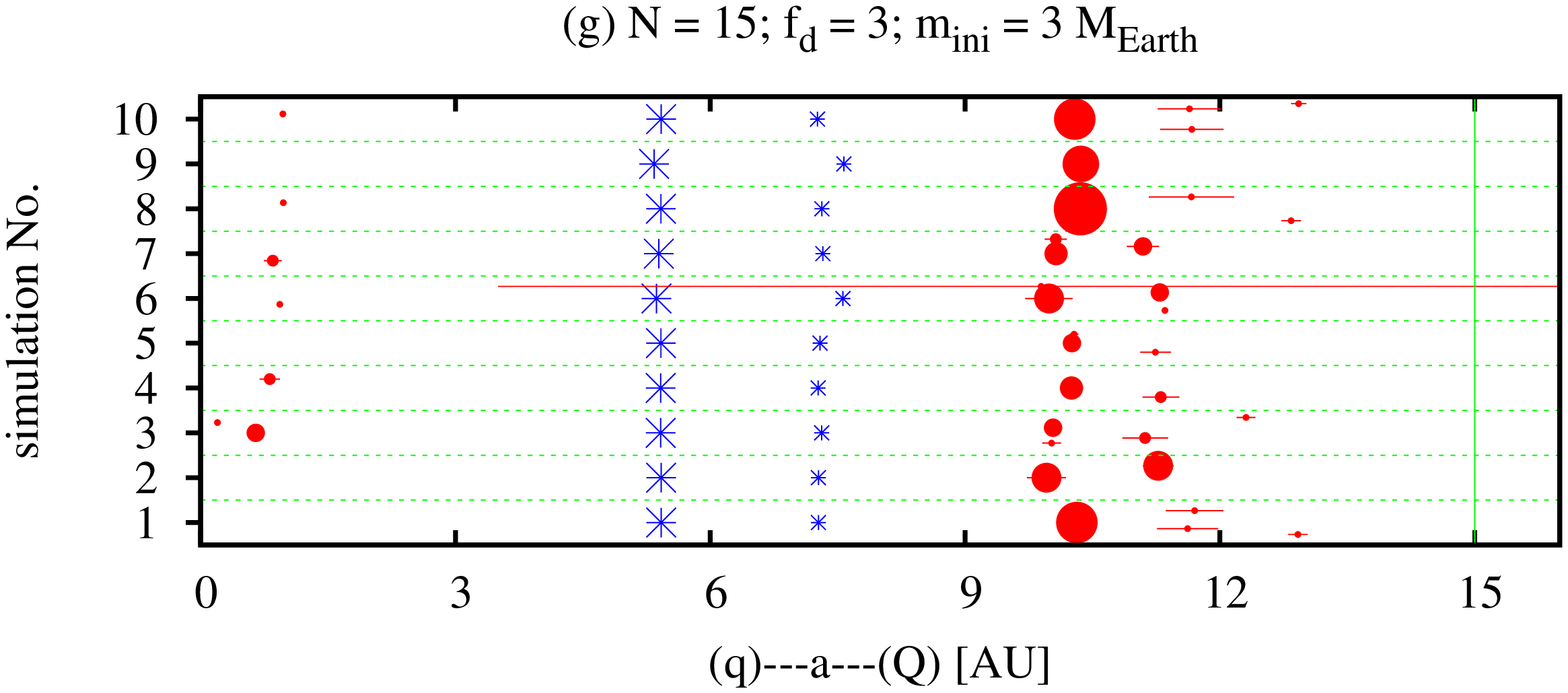}
    \includegraphics[width=6.5cm, angle=-90]{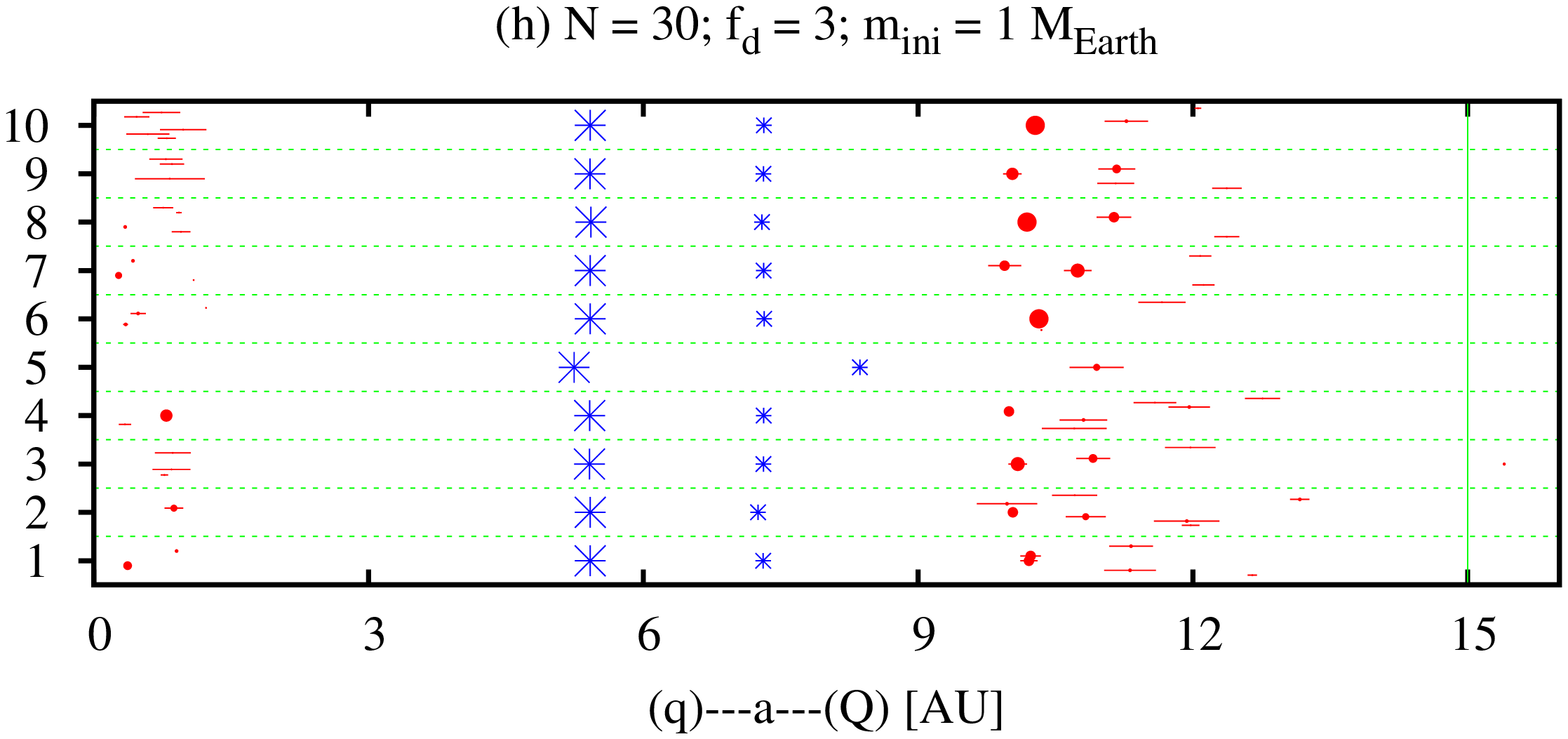}}
\caption{
The same as Fig.~A1 but for the simulations
in which the planet trap was considered. 
The plots (a,b,c) show the end-states starting with initial embryo-mass 
of $3\,$M$_{\oplus}$ when the turbulence is not taken into account, 
plots (d,e,f) are related to the cases with the turbulence  
characterized by $\gamma = 3\times10^{-4}$, $1\times10^{-3}$, and $3\times10^{-3}$, 
respectively. At the start, 10 embryos are assumed in these simulations.
Plot (g) shows the end-states of simulations starting with a higher 
total mass (15 embryos of $3\,$M$_{\oplus}$) and plot (h) does this for 
a larger number of initally less massive embryos (30 embryos of $1\,$M$_{\oplus}$).
(For the discussion - see Sect.~5).
}
\label{smFIG3}
\end{figure*}

\newpage

\begin{figure*}
\centerline{\includegraphics[width=6.5cm, angle=-90]{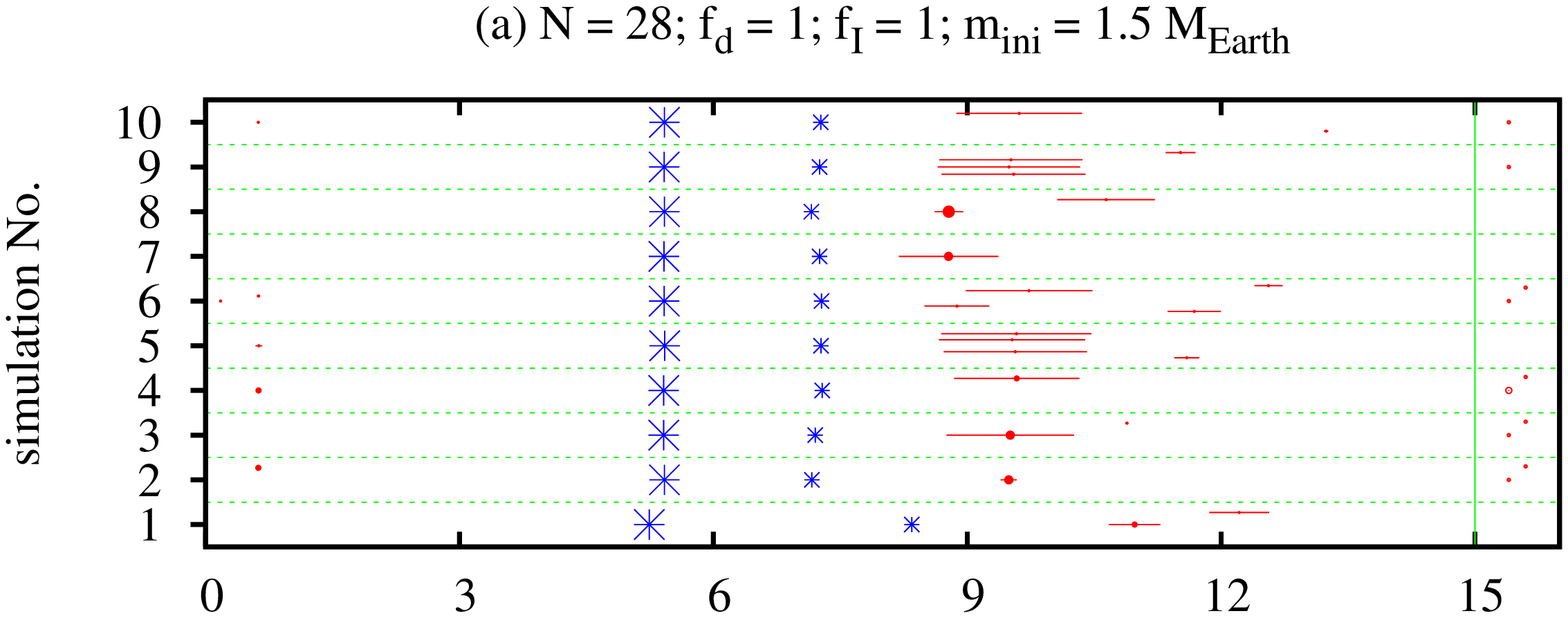}
    \includegraphics[width=6.5cm, angle=-90]{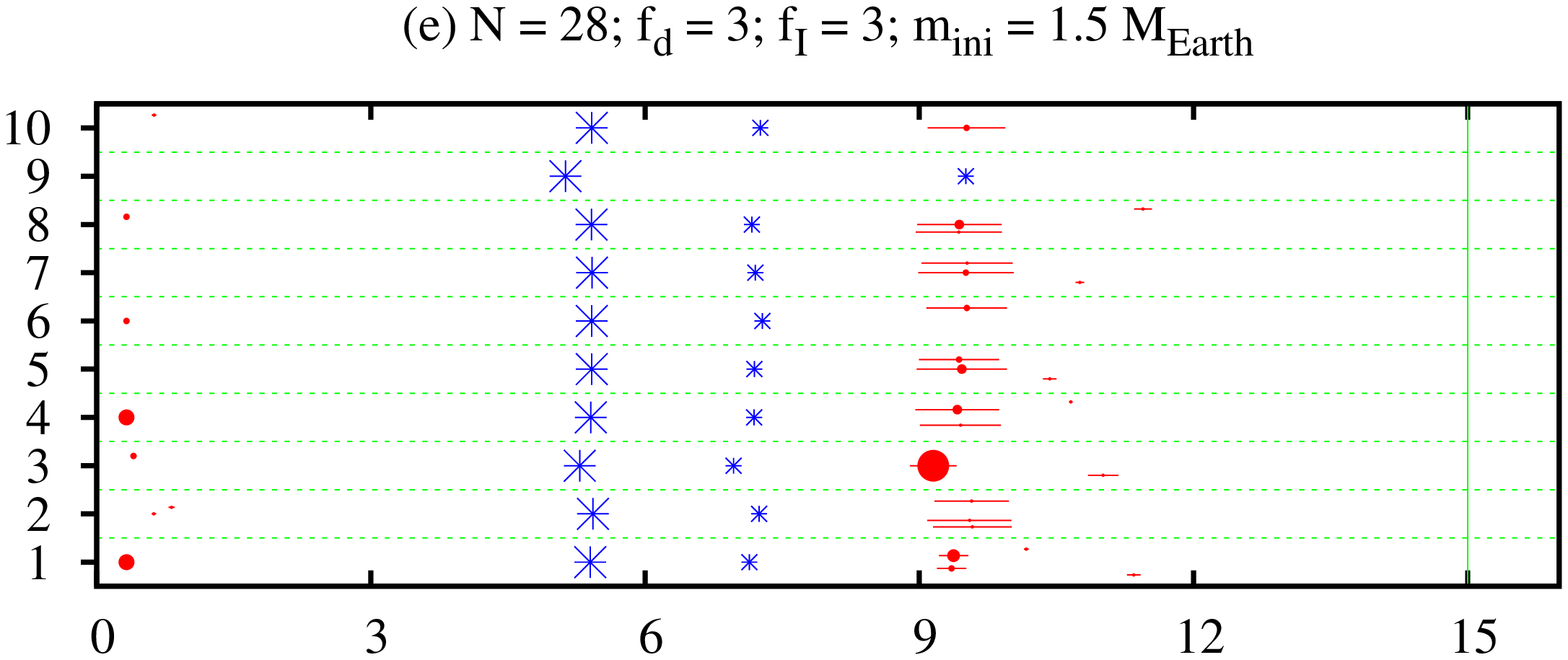}}
\vspace{-2.7cm}
\centerline{\includegraphics[width=6.5cm, angle=-90]{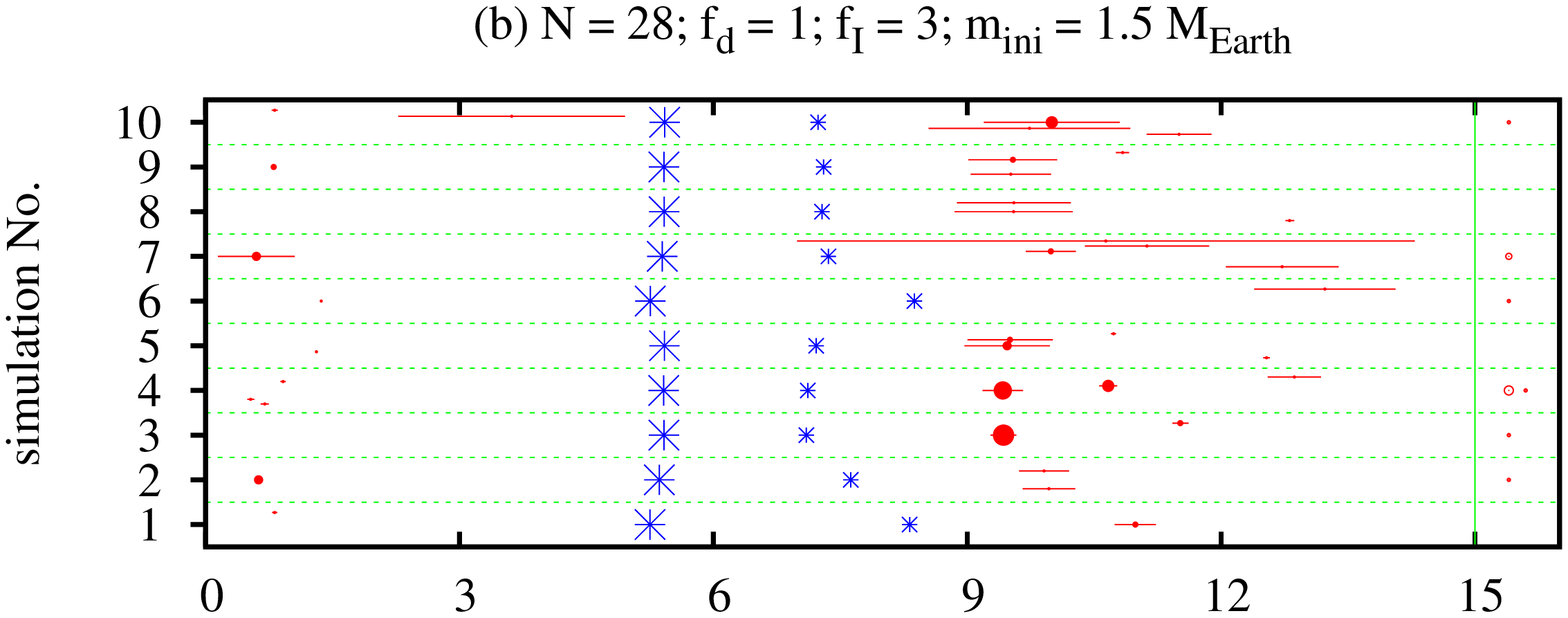}
    \includegraphics[width=6.5cm, angle=-90]{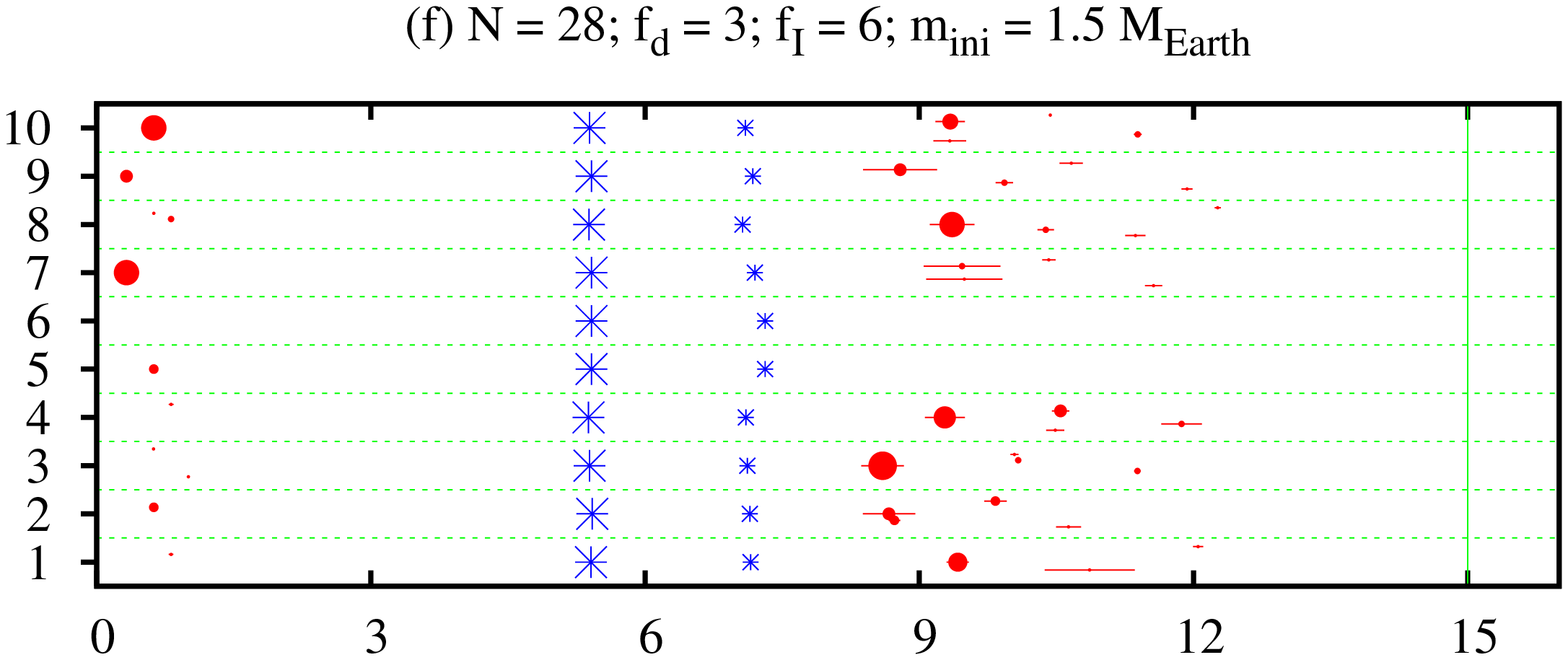}}
\vspace{-2.7cm}
\centerline{\includegraphics[width=6.5cm, angle=-90]{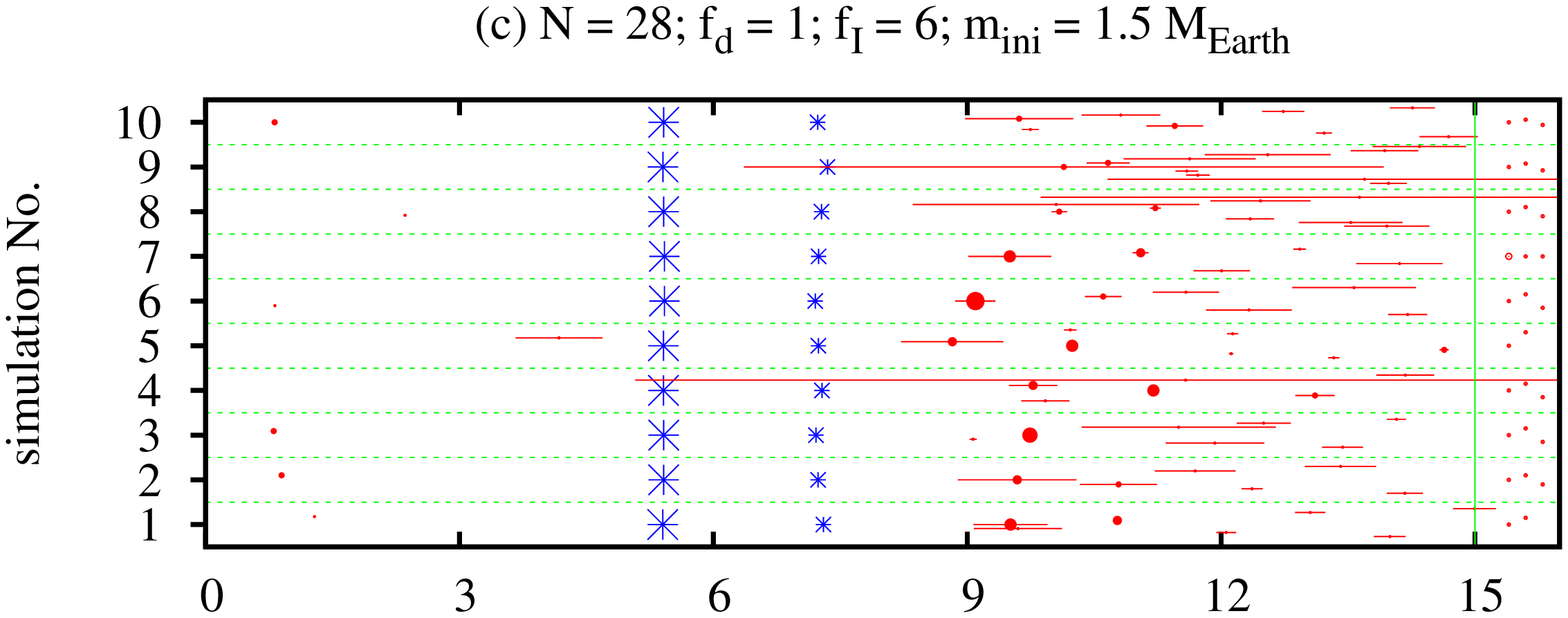}
    \includegraphics[width=6.5cm, angle=-90]{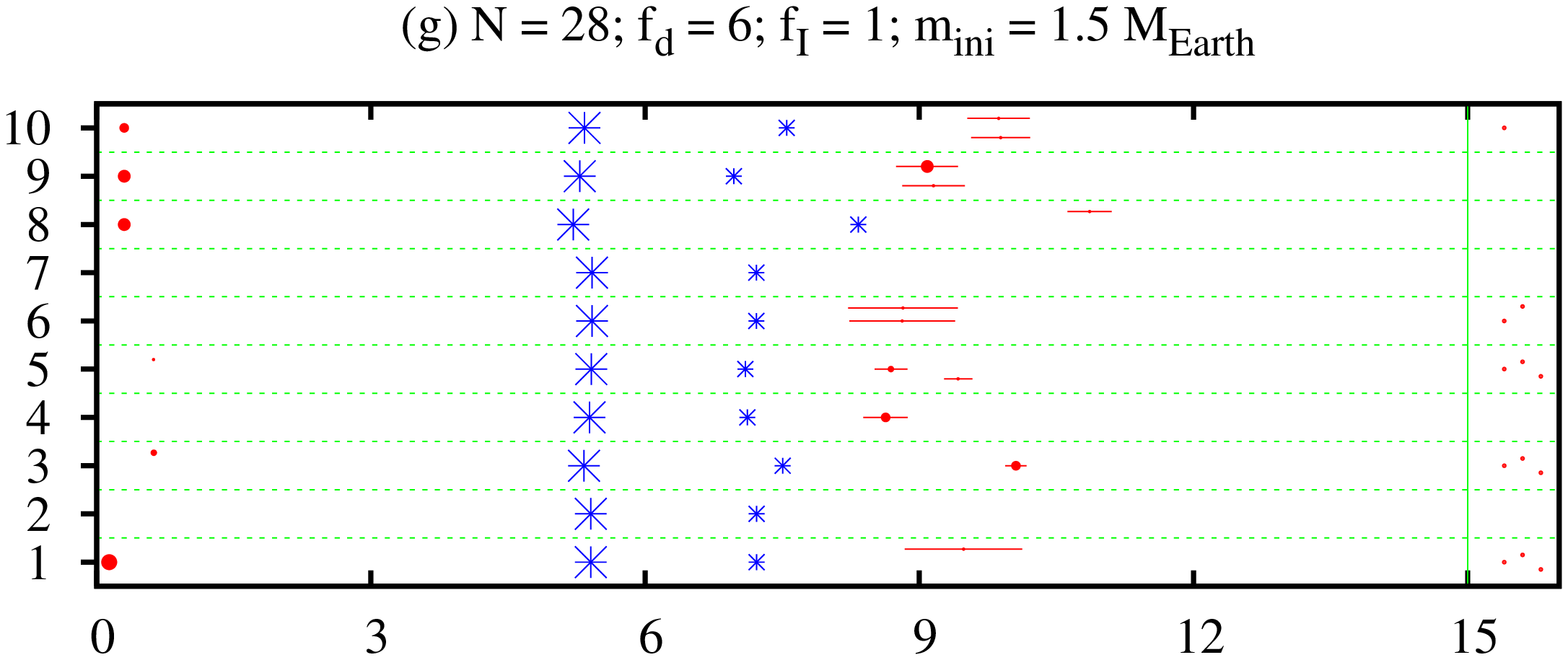}}
\vspace{-2.7cm}
\centerline{\includegraphics[width=6.5cm, angle=-90]{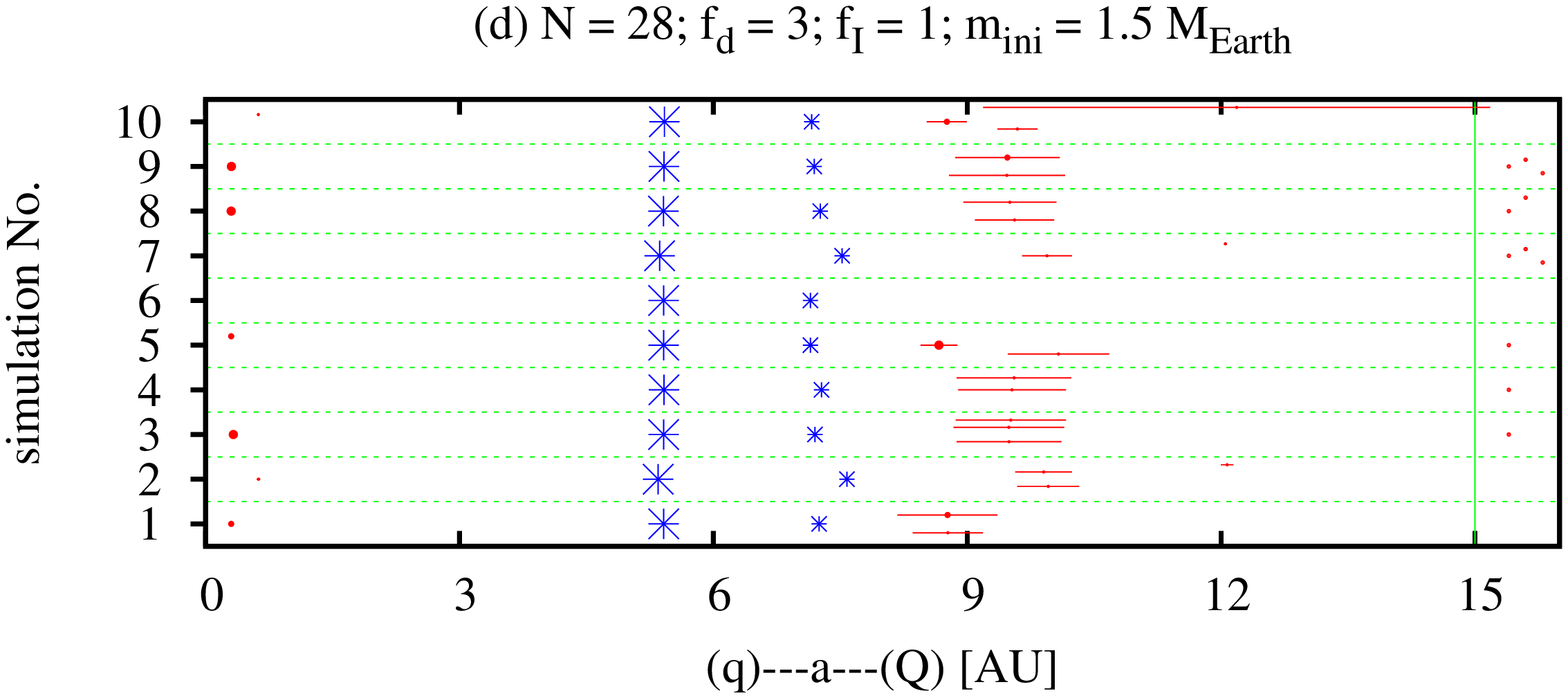}
    \includegraphics[width=6.5cm, angle=-90]{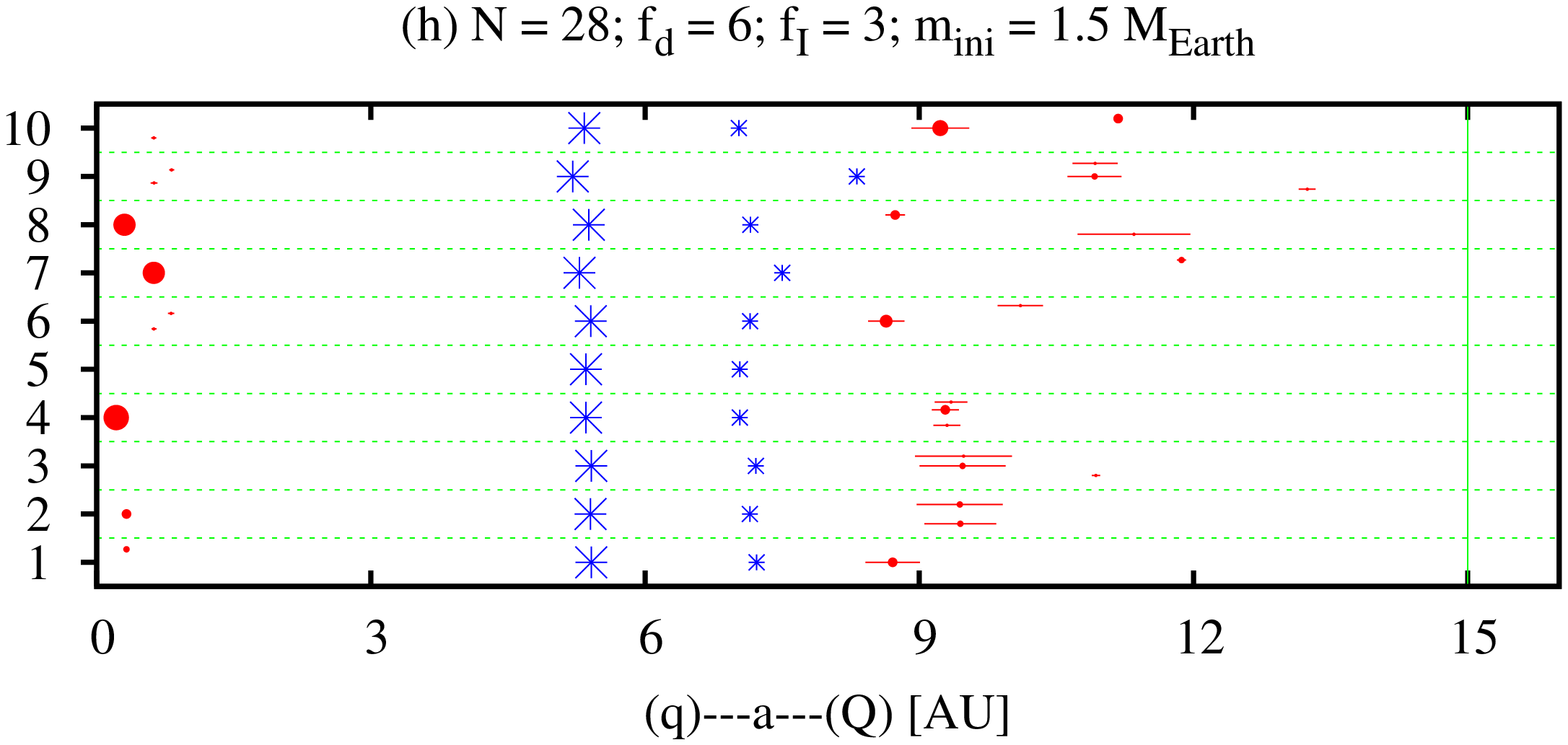}}
\vspace{-2.7cm}
\centerline{\includegraphics[width=6.5cm, angle=-90]{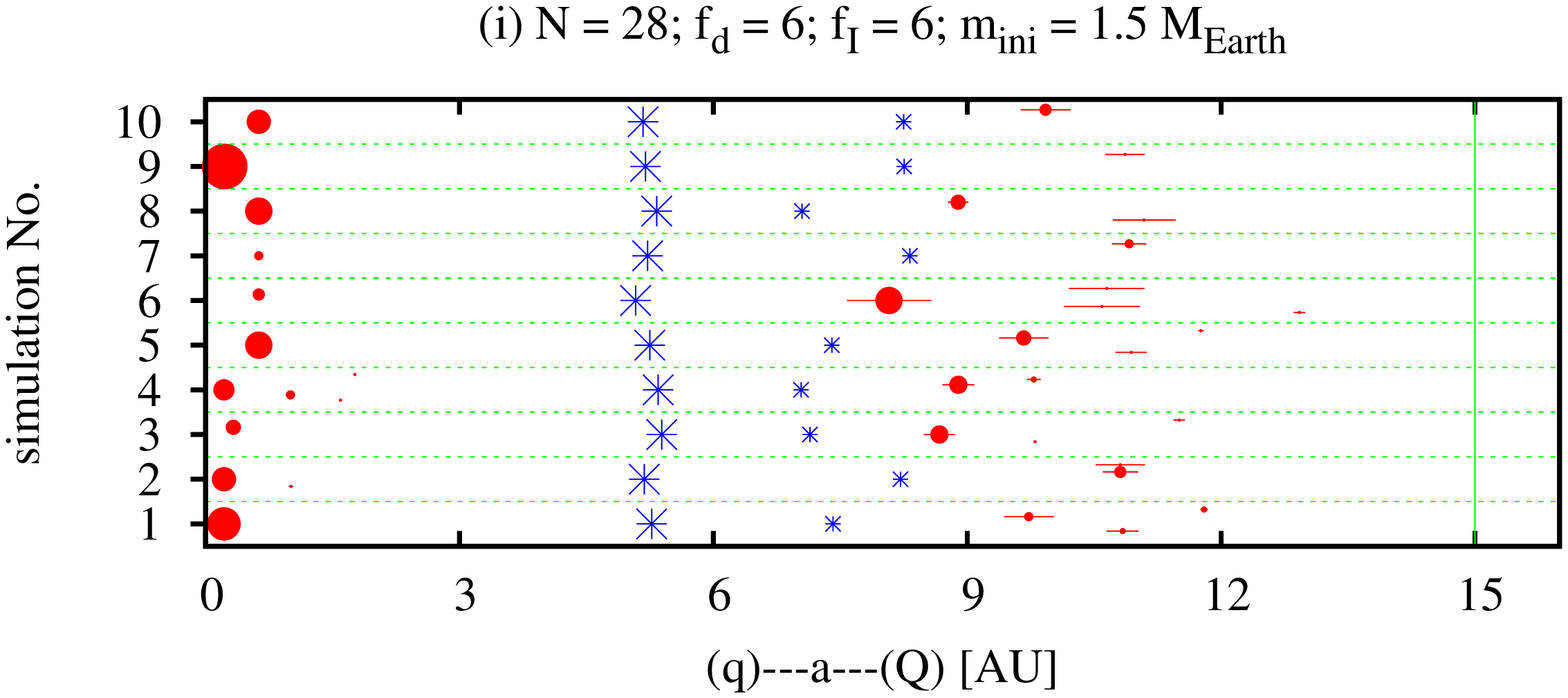}}
\vspace{-1 cm}
\caption{The same as Fig.~A3 but with the initial mass of 
each embryo of $1.5\,$M$_{\oplus}$.
(For the discussion - see Sect.~6.1).}
\label{smFIG4}
\end{figure*}

\newpage

\begin{figure*}
\centerline{\includegraphics[width=6.5cm, angle=-90]{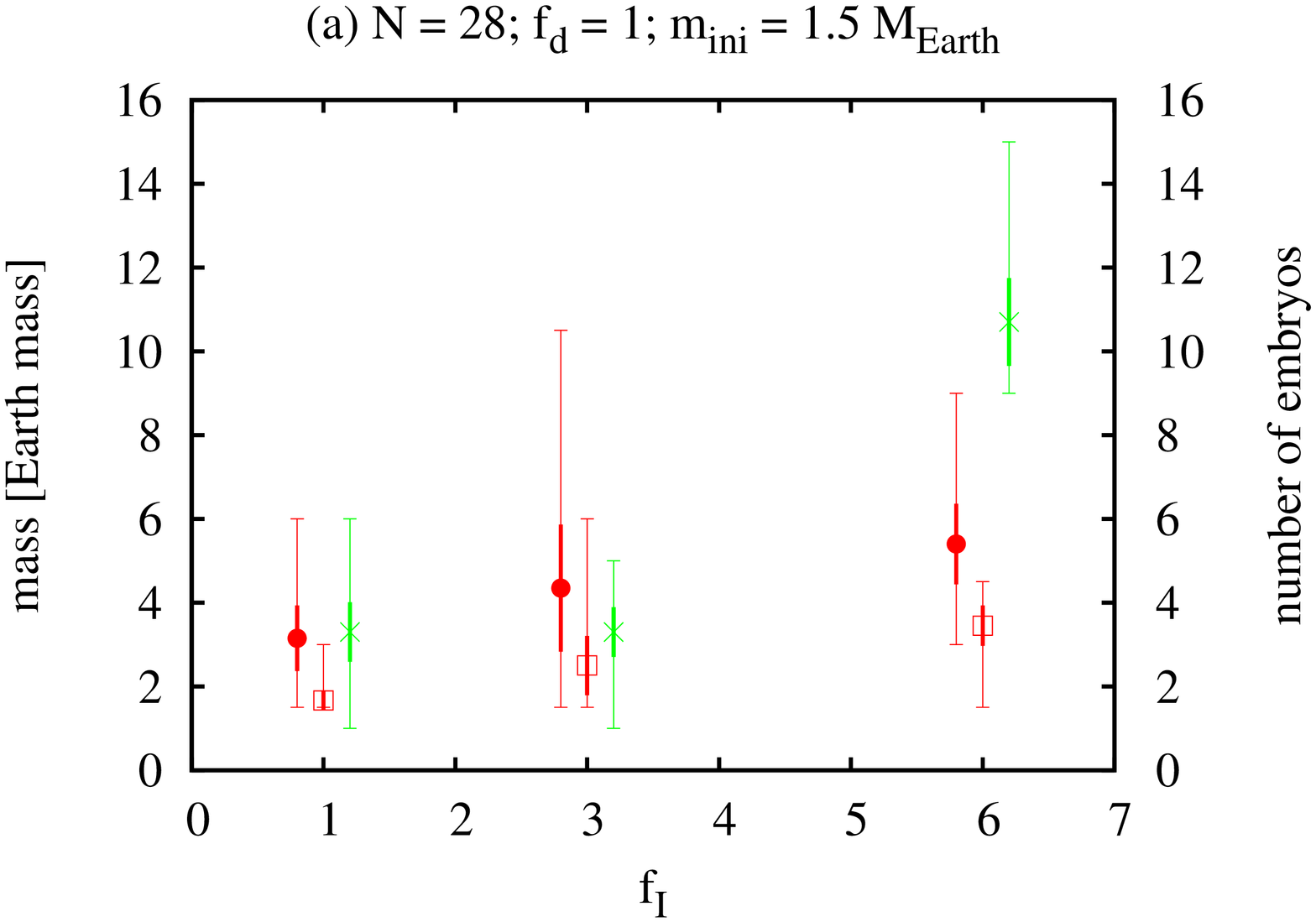}
    \includegraphics[width=6.5cm, angle=-90]{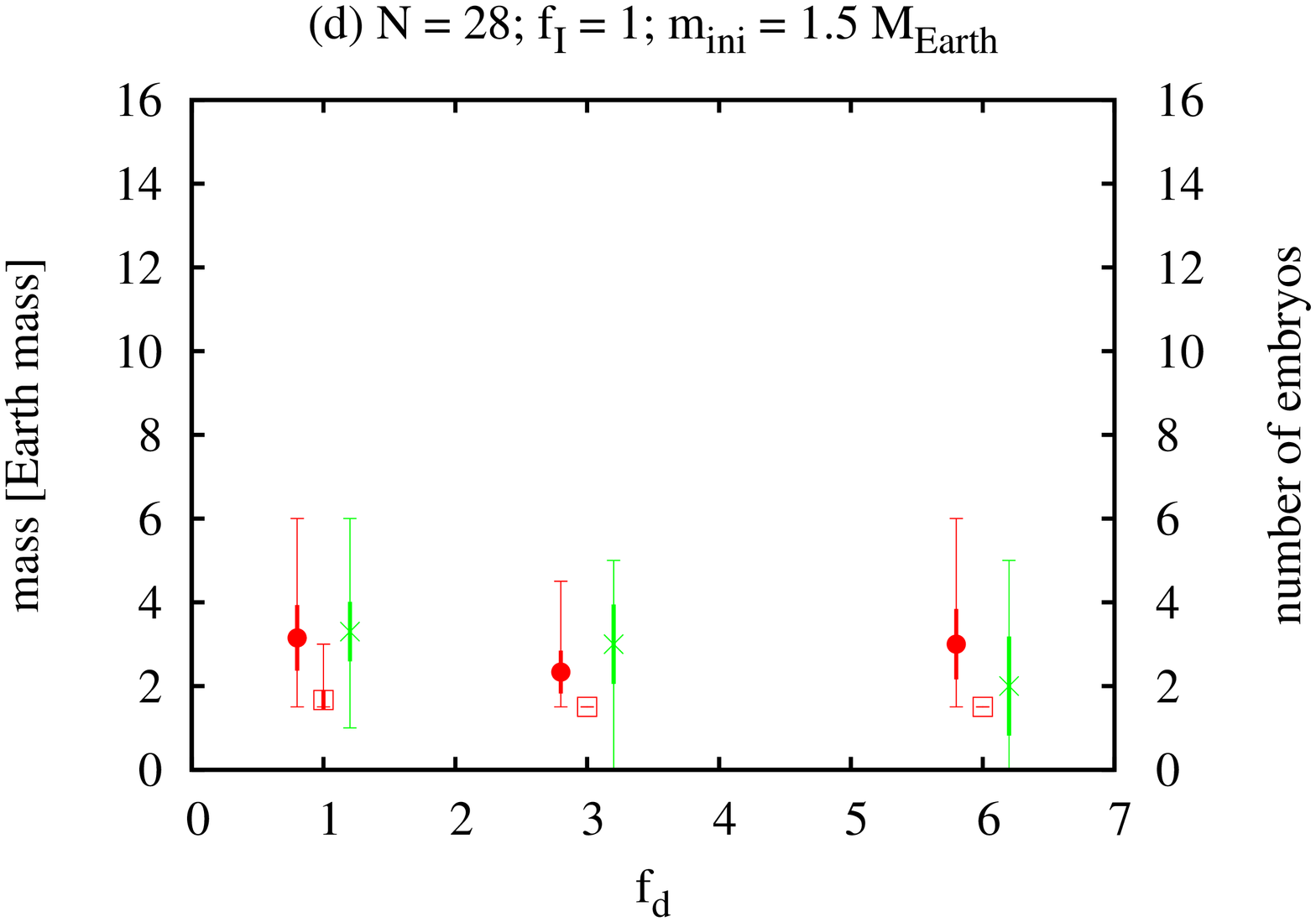}}
\centerline{\includegraphics[width=6.5cm, angle=-90]{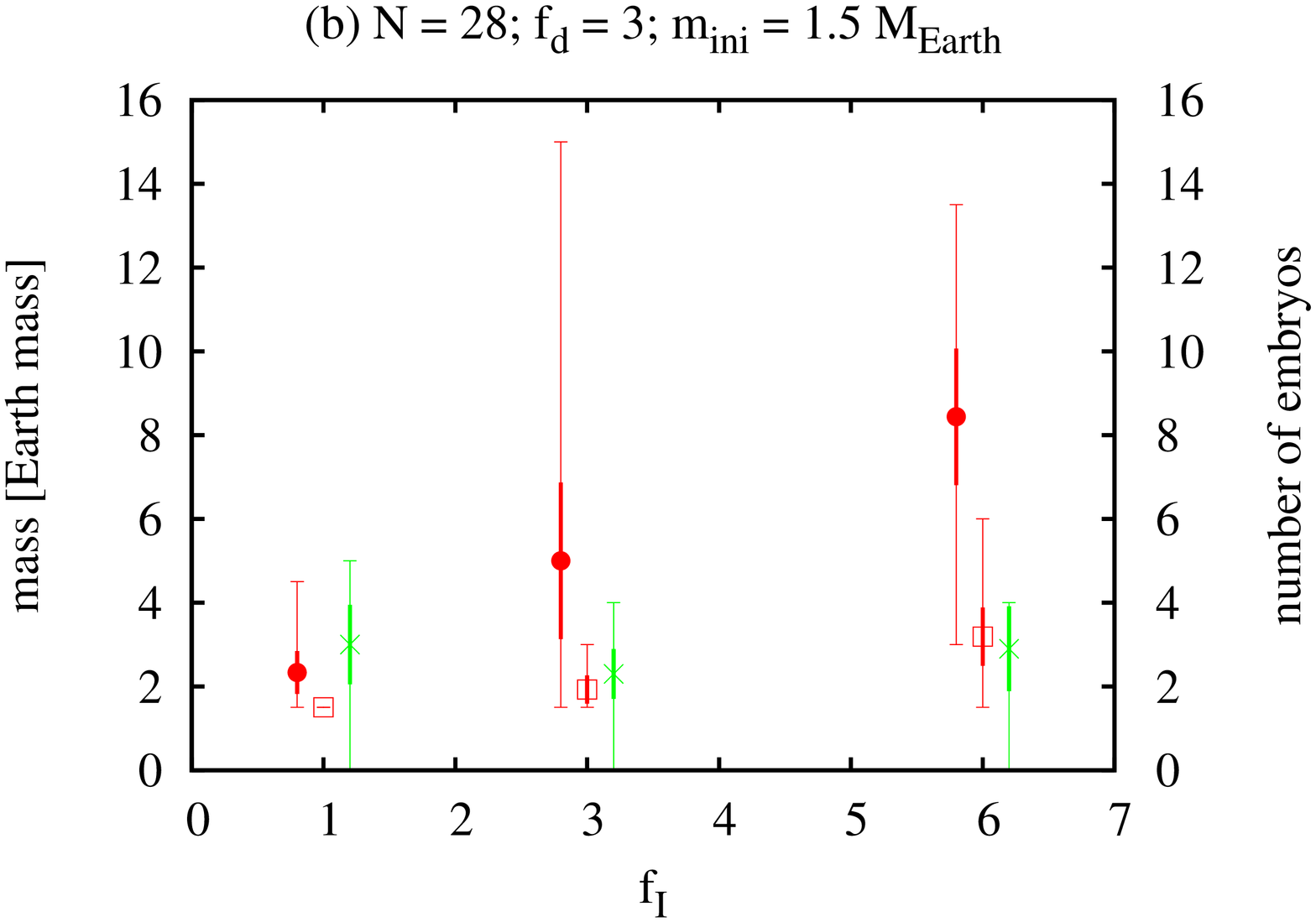}
    \includegraphics[width=6.5cm, angle=-90]{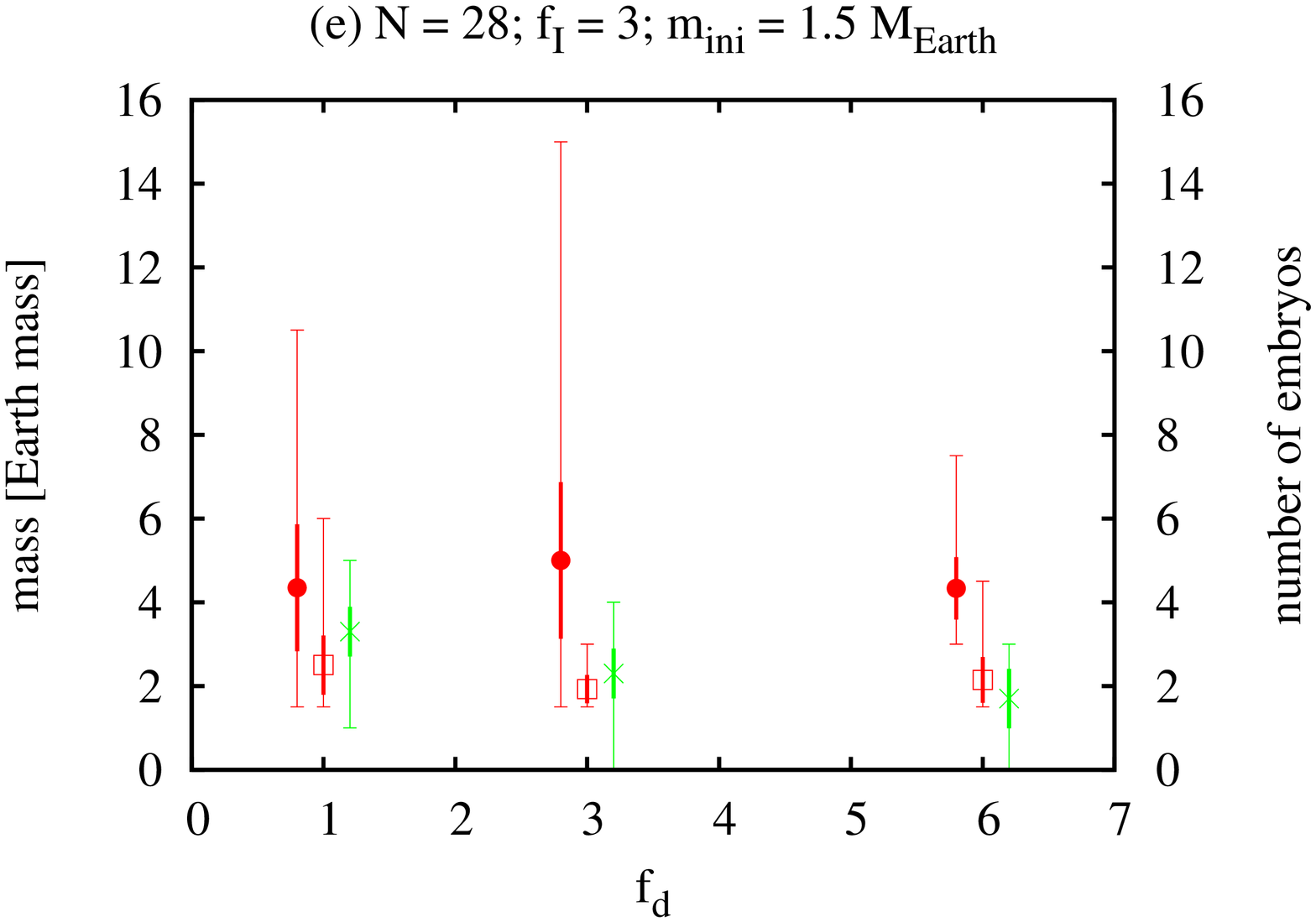}}
\centerline{\includegraphics[width=6.5cm, angle=-90]{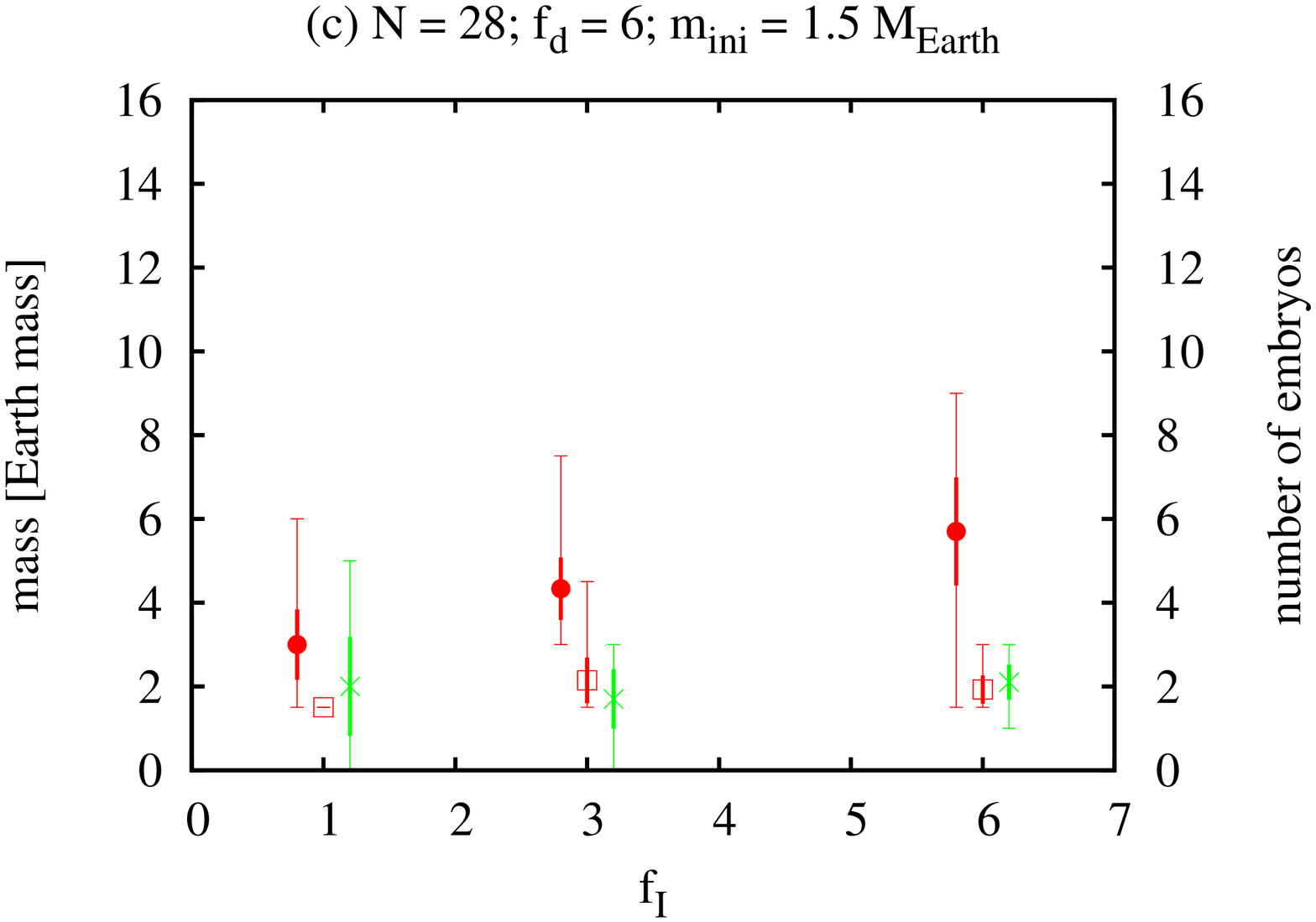}
    \includegraphics[width=6.5cm, angle=-90]{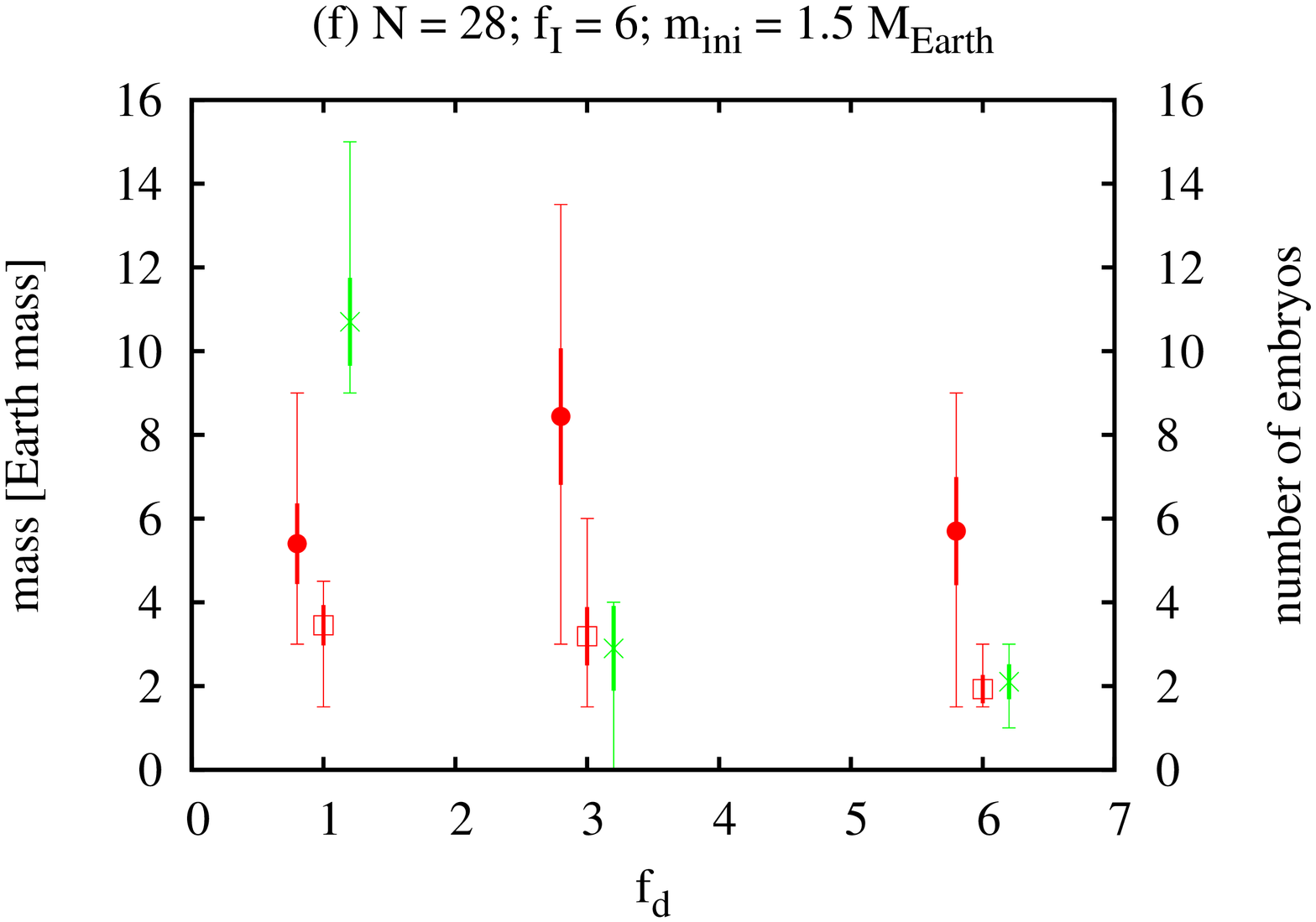}}
\caption{The same as Fig.~A2 but with the initial mass of
each embryo of $1.5\,$M$_{\oplus}$.
(For the discussion - see Sect.~6.1).
}
\label{smFIG5}
\end{figure*}

\end{document}